\documentclass[useAMS,usenatbib]{mn2e}

\def\Msol{\hbox{M$_\odot$}}
\def\Zsol{\hbox{Z$_\odot$}}

\def\kms{\hbox{km$\,$s$^{-1}$}}
\def\cmt{\hbox{cm$^{-3}$}}

\def\one{\,{\sc i}}             
\def\two{\,{\sc ii}}
\def\three{\,{\sc iii}}
\def\four{\,{\sc iv}}
\def\five{\,{\sc v}}

\usepackage{amsmath}
\usepackage{amssymb}
\usepackage{mathrsfs}
\usepackage{color}
\usepackage{soul} 
\usepackage{upgreek} 
\usepackage{graphicx}
\usepackage{overpic}
\usepackage[compatibility=false]{caption}

\usepackage{rotating}

\defcitealias{bendo09}{Paper~I}
\title[Local ULIRG feedback with VLT/VIMOS]{Spatially resolved observations of warm ionized gas and feedback in local ULIRGs\thanks{Based on observations collected at the European Organisation for Astronomical Research in the Southern Hemisphere, Chile under programmes 080.B-0085 and 383.B-0372.}}
\author[M.S.\ Westmoquette et al.] {M.\ S.\ Westmoquette$^1$\thanks{E-mail: mwestmoq@eso.org}, D.\ L.\ Clements$^{2}$, G.\ J.\ Bendo$^{2,3}$, and S.\ A.\ Khan$^{2}$ \\
$^1$European Southern Observatory, Karl-Schwarzschild-Str. 2, 85748 Garching bei M\"{u}nchen, Germany\\
$^2$Astrophysics Group, Imperial College London, Blackett Laboratory, London SW7 2AZ\\
$^3$UK ALMA Regional Centre Node, Jodrell Bank Centre for Astrophysics, School of Physics and Astronomy, University of \\Manchester, Oxford Road, Manchester M13 9PL\\
}
\date{Accepted. Received; in original form}
\pagerange{\pageref{firstpage}--\pageref{lastpage}}
\pubyear{2012}
\begin{document}
\maketitle
\label{firstpage}
\begin{abstract}
We present VLT/VIMOS-IFU emission-line spectroscopy of a volume limited sample of 18 southern ULIRGs selected with $z < 0.09$ and $\delta < 10$. By covering a wide range of ULIRG types, including many systems that have received very little previous attention, this dataset provides an important set of templates for comparison with high-redshift galaxies. We employed an automated Gaussian line fitting program to decompose the emission line profiles of H$\alpha$, [N\two], [S\two], and [O\one] into individual components, and chart the H$\alpha$ kinematics, and the ionized gas excitations and densities. 11/18 of our galaxies show evidence for outflowing warm ionized gas with speeds between 500 and a few 1000~\kms, with the fastest outflows associated with systems that contain an AGN. Our spatially resolved spectroscopy has allowed us to map the outflows, and in some cases determine for the first time to which nucleus the wind is associated. In three of our targets we find line components with widths $>$2000~\kms\ over spatially extended regions in both the recombination and forbidden lines; in two of these three, they are associated with a known Sy2 nucleus. Eight galaxies have clear rotating gaseous disks, and for these we measure rotation velocities, virial masses, and calculate Toomre $Q$ parameters. We find radial gradients in the emission line ratios in a significant number of systems in our study.
We attribute these gradients to changes in ionizing radiation field strength, most likely due to an increasing contribution of shocks with radius. We conclude with a detailed discussion of the results for each individual system, with reference to the existing literature.

Our observations demonstrate that the complexity of the kinematics and gas properties in ULIRGs can only be disentangled with high sensitivity, spatially resolved IFU observations. Many of our targets are ideal candidates for future high spatial resolution follow-up observations.
\end{abstract}


\section{Introduction}\label{intro}
Ultra-Luminous InfraRed Galaxies (ULIRGs) are the most IR luminous population in the local universe (Sanders \& Mirabel 1996). While rare locally, the appearance of large numbers of luminous objects in far-IR/submm blank field surveys \citep[e.g.,][]{smail97, hughes98, barger98, eales00, coppin06} suggests that they were more important in the high-$z$ universe. The process that triggers ULIRG activity appears to be a major merger between two disk galaxies \citep{clements96, duc97, borne99, borne00}, which consequently initiates a (at least partially heavily obscured) starburst and/or AGN. In fact ULIRGs are implicated as hosting $\sim$50\% the star formation in the universe at $z\sim1$ and above \citep[][Clements et al.\ in prep.]{perez05, le-floch05, bethermin11}. Since most massive galaxies are thought to have experienced at least one major merger \citep{man11}, it is therefore essential to study the physical mechanisms at work in local ULIRGs where the best spatial resolution is achievable.

Hydrodynamic simulations predict that as galaxies collide and coalesce, gravitational forces and dissipative shocks funnel gas toward the centre, triggering a nuclear starburst and fuelling the growth of the central black hole \citep{barnes92, cox08}. The end product of such a ``wet'' (i.e.\ gas-rich) merger is likely to be an elliptical galaxy, as predicted by theory \citep{barnes96} and found by observations of local ULIRGs \citep[e.g.,][]{wright90, genzel01, dasyra06b}. The high IR luminosities of ULIRGs therefore arise through the dust re-processing of UV radiation from the concomitant starburst/AGN.
Recent studies, however, are challenging this idea \citep[e.g.,][]{forster09, chien10, cisternas11}, and the emerging view is that merger-induced starbursts will often have both a nuclear and extended component in the form of massive star-forming clumps in the disk \citep{bournaud11}.

Although some ULIRGs are known to host an AGN, their role, and the relationship between ULIRGs and quasars, is still unclear \citep{alexander05}. The prevalence of optically visible AGNs increases with increasing infrared luminosity \citep{veilleux99} and merger stage \citep{veilleux02}, supporting the canonical view that the dust obscuring an AGN is removed through feedback processes, revealing an optical quasar \citep{sanders88}. However, this evolutionary sequence has also been brought into question by recent work that, for example, shows that many ULIRGs are evolving into ellipticals of lower mass than known quasar hosts \citep{genzel01, veilleux02, tacconi02, dasyra06a, dasyra06b, dasyra07}. Evidence for feedback that is clearly AGN-driven in LIRGs or ULIRGs has thus far remained elusive \citep{rupke05b, rupke11, muller-sanchez11}

Nevertheless, winds with significant mass and energy flux have been observed in ULIRGs \citep{clements02, martin05, martin06, rupke05a, rupke05b, fu09}. Theory predicts that this feedback regulates the growth of supermassive black holes and their accompanying starbursts \citep{dimatteo05, veilleux05, narayanan08, hopkins10}, making them an indispensable element in models of galaxy evolution. However, exactly how this feedback regulation occurs is still very poorly understood.

Warm ionized and cool neutral gas is thought to reside in clouds advected into the outflow at the shear interface between the hot wind and the quiescent disk gas \citep[e.g.][]{t-t98}. Emission line observations have the ability to trace this warm ionized phase, and studies of local starbursting galaxies have shown that superwinds are common \citep[e.g.,][]{heckman90, marlowe95, martin98, meurer04}. Detailed studies of individual nearby cases highlight their complexity, both morphologically and kinematically \citep[e.g.,][]{westm07a,westm09a,westm11}. Absorption-line spectroscopy, however, has proved a much easier method for identifying and studying outflows since this technique relies on the background continuum light from the galaxy, which is much brighter than the emission lines.
Up to now much of our knowledge about ULIRG outflows has therefore come from (mostly long-slit) studies of the cool gas component via the Na\one\,$\lambda\lambda$5890,5896 interstellar absorption line. These studies have found that $\sim$80\% of ULIRG galaxies show evidence for neutral outflows with speeds of 300--700~\kms\ and high mass outflow rates of order 10--100~\Msol~yr$^{-1}$ that appear to scale with L$_{\rm IR}$, SFR and galaxy mass \citep{heckman00a, rupke02, martin05, rupke05a, rupke05b, martin06}. 

The major limitation of such absorption line studies is the limited radial extent over which the absorption lines can be probed. At large radii the surface brightness of the continuum source (starlight) becomes too faint to probe any absorbing material, whereas ionized gas emission is independent of the stellar surface brightness and can often be observed to much larger distances. Thus the last decade has seen a steady flux of spatially resolved optical emission line studies of (U)LIRGs, both on a detailed case-by-case basis \citep[e.g.,][]{colina00, murphy01, arribas03, lipari04b, lipari04c, bedregal09, bendo09, shih10, rupke11} and as part of small surveys \citep{colina05, arribas08, garciamarin09a, garciamarin09b, alonsoherrero09, alonsoherrero10, monreal10}. To date no integral field survey of the emission-line characteristics of local ULIRGs has been carried out with spatial resolutions better than $1''$ or spectral resolutions better than R$\approx$1000. For these reasons, and to capitalise on the recent interest, we have obtained VLT/VIMOS integral field spectroscopy of a volume limited sample of 18 southern ULIRGs, with spatial and spectral resolutions of $\sim$$0\farcs7$ and R=3100. Our sample covers a range of ULIRG types, including starburst dominated, LINER, Sy2, starburst+Sy2, LINER+Sy2, (but not the rare Sy1 class of which there are no low-$z$ southern hemisphere examples), distant pairs, close pairs, and fully-coalesced systems, and has allowed us to measure detailed gas dynamics, excitations and densities in a large sample of local ULIRG systems for the first time. This paper forms part of an ongoing programme to study the gas dynamics in ULIRGs; a detailed investigation of IRAS 19254-7245 (``The Superantennae'') was the subject of our first paper \citep[][hereafter \citetalias{bendo09}]{bendo09}.

The remainder of this article is organised as follows: In Section~\ref{sect:data} we briefly describe the sample selection, data acquisition and reduction, and describe how we modelled the emission line shapes using multi-Gaussian fitting. In Section~\ref{sect:results} we present our emission line maps together with line diagnostic plots, and describe a number of measured and derived properties of the galaxies in the whole sample (Section~\ref{sect:props_sample}). We then discuss our results with respect to the gas dynamics, line ratios and evidence for outflows in Section~\ref{sect:disc}, and summarise our main conclusions in Section~\ref{sect:summary}. In the Appendix we give a detailed case-by-case description of the emission line maps for each target, with reference to the existing literature.

\section{Observations and Data Reduction} \label{sect:data} 

\subsection{Data acquisition and reduction}
Our sample was selected from the the \textit{IRAS} 1~Jy survey of ULIRGs \citep{kim98}. Observations were performed with the VIsible MultiObject Spectrograph \citep[VIMOS;][]{le-fevre03} integral field unit (IFU) at the Very Large Telescope in 2007 (080.B-0085) and 2009 (383.B-0372). A detailed description of the data reduction procedure was given in \citetalias{bendo09}, so we only give a brief recap here. The data were selected to be nearby ($z < 0.09$) and accessible from the VLT ($\delta < 10$). They were taken using the HR red grism (6300--8700 \AA) giving a spectral resolution of R=3100, and with the $0\farcs67$~spaxel$^{-1}$ spatial resolution. A 5-point dither pattern was used, which provided redundancy within the central region and ensured the availability of blank sky for background measurements. Four exposures were taken at each dither position; the total exposure times are given in Table~\ref{tbl:sample}. The seeing during the observations was $\sim$0.7--1~arcsec. At the time of observation, two of the complete sample of ULIRGs, IRAS 05189-2524 and IRAS 06035-7102, had already been observed with the VIMOS IFU as part of the \citet{arribas08} study. The reduced datacubes for these galaxies were kindly made available to us for inclusion in this study (S. Arribas, private communication)\footnote{Note the observations in our programme are a factor of four deeper than those in the study of \citet{arribas08}.}.

\begin{figure*}
\centering
\includegraphics[width=0.95\textwidth]{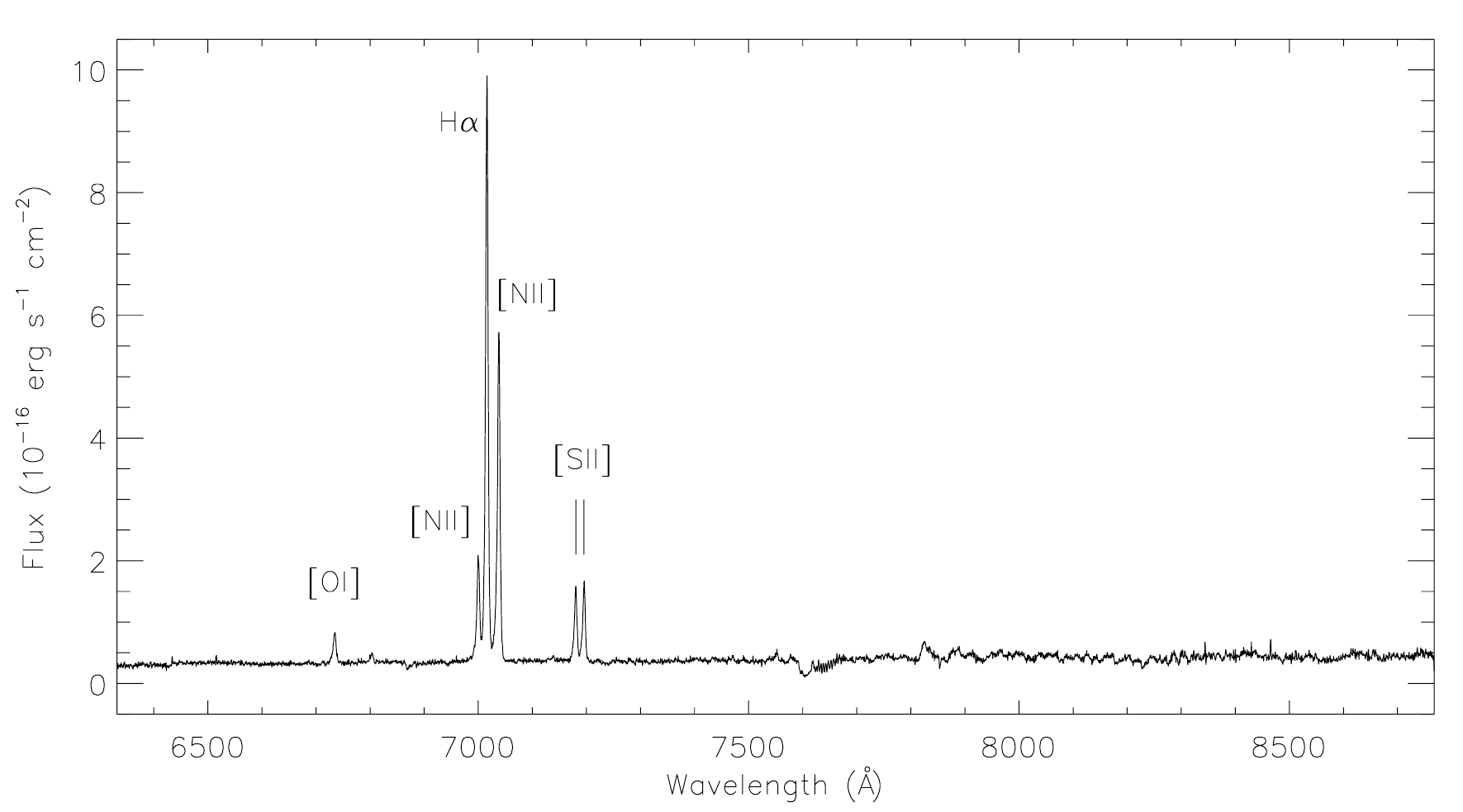}
\caption{High S/N example spectrum from IRAS 00335-2732 showing the full wavelength coverage of the HR Red grism. Wavelengths are not corrected to the rest-frame. The emission lines examined in this paper are labelled.}
\label{fig:spec_plot}
\end{figure*}

Basic data reduction was performed following the standard ESO reduction pipeline tasks implemented in \textsc{gasgano}. For each pointing and for each quadrant, the median background spectrum was calculated and subtracted from the data. Following this, the spectra from each pointing were mapped into individual spectral cubes, and then the cubes were median combined to produce the final spectral cube. An example full HR Red spectrum from IRAS 00335-2732 is shown in Fig.~\ref{fig:spec_plot} with the emission lines examined in this paper labelled. Finally, a S/N cut was applied to each cube based on the integrated line flux in the H$\alpha$-[N\two] region to remove spaxels with no detected line emission.

\begin{table*}
\begin{minipage}{17.5cm}
\begin{center}
\caption{VIMOS ULIRG galaxy sample.}
\label{tbl:sample}
\begin{tabular}{l c c c c l c c}
\hline
Galaxy & $\alpha$ (J2000), $\delta$ (J2000) & $z^{a}$ & $D^{a}$ & log($L_{\rm IR}/L_{\odot}$)$^{b}$ & Sp type$^{c}$ & SFR$^{d}$ & Exposure \\
& & & (Mpc) & & & (\Msol\,yr$^{-1}$) & time (s) \\
\hline
IRAS 00198-7926&  00 21 52.9, $-$79 10 08.0 & 0.0728 & 300 & 12.01 & Sy2 (AGN in N$_{\rm N}$$^{e}$) & 53 & 12000 \\
IRAS 00335-2732 & 00 36 00.5, $-$27 15 34.5 & 0.0693 & 282 & 11.93 & SB & 117 & 12000 \\
IRAS 03068-5346 & 03 08 21.3, $-$53 35 12.0 & 0.0745 & 304 & 11.96 & SB & 126 & 12000 \\
IRAS 05189-2524 & 05 21 50.5, $-$25 21 46.3 & 0.0426 & 173 & 12.19 & Sy2 & 80 & 3000  \\
IRAS 06035-7102 & 06 05 11.2, $-$71 03 05.5 & 0.0795 & 325 & 12.26 & SB (AGN in N$_{\rm W}$?$^{e}$) & 182 & 2880  \\
IRAS 09111-1007W & 09 13 38.8, $-$10 19 20.3 & 0.0541 & 220 & 11.98 & SB+Sy2 & 66 & 12000 \\
IRAS 12112+0305 & 12 13 46.0, +02 48 38.0 & 0.0733 & 300 & 12.28 & LINER & 263 & 12000 \\
IRAS 14348-1447 & 14 37 38.4, $-$15 00 22.8 & 0.0827 & 345 & 12.30 & LINER (N$_{\rm SW}$ = Sy2$^{f}$) & 275 & 9600 \\
IRAS 14378-3651 & 14 40 58.9, $-$37 04 33.0 & 0.0676 & 280 & 12.24 & LINER+Sy2 & 120 & 12000 \\
IRAS 17208-0014 & 17 23 21.9, $-$00 17 00.4 & 0.0428 & 175 & 12.25 & SB (LINER$^{g}$) & 245 & 12000 \\
IRAS 19254-7245 & 19 31 21.6, $-$72 39 21.7 & 0.0617 & 250 & 11.91 & Sy2 & 56 & 12000  \\
IRAS 19297-0406 & 19 32 20.7, $-$04 00 06.0 & 0.0857 & 353 & 12.36 & SB & 316 & 12000 \\
IRAS 20046-0623 & 20 07 19.4, $-$06 14 26.0 & 0.0844 & 348 & 11.97 & SB & 129 & 12000 \\
IRAS 20414-1651 & 20 44 18.2, $-$44 33 37.7 & 0.0871 & 358 & 12.26 & SB & 251 & 12000 \\
IRAS 20551-4250 & 20 58 27.4, $-$42 39 03.1 & 0.0428 & 175 & 11.98 & Sy2$^{h}$ & 132 & 6000 \\
IRAS 21504-0628 & 21 53 05.5, $-$06 14 49.9 & 0.0776 & 320 & 11.92 & Sy2? & 57 & 12000 \\
IRAS 22491-1808 & 22 51 49.3, $-$17 52 23.4 & 0.0778 & 320 & 12.09 & SB & 170 & 12000 \\
IRAS 23128-5919 & 23 15 47.0, $-$59 03 16.9 & 0.0446 & 182 & 11.96 & SB+Sy2 & 63 & 6000\\
\hline
\end{tabular}
\end{center}

$^{a}$ Reference: NED.\\
$^{b}$ 8--1000~$\micron$ infrared fluxes from the IRAS Faint Source Catalog (FSC) version 2 \citep{moshir90}, recalculated with updated redshifts where required.\\
$^{c}$ References: \citet{farrah03}, NED, except where otherwise indicated. \\
$^{d}$ Star formation rate derived from $L_{\rm IR}$ and scaled according to AGN input (see text for scaling factors)\\
$^{e}$ N$_{\rm N}$ = northern nucleus. Reference: this study \\
$^{f}$ \citet{sanders88, nakajima91} \\
$^{g}$ \citet{arribas03} re-classify this galaxy as a LINER due to the effect of differential extinction on the optical nuclear spectrum.
$^{h}$ \citet{franceschini03, ptak03, farrah07}
\end{minipage}
\end{table*}

\subsection{Decomposing the emission line profile shapes} \label{sect:line_profiles}

Following the methodology first presented by \citet{westm07a}, we fitted multiple Gaussian profile models to the H$\alpha$, [N\two], [S\two] and [O\one] lines using an \textsc{idl} $\chi^{2}$ fitting package called \textsc{pan} \citep[Peak ANalysis;][]{dimeo, westm07a} to quantify the gas properties observed in each IFU field. The high S/N and spectral resolution of these data have allowed us to quantify the line profile shapes of these lines to a high degree of accuracy. Our fitting methodology differs slightly to that employed in \citetalias{bendo09} since we have had to optimise the process for the automatic multi-Gaussian decomposition of the emission lines in our entire sample.

Each line in each spectrum was fitted using a single, double, and triple Gaussian component initial guess (no lines were found to need more than three components). Line fluxes were constrained to be positive and widths to be greater than the instrumental contribution to guard against spurious results. We made no attempt to fit an H$\alpha$ stellar absorption component since it was never needed to obtain a good fit.

The H$\alpha$ and [N\two]$\lambda 6583$ lines were fit simultaneously, where for each component the wavelength difference between the two Gaussian models was constrained to be equal to the laboratory difference, and the FWHMs equal to one another. This constrained approach dramatically improves the quality of the fits, particularly for spectra with lower S/N. Separately, we fitted the [S\two] doublet also simultaneously, and the [OI] line individually. Multi-component fits were run several times with different initial guess configurations (widths and wavelengths) in order to account for the varied profile shapes, and the one with the lowest $\chi^2$ fit statistic was kept. However, we note that the $\chi^2$ minimisation routine employed by PAN is very robust with respect to the initial guess parameters. 

To determine how many Gaussian components best fit an observed profile (one, two or three), we used a likelihood ratio test to determine whether a fit of $n$+1 components was more appropriate than an $n$-component fit. This test says that if the ratio of the $\chi^{2}$ statistics for the two fits falls above a certain threshold, then the fits are considered statistically distinguishable, and the one with the lower $\chi^{2}$ can be selected. Here we determine the threshold ratio by visual inspection of a range of spectra and fits. We are aware that \citet{protassov02} show that using any kind of likelihood ratio test (including the F-test) in this kind of situation is statistically incorrect, but given the absence of a statistically correct alternative that could be applied sensibly to the volume of data presented here, we have chosen to opt for this generalised $\chi^{2}$ ratio test approach.

This test, however, only tells us which of the fits (single, double or triple component) is most appropriate for the corresponding line profile. Experience has taught us that we need to apply a number of additional tests to filter out well-fit but physically improbable results. We specified that the measured FWHM had to be greater than the associated error on the FWHM result (a common symptom of a bad fit), the fluxes of all components should be $>$0, and we rejected any fits where $\chi^{2}_{\rm single}/\chi^{2}_{\rm double}=0$ (another symptom of a bad fit).

To estimate the uncertainties on the fit parameters, we used a bootstrap Monte Carlo technique, assuming that the noise is uncorrelated. Using the fit results determined from the above step as the initial guess, we re-fitted each profile 100 times, each time seeding the spectrum with randomly generated noise based on the standard deviation of the continuum. The standard deviation on each parameter (flux, FWHM and radial velocity) over the 100 results was then adopted as the associated uncertainty for that parameter, while the mean of those results was adopted as the actual value. Uncertainties range from 1--30\% depending on the S/N of the spectrum and how many components are fit.

The final step of the fitting procedure was to assign each fit component to a particular map in such a way as to limit the confusion that might arise during analysis of the results, such as discontinuous spatial regions arising from incorrect component assignments. For triple fits, we specified C1 to be the brightest and C3 the faintest (regardless of the relative velocities of the components), except for the galaxies that exhibit a very broad H$\alpha$ component. For IRAS 05189-2524 we specified C1 to be the reddest and C3 the bluest (regardless of their relative flux), and for IRAS 06035-7102 we specified C3 to be the broadest then C1 the brightest of the remaining two components (regardless of their relative velocities). And for IRAS 19254-7245, we set C3 to be the broadest component then C1 as the narrowest (regardless of their relative flux), and for a few double fits we manually set a few very broad components to C3 to make the maps more consistent. In all other double-component fits we specified C1 to be the brightest of the two.

Many galaxies show 2--3 line components in nuclear regions. However, in some cases the difference between the components in the FWHM and/or radial velocity plots can be difficult to see due to the small differences in values compared to the range of the colour bar. This is an inherent limitation in the dynamic range that can be represented in these type of plots. To highlight this we have chosen an example from the IRAS 20046-0623 map. In Fig.~\ref{fig:egfit} we show the H$\alpha$+[N\two] line profiles and best-fitting single and double-Gaussian models for the spaxel indicated with a white cross in the H$\alpha$ C2 FWHM map in Fig.~\ref{fig:iras20046-0623}. Both the $\chi^{2}$ and residuals demonstrate that a two-component fit is the most appropriate choice, although on the maps it is difficult to see the difference between the two components.

The emission line maps are shown in Figs.~\ref{fig:iras00198-7926}--\ref{fig:iras23128-5919}. Each figure includes example H$\alpha$+[N\two] line profiles and best-fitting Gaussian models from selected spaxel(s) within the IFU field, together with the integrated spectrum of the whole system. These examples have been chosen to demonstrate the quality of the spectra and the accuracy of the line-fitting.

\section{Results} \label{sect:results}

Emission line maps created from the line profile fits described above are shown in Figs.~\ref{fig:iras00198-7926}--\ref{fig:iras23128-5919}. For each system we show the \textit{HST} F814W image (or DSS image if \textit{HST} imaging is not available), and for each identified line component, the H$\alpha$ flux, FWHM and radial velocity maps, the [N\two]$\lambda$6583/H$\alpha$, [S\two]($\lambda$6717+$\lambda$6731)/H$\alpha$ and [O\one]$\lambda$6300/H$\alpha$ flux ratio maps, and the electron density map derived from the [S\two] doublet ratio (where we have summed all line components together) using the IRAF \textsc{nebular} task and assuming T$_{\rm e}=10^4$~K.

[N\two]$\lambda$6583/H$\alpha$ vs.\ [S\two]($\lambda$6717+$\lambda$6731)/H$\alpha$ flux ratios are also plotted for each galaxy on a spaxel-to-spaxel basis in Figs.~\ref{fig:ratio_plots1}--\ref{fig:ratio_plots3}, together with predictions from the photoionization and shock models of \citet{dopita06b} and \citet{allen08}, and (where appropriate) the dusty AGN models of \citet{groves04}. Without the [O\three] line we cannot plot any of the traditional BPT diagrams \citep{baldwin81}, however the [N\two]/H$\alpha$ vs.\ [S\two]/H$\alpha$ plot can still be instructive to investigate the contribution of harder radiation fields (AGN/shocks) to the ionization. 

We discuss these maps in the context of the whole sample in Section~\ref{sect:disc} and describe them in detail for each galaxy individually in Appendix~\ref{sect:gal_desc}.

\subsection{Measured and derived properties of the galaxies in the whole sample} \label{sect:props_sample}

Integrated SFRs for each of our galaxies are listed in Table~\ref{tbl:sample}, and are derived from a modified version of the \citet{kennicutt98} calibration of IR luminosity to SFR:
\begin{equation}
{\rm SFR} = \alpha\, \frac{L_{\rm IR}}{5.8\times 10^9 \, L_{\odot}},
\end{equation}
where $\alpha$ is a factor equal to the fraction of the IR luminosity powered by star formation. Following \citet{rupke05a,rupke05b}, we assume $\alpha = 0.8$ for SB ULIRGs, $\alpha = 0.3$ for Sy2 galaxies that show either broad lines in the near-IR or faint broad lines in the optical (e.g.\ IRAS 05189-2524), and $\alpha = 0.4$ for all other Sy2s.

Following \citet{monreal10}, we morphologically classify the galaxies into three classes (based loosely on the \citealt{veilleux02a} scheme), using the \textit{HST} F814W imaging (or DSS imaging where \textit{HST} imaging is not available). Our classifications are shown in Table~\ref{tbl:measured_gal}. Class 0 represents objects that appear to be single isolated galaxies with a relatively symmetric morphology and no obvious evidence for strong past or ongoing interaction. Class 1 includes objects in a pre-coalescence phase with two nuclei separated by a projected distance of $D>1.5$~kpc. This limit was chosen because theoretical models predict a fast coalescence phase after the nuclei become closer than 1.5~kpc \citep{mihos96, naab06}. Class 2 includes objects with relatively asymmetric morphology suggesting a post-coalescence phase. These systems may have nuclei separated by $D<1.5$~kpc.

For those systems with what appears to be a rotating disk component, we give in Table~\ref{tbl:measured_gal} their H$\alpha$-derived rotation velocities (uncorrected for the unknown inclinations). For all these systems but IRAS 03068-5346 and 19297-0406, \citet{dasyra06a, dasyra06b} measure stellar-derived rotation velocities from fitting the CO bandheads in their H-band long-slit data. Their measurements are systematically lower than ours by $\sim$30--50~\kms, suggesting that the kinematics of the stellar and gaseous components of these rotating disks may actually be de-coupled from one another. This deserves further investigation, particularly with spatially resolved stellar velocities at a comparable resolution to our gas measurements.

Assuming virial equilibrium, our gas rotational velocities allow us to estimate a mass ($M_{\rm virial}=(2 R v^{2})/G$), which is given in the following column in Table~\ref{tbl:measured_gal}. This, in turn, allows us to calculate the Toomre $Q$ \citep{toomre64} parameter for characterising the stability of a disk supported by differential rotation and random motion against local axisymmetric perturbations. \citep[A review of the criterion applied to gravitationally coupled stars and gas in a disk is given in][see also \citealt{downes98}]{jog96}. This criterion says whether a disk is stable to collapse ($Q>1$) or not ($Q<1$), and is calculated as following:
\begin{equation}
Q=\sigma_{r} \frac{\kappa}{\pi \, G\, \Sigma_{\rm mass}}
\end{equation}
where $\sigma_{r}$ is the uniformly-weighted velocity dispersion of the galaxy \citep{davies11}, $\kappa = 1.5 \, v_{\rm rot}/R$, and $\Sigma_{\rm mass}$ is the surface mass density (= mass/surface area, where the mass is estimated from the Virial mass, $M_{\rm virial}=(2 R v^{2})/G$).
Our results are discussed in Section~\ref{sect:toomreq}.

In Table~\ref{tbl:measured_wind} we indicate whether evidence for warm ionized (wi), neutral atomic (na) or molecular (mol) gas outflows in each galaxy are known either from the literature or from this study. A discussion of the differences between the outflows measured in the different phases is presented in Section~\ref{sect:outflows}. The next two columns in Table~\ref{tbl:measured_wind} show the average centroid velocity (relative to the systemic velocity) and line width (FWHM) of the outflow components (typical errors of $\pm$30\%). The final two columns show the two commonly used ways of calculating the ``maximum'' wind speed: $\Delta v_{\rm max,1} = \left| \langle v \rangle_{\rm broad} - \Delta v_{\rm broad(FWHM)}/2 \right|$ \citep[e.g.][]{veilleux05, rupke05a}, and $\Delta v_{\rm max,2} = \left| \langle v \rangle_{\rm broad} - 2\, \sigma_{\rm broad} \right|$ \citep[e.g.][]{martin05, genzel11}, where we use $v_{\rm outflow} = \langle v \rangle_{\rm broad}$.

\begin{table*}
\begin{minipage}{17.5cm}
\begin{center}
\caption{Galaxy properties of the ULIRG sample.}
\label{tbl:measured_gal}
\begin{tabular}{l c c c c c}
\hline
Galaxy & Morph & Separation of & $v_{\rm rot}$$^{b}$ & Virial & Toomre \\
& classif.$^{a}$ & nuclei (kpc) & (\kms) & mass (\Msol)& $Q$ \\
\hline 
IRAS 00198-7926        & 1 & 6 \\
IRAS 00335-2732        & 2 \\
IRAS 03068-5346        & 2 && 60 & $6.0\times 10^9$ & 2.08 \\
IRAS 05189-2524        & 0 \\
IRAS 06035-7102        & 1 & 6 \\
IRAS 09111-1007W   & 1 & 40 & 110 & $3.0\times 10^{10}$ & 0.93 \\
IRAS 12112+0305       & 1  & 4 \\
IRAS 14348-1447        & 1 & 4.8$^{c}$ \\
IRAS 14378-3651         & 2 \\
IRAS 17208-0014         & 2 && 140 & $3.3\times 10^{10}$ & 1.32 \\
IRAS 19254-7245         & 1 & 10 \\
IRAS 19297-0406         & 2 & & 175 & $7.1\times 10^{10}$ & 1.32 \\
IRAS 20046-0623          & 2 && 185 & $1.3\times 10^{11}$ & 0.87 \\
IRAS 20414-1651         & 2 && 150 & $5.3\times 10^{10}$ & 0.79 \\
IRAS 20551-4250         & 2 && 90 & $5.5\times 10^{10}$ & 1.89 \\
IRAS 21504-0628         & 0 && 175 & $1.4\times 10^{11}$ & 1.01  \\
IRAS 22491-1808         & 2 & 1.2, 2.4$^{d}$ \\
IRAS 23128-5919         & 1 & 4 \\
\hline
\end{tabular}
\end{center}

$^{a}$ Class 0: single isolated galaxies with no obvious evidence for strong past or ongoing interaction. Class 1: pre-coalescence phase with two nuclei. Class 2: relatively asymmetric morphology, or nuclei separated by $<$1.5~kpc, suggesting a post-coalescence phase.\\
$^{b}$ Disk rotation velocity (for objects with clear disks and rotation)\\
$^{c}$ \citet{carico90, scoville99, surace00} \\
$^{d}$ Triple nuclei \citep{carico90, surace00, cui01}
\end{minipage}
\end{table*}

\begin{table*}
\begin{minipage}{17.5cm}
\begin{center}
\caption{Outflow properties of the ULIRG sample.}
\label{tbl:measured_wind}
\begin{tabular}{l c c c c c}
\hline
Galaxy & Known evidence & $v_{\rm outflow}$$^{b}$ & $\Delta v_{\rm broad(FWHM)}$$^{c}$ & $\Delta v_{\rm max,1}$$^{d}$ & $\Delta v_{\rm max,2}$$^{d}$ \\
& for outflows$^{a}$ & (\kms) & (\kms) & (\kms) & (\kms) \\
\hline 
IRAS 00198-7926        & wi & $-$400 & 2000 & $-$1400 & $-$2100 \\
IRAS 00335-2732         \\
IRAS 03068-5346        \\
IRAS 05189-2524        & wi, mol & $-$3000 & 4000 & $-$5000 & $-$6400 \\
IRAS 06035-7102        & wi & $-$550 & 2500 & $-$1800 & $-$2700 \\
IRAS 09111-1007W   & wi & $-$250 & 700 & $-$600 & $-$850 \\
IRAS 12112+0305       & wi, mol & $-$300 & 500 & $-$550 & $-$725 \\
IRAS 14348-1447        & wi, mol & $-$150 & 750 & $-$525 & $-$790 \\
IRAS 14378-3651         & wi, mol & $-$300 & 700 & $-$650 & $-$900 \\
IRAS 17208-0014         & na, mol &&& \\
IRAS 19254-7245         & wi & $>$500 & 2000 & $>$$-$1500 & $>$$-$2200 \\
IRAS 19297-0406         & wi, na, mol & $\sim$0 & 800 & $-$400 & $-$680 \\
IRAS 20046-0623          & na &&& \\
IRAS 20414-1651         & na &&& \\
IRAS 20551-4250         & wi, mol & $-$150 & 700 & $-$500 & $-$750 \\
IRAS 21504-0628         \\
IRAS 22491-1808         \\
IRAS 23128-5919         & wi & $>$$-$450 & 700 & $>$$-$800 & $>$$-$1050 \\
\hline
\end{tabular}
\end{center}

$^{a}$ wi = warm ionized (ref: this paper; \citealt{lipari03, spoon09}; \citealt{dasyra11} for IRAS 05189-2524); na = neutral atomic \citep[ref:][]{martin05, rupke05a}; mol = molecular (\citealt{sturm11}; E.\ Sturm, priv.\ comm.).\\
$^{b}$ $v_{\rm outflow} = \langle v \rangle_{\rm broad}$ = average centroid velocity (relative to systemic velocity) of the outflow components. Typical errors are $\pm$30\%. \\
$^{c}$ Average FWHM of the outflow components. Typical errors are $\pm$30\%. \\
$^{d}$ The ``maximum'' wind speed calculated in the two ways used in the literature: $\Delta v_{\rm max,1} = \left| \langle v \rangle_{\rm broad} - \Delta v_{\rm broad(FWHM)}/2 \right|$ \citep[e.g.][]{veilleux05, rupke05a}, and $\Delta v_{\rm max,2} = \left| \langle v \rangle_{\rm broad} - 2\, \sigma_{\rm broad} \right|$ \citep[e.g.][]{martin05, genzel11}. \\
\end{minipage}
\end{table*}

\section{Discussion} \label{sect:disc}

\subsection{Gas dynamics and the Toomre $Q$ parameter}\label{sect:toomreq}


One of the primary goals of studying local ULIRGs is to gain insights into the physical processes at work in high-$z$ intensely star forming galaxies. What determines the structure and nature of the star formation? In the merger process, does the gas become immediately unstable and collapse in clumps to form stars, or does it get driven to the centre to fuel a nuclear starburst and/or AGN? What governs these processes and their timescales? For example, contrary to what was expected, the rest-frame UV/optical morphologies of $z > 1$ ``normal'' star-forming galaxies have often been found to be dominated by several giant (kpc size) star-forming clumps \citep{cowie95, vandenbergh96, elmegreen07, elmegreen09, elmegreen05, elmegreen06a, forster09, forster11b}, implying that the gas in these disks forms stars in-situ, rather than being driven into the centre to fuel a nuclear starburst. This is of interest here as these clumpy, asymmetric structures often resemble local ($z\sim 0$) mergers \citep{conselice03, lotz04}.

One way of probing what physical mechanisms are at work is to measure the stability of a system's disk to gravitational collapse via the Toomre $Q$ parameter \citep{toomre64}. In order for star formation to commence, the ISM within a rotating disk must become globally gravitationally unstable whilst under the influence of shear resulting from differential rotation. Regions that become too big will be torn apart by shear before gravitational collapse can occur. \citet{elmegreen99} showed that large-scale gravitational collapse must set in whenever the gas density becomes a significant fraction of the ``virial'' density, so it follows that large, turbulent, gas-rich disks are prone to developing star forming clumps. SMGs \citep{chapman03}, for example, with SFRs of many 100s \Msol~yr$^{-1}$, have been shown to have $Q\ll1$ \citep{tacconi06}. This is unsurprising given that they are known to fulfil the criteria for being ``maximal starbursts'' \citep{tacconi06}. The situation in local ULIRGs is, however, less polarised. $Q$ has been found to vary from system to system with some $<$1 and some $>$1 \citep{downes98}.

In our sample, 3/8 systems with measurable $Q$s (i.e. that exhibit a gaseous disk) have $Q<1$ (IRAS 09111-1007W , IRAS 20046-0623 and IRAS 20414-1651), whereas the remaining five appear to have gravitationally stable disks ($Q>1$). We note here that our measurements assume that the stars and gas are gravitationally coupled \citep{jog96, downes98}. This may not be the case (in fact, given the disturbed nature of these systems, is highly unlikely), however, we do not have measurements of the stellar kinematics to quantify what effect this assumption has (the Ca\two\ stellar absorption lines fall just outside of our wavelength range at the redshift of our sources).

This variation in $Q$s can be interpreted in terms of the concept of self-regulated star formation \citep[e.g.,][]{gammie01, thompson05, tacconi06}. Gravitationally unstable disks with $Q<1$ collapse and start forming stars vigorously. The resulting feedback increases the local velocity dispersion which in turn increases $Q$ and has a negative effect on the star formation rate. Like this, $Q$ self-regulates at $Q \sim 1$. In each individual case we are likely seeing the system simply at a particular point in this self-regulation cycle.

Using detailed IFU observations of star forming galaxies at $z\sim1$, \citet{genzel11} mapped out the $Q$ parameter on a spaxel-by-spaxel basis and found that giant star forming clumps are found where locally $Q<1$, even if the global average $Q>1$. While it is outside the scope of this paper to derive spatially resolved $Q$ maps for our targets, it is very likely that even systems that we measure to have $Q>1$ will have locally unstable regions that are forming stars. For example, IRAS 20551 has a disk $Q$ parameter of 2.08, but strong H$\alpha$ emission is observed in tidal features implying strong extra-nuclear star formation.

The emerging view is that merger-induced starbursts have a nuclear component, and also an extended component in the form of massive star forming clumps in the disk \citep{bournaud11}. What we see in any particular case all depends on the exact merger stage and individual characteristics of the system.

\subsection{Line ratios in the nuclear and outer regions}\label{sect:line_ratios}
Forbidden-to-recombination line ratios can be useful in diagnosing the strength of the ionizing field strength. In this study we have examined the [N\two]/H$\alpha$, [S\two]/H$\alpha$ and [O\one]/H$\alpha$ ratios, chosen to be close in wavelength so as to reduce the effects of reddening as much as possible. [S\two] and [O\one] are produced in the partially ionized zone at the edge of an ionized nebula. This zone is large and extended for hard radiation fields so the [S\two]/H$\alpha$ and [O\one]/H$\alpha$ ratios become stronger for Seyferts compared to SB/LINER galaxies \citep{yuan10}, or in the presence of shocks \citep{allen08}. [S\two]/H$\alpha$ is particularly sensitive to shock ionization because relatively cool, high-density regions form behind shock fronts which emit strongly in [S\two] \citep{dopita97, oey00}. The [N\two]/H$\alpha$ ratio is more sensitive to the presence of a low-level AGN than the other ratios because [N\two]/H$\alpha$ is a linear function of nebular metallicity until high metallicities where the ratio plateaus. Beyond this, an AGN contribution moves the line ratio higher \citep{yuan10}. In general, the highest line ratios in all three combinations are found in the faintest regions, on the outskirts of the galaxies, and depressed line ratios in the nuclear regions is commonly observed.

7/18 systems in our sample show a nuclear depression in all three line ratios ([N\two]/H$\alpha$, [S\two]/H$\alpha$ and [O\one]/H$\alpha$), and a further 7/18 exhibit a nuclear depression just in [S\two]/H$\alpha$ and [O\one]/H$\alpha$. IRAS 06035, 12112, 19254 and 20046 do not show a correlation. What is the cause of this? The existence of cases where only two of the line ratios are low excludes effects due to H$\alpha$ stellar absorption artificially reducing the ratios, and artefacts due to multi-component fitting, since H$\alpha$ is part of all three ratios. Furthermore, differential extinction across the face of the galaxies should also not affect the ratios as they are specifically chosen to be reddening free \citep{baldwin81, veilleux87}. Thus we must be seeing a real physical effect due to either ionization gradients or abundance changes (total metallicity and/or relative abundances) between the nuclear and outer parts of these systems.

\begin{figure}
\centering
\begin{minipage}{8cm}
\includegraphics[width=8cm]{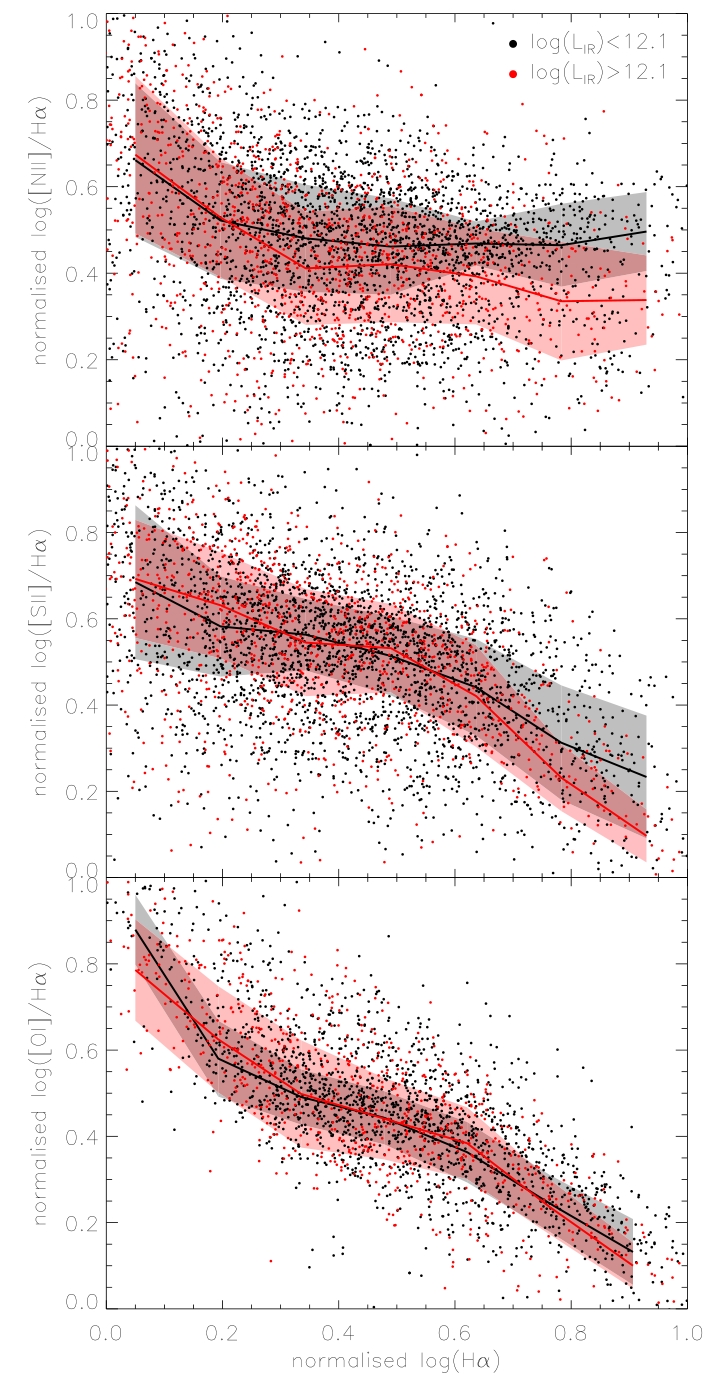}
\end{minipage}
\caption{[N\two]/H$\alpha$, [S\two]/H$\alpha$, and [O\one]/H$\alpha$ flux ratios vs.\ H$\alpha$ flux for all spaxels in the subsample of systems that show some correlation between these measurements (see text for more details). The values in each axis have been normalised to the maximum and minimum of the range found in each individual system for ease of comparison. The black points/lines belong to systems with log($L_{\rm IR}$)$<$12.1 and the red to log($L_{\rm IR}$)$<$12.1. The thick lines represent the sigma-clipped mean ratios in bins of width 0.15, and the filled areas the $\pm$1$\sigma$ on this mean.}
\label{fig:diagnostic_correlations}
\end{figure}

In order to investigate this phenomenon further, we have plotted all three line ratios for C1 and C2 in every spaxel in each system that shows a nuclear depression (i.e.\ excluding IRAS 06035, 12112, 19254 and 20046) vs.\ H$\alpha$ flux in Fig.~\ref{fig:diagnostic_correlations}. Here we use the H$\alpha$ flux as a proxy for distance, since this avoids any problems due to absolute differences in ``distance'' from system-to-system. Generally H$\alpha$ peaks at the centre and falls of with radius. To aid the eye, we have also plotted the sigma-clipped mean ratio and its $\pm$1$\sigma$ error in bins of 0.15. A clear correlation can be seen between the [S\two]/H$\alpha$ and [O\one]/H$\alpha$ ratios and H$\alpha$ flux (distance from the nucleus), and a weaker one in [N\two]/H$\alpha$. The black and red points/lines differentiate those systems with log($L_{\rm IR}$)$<$12.1 and $>$12.1. Although no significant difference can be seen between the two $L_{\rm IR}$ groups in [S\two]/H$\alpha$ and [O\one]/H$\alpha$, this shows that systems with depressions in all three ratios (including [N\two]/H$\alpha$) tend to have higher $L_{\rm IR}$.

All four ULIRGs that do not show any nuclear line ratio depressions are interacting double-nucleus or highly disturbed systems of class 1 (pre-coalescence phase). Those that show depressions in all three ratios are predominantly of class 2 (asymmetric morphology and/or nuclei separated by $D<1.5$~kpc, suggesting a post-coalescence phase). For the group of seven that show depressed nuclei in all three ratios, five are centrally concentrated systems. There is no particular correlation with spectral classification (SB, LINER or Sy2).

In summary, nuclear line ratio depressions appear to be found in the less disturbed systems (either isolated class 0 or post-coalescence class 2), and a depression in all three ratios is associated with high IR luminosity and a central nuclear concentration. This implies that the nuclear-to-outskirts line ratio changes are predominantly an ionization effect. If the system is centrally concentrated then it is likely that the nuclear regions are dense and very dusty (therefore highly extinguished), so in the optical we see only the surface layers of the gas. Thus one possibility is that the low line ratios reflect the fact that this gas is not experiencing the full radiation field of the starburst/AGN. Inclination effects may also play a role, since the nucleus may appear obscured along our line-of-sight, but not to gas located further out in the plane of the sky. It would be illuminating to repeat this experiment with near/mid-IR observations where the effects of obscuration are mitigated.

Instead of asking why the nuclear regions have low line ratios, one might instead ask why the outer regions exhibit higher line ratios. The diagnostic diagrams of Fig.~\ref{fig:ratio_plots1} show that shocks may play a strong part in the ionization of the outer (fainter) regions of very many of our sample galaxies. Radially increasing line ratios may therefore also result from the dilution of the nuclear (SB or AGN) radiation field with radius and/or an increasing effect of shock ionization. 

In their spatially-resolved study of LIRG galaxies, \citet{monreal10} found a correlation between the gas excitation (as measured by the [N\two]/H$\alpha$, [S\two]/H$\alpha$ and [O\one]/H$\alpha$ line ratios) and the velocity dispersion of the emission lines in interacting/merging systems \citep[see also][]{monreal06}, and interpreted this relationship as arising from an increasing importance of shocks as ionizing sources in the extranuclear regions, since shocks can be a cause of increased line widths. We, however, find no correlation between the three line ratios and the H$\alpha$ FWHM in our ULIRG sample.

It is not altogether unsurprising that we do not see a correlation of this kind (see discussion above). We often see the highest line widths in the nuclear regions, precisely where the line ratios have a tendency to be at their lowest. Given the lack of a clear correlation between the line ratios and H$\alpha$ FWHM, the LIRG relationship found by \citet{monreal10} does not straightforwardly seem to apply to ULIRGs.

The other explanation for the line ratio gradients is changes in metal abundance, although we have no real means to test this with our dataset. Compared to galaxies of the same luminosity and mass, LIRGs and ULIRGs are known to have an oxygen under-abundance in their nuclei \citep{rupke08}. This is attributed to the merger-induced radial inflow of relatively low metallicity gas from large radii to the centre, thus diluting the abundance of the nuclear gas \citep{rupke08, soto10}. A decreased nuclear metallicity would indeed reduce the [N\two]/H$\alpha$, [S\two]/H$\alpha$ and [O\one]/H$\alpha$ emission line ratios depending on the relative abundance of the infalling gas. However, it is difficult to reconcile the fact that of the three ratios [N\two]/H$\alpha$ is most sensitive to metallicity yet is the one with the weakest correlation with radius, with an abundance explanation. Furthermore, if unstable disks ($Q$$<$1) are prone to forming massive in-situ star forming clumps rather than funnelling most of this gas to the centre, then systems with $Q$$<$1 might be expected not to show depressed nuclear line ratios. This is not the case; for example the disk in IRAS 20414-1651 has $Q$=0.79 (the lowest in our sample), but very clearly depressed line ratios in both [S\two]/H$\alpha$ and [O\one]/H$\alpha$.

Thus we conclude that the line ratio gradients are due to ionization differences, and very likely reflect obscuration in the central regions mitigating the full effects of the starburst/AGN radiation field and/or the increasing dominance of shock ionization with radius.


\subsection{Ionized gas outflows}\label{sect:outflows}

11/18 of our galaxies show evidence for outflowing warm ionized gas with $\Delta v_{\rm max}$ between 500 and a few 1000~\kms. Of the SB galaxies in our sample with outflow detections, $\Delta v_{\rm max}$ ranges between 400--800~\kms\ (depending on the method used to calculate $\Delta v_{\rm max}$), and some classified as Sy2 (i.e.\ contain an AGN) have $\Delta v_{\rm max} > 1000$~\kms. The fact that our Sy2 ULIRGs exhibit faster outflows than the pure starburst systems is consistent with previous studies that suggest AGN are required to drive winds to velocities $\gtrsim$500~\kms\ \citep{lipari03, rupke05b}. 

Systems that exhibit line components with widths $>$1500~\kms\ include IRAS 00198-7926, 05189-2524, 06035-7102, and 19254-7245. In each of these cases the very broad components are detected over spatially extended regions (for more details see the Appendix). Fig.~\ref{fig:broad_fits} shows the H$\alpha$+[N\two] line profiles and best-fitting Gaussian models for all spaxels in these four systems where one component has FWHM$>$1500~\kms. In IRAS 14348-1447 we also find line widths $>$700~\kms. Since all these systems (except 06035-7102, but see the Appendix) are known to host an AGN, these kind of line widths might suggest we are directly seeing evidence of the AGN broad line region (BLR; despite there being no type 1 Seyferts in our sample). This cannot be the case for two reasons. Firstly, as already mentioned, the broad components are detected over spatially extended regions, whereas the BLR would be unresolved in our observations and therefore its contribution would only be seen in $\sim$one spaxel (equivalent to the seeing disk). Secondly, the broad components are identified in both the recombination and forbidden lines, as can be seen in Fig.~\ref{fig:fitall}, where we show simultaneous fits to the [O\one]$\lambda$6300, H$\alpha$, [N\two]$\lambda\lambda$6548,6583, and [S\two]$\lambda\lambda$6717,6731 lines from two example spaxels in IRAS 14348-1447 and 19254-7245. The case is the same for the other systems. This indicates that any BLR contribution must be small since forbidden line emission in the BLR is collisionally de-excited. These line widths must therefore reflect very high velocity gas in the narrow line region or halo of the galaxy, and is a very important result of this study.

In Fig.~\ref{fig:lir_wind}, we place our outflow measurements in context with others from the literature by plotting them as a function of $L_{\rm IR}$. We include data on ULIRGs compiled from Na\one\ absorption line \citep{rupke05a, rupke05b, martin05}, mid-IR [Ne\three] emission line \citep{spoon09}, and far-IR OH~79~\micron\ \citep{sturm11} measurements, and Na\one\ absorption line measurements of LIRGs \citep{rupke05a} and local starburst galaxies \citep{schwartz04}. Where possible, we have recalculated the measurements given in these papers to correspond to our definition of $\Delta v_{\rm max,1}$ (see Section~\ref{sect:props_sample}). This plot shows the now well-known trend of increasing outflow velocity with IR luminosity \citep{rupke05a, martin05}, despite the inhomogeneous tracers used. The outflow speeds of the starbursts in our sample are consistent with those found by other authors, except for IRAS 06035-7102 whose outflow speed is closer to the Sy2 measurements, suggesting an AGN contribution. The outflow speeds of our Sy2 systems are consistent with the upper end of those found by \citet{rupke05b}; IRAS 05189-2524 is notable for having an outflow speed significantly faster than the average.

The warm (and cool) phases of winds are thought to be driven by (1) the momentum of expanding hot gas heated by SNe and stellar winds \citep[ram pressure;][]{lehnert96, martin05, veilleux05}, shocks, or X-ray Compton heating if an AGN is present, and/or (2) radiation pressure from the OB stars in the starburst or AGN accretion disk on dust grains mixed with the gas \citep{martin05, murray05, murray11}, and/or (3) magnetocentrifugal forces within an AGN \citep{everett05}. \citet{cox04} hypothesised that the collision of two gas-rich galaxies can also produce outflows driven by the shock heating of gas resulting from the transfer of orbital energy rather than supernovae. However, since the shock heating should occur in the central few kpc region, its effect is likely indistinguishable from that of the SNe without detailed X-ray observations, although some evidence for the role of merger-induced shock heating has been found in the nearest ULIRG, Arp220. Chandra observations of this source \citep{mcdowell03} revealed two distinct X-ray structures with different spatial scales; one suggesting the presence of a superwind, and another much larger one suggesting shock heating from the merger process.


The theory of ram pressure exerted by a supernova heated wind predicts a maximum outflow velocity determined by the temperature of the momentum-carrying component of the hot wind. In their samples of starburst dominated ULIRGs, \citet{martin05} found values of $\Delta v_{\rm max,2}$ between 300--750~\kms, and \citet{rupke05a} found $\langle \Delta v_{\rm max,1} \rangle = 300$--400~\kms, both of which are consistent with a maximum wind speed of order 500~\kms. These two studies, however, measured the cool neutral phase via the Na\one\ absorption line, thus some caution is required since a detailed comparison of column-density-weighted mean absorption profiles with rms-density-weighted emission profiles may be misleading (even if the absorption and emission lines were probing the same gas phase). Radiation pressure theory does not predict such a maximum wind speed, and it has been shown that ULIRGs have continuum luminosities that are easily sufficient to accelerate significant columns of cool/warm gas to the observed velocities via radiation pressure \citep{martin05}.


It is very likely that the different mechanisms play a larger or smaller role in accelerating different material at different radii, times and temperature phases, and that if we observe a neutral outflow, it may not necessarily follow that we also find an ionized gas outflow. For example, evidence for a molecular (via OH 79~$\micron$; $v_{\rm max}=-370$~\kms) and neutral (Na\one\ D; $v_{\rm max}=-690$~\kms) gas outflow has been found in IRAS 17208-0014 \citep{martin05, sturm11}, but we find no evidence for it in our optical spectroscopy. Neutral gas outflows have also been identified in IRAS 20414-1651 with $v_{\rm max}=-230$~\kms\ \citep{rupke05a}, for which we do not see an ionized phase counterpart in our data. In e.g.\ IRAS 19297-0406, however, we do find an ionized outflow component that is consistent with the neutral component found by \citet[][with $v_{\rm max}=-700$~\kms]{martin05}.

As a final note, our spatially resolved data show that spatial and spectral properties of ULIRGs are clearly complex and dependent on the interaction/merger state of the system. Furthermore, the extinction in ULIRGs is known to be high and very patchy \citep{garciamarin09b}. One must therefore be careful when making a comparison with spatially integrated or luminosity-weighted measurements \citep[e.g.][]{martin05, rupke05a, rupke05b, spoon09, sturm11}. However, for the case of the outflow properties, in the majority of systems in our study we find outflow components coincident with the nucleus, meaning that integrated slit spectra of the nucleus would generally be sufficient. One caveat to this is that in some multiple-nucleus systems with even relatively close separations between the nuclei, we find that one nucleus is driving an outflow and the other not, meaning it would be important to know which nucleus a long-slit observation covered.

\section{Conclusions/Summary} \label{sect:summary}

We have obtained VLT/VIMOS IFU emission-line spectroscopy of a volume limited sample of 18 southern ULIRGs selected with $z < 0.09$ and $\delta < 10$. Our sample covers a range of ULIRG types starburst dominated, LINER and Sy2 --- but not the rare Sy1 class of which there are no low-$z$ southern hemisphere examples ---, distant pairs, close pairs, and fully-coalesced systems). Some systems, despite their proximity and ULIRG classification, have received very little specific attention outside of the more statistical analyses of the local ULIRG population in the literature. By employing an automated Gaussian line fitting program, we decomposed the emission line profiles of H$\alpha$, [N\two], [S\two], and [O\one] into individual components, and have presented here for each galaxy spatially resolved maps of the H$\alpha$ line flux and kinematics (FWHM and radial velocity), together with line ratio maps of [N\two]/H$\alpha$, [S\two]/H$\alpha$ and [O\one]/H$\alpha$ that can be used to diagnose the radiation field strength, and electron density maps derived from the [S\two] doublet ratio. Together with these VIMOS data, we also present corresponding \textit{HST} F814W imaging (or DSS imaging if \textit{HST} data is not available). 

We discuss the results for each galaxy in some detail in the Appendix. Our sample contains galaxies with a wide range of morphologies and merger states. Only two are classified as single isolated galaxies with no obvious evidence for strong past or ongoing interaction, seven are classified as being in the pre-coalescence phase with two distinct nuclei, and nine are classified as having relatively asymmetric morphologies and/or nuclei separated by $<$1.5~kpc, suggesting they are in a post-coalescence phase. As is expected, we find a large range in gas dynamics and excitations.

Our overall conclusions are summarised in the following:
\begin{itemize}
  \item 11/18 of our galaxies show evidence for outflowing warm ionized gas with $\Delta v_{\rm max}$ between 500 and a few 1000~\kms. Of the starburst galaxies in our sample with outflow detections, $\Delta v_{\rm max}$ ranges between 400--800~\kms\ (depending on the method used to calculate $\Delta v_{\rm max}$). Some systems classified as Sy2 (i.e.\ contain an AGN) have $\Delta v_{\rm max}$ $>$ 1000~\kms. These fast outflows cannot be related to the AGN broad line region (BLR) since in all cases they are spatially extended and seen in both the recombination and forbidden lines. Four of the fastest outflows are found in:
  \begin{itemize} 
    \item IRAS 00198-7926: we find line widths $>$1000~\kms\ over a $\sim$3\,$\times$\,3 spaxel region ($\sim$3\,$\times$\,3~kpc) in a region to the west of the northern nucleus. These very broad lines are redshifted by 300--450~\kms\ with respect to systemic. We also detect broad line components (FWHM=400--500~\kms) that are blueshifted by a similar amount (300--450~\kms), implying the presence of a bipolar wind emerging from behind obscuring material associated with the northern nucleus.
    \item IRAS 05189-2524: in this known Sy2 system we identify four spaxels in H$\alpha$ C3 with line widths $>$2000~\kms, which are all blueshifted by few 1000~\kms. Blue wings to the [S\two] doublet line profile are also seen. These results, together with those at other wavelengths from the literature, strongly suggest that this emission originates in a fast nuclear wind.
    \item IRAS 06035-7102: we find line widths $>$2000~\kms\ over nine contiguous spaxels ($\sim$2.5\,$\times$\,2.5~kpc) coincident with the western nucleus. These are all strongly blueshifted. The classification of this system is uncertain, however these results suggest that this western nucleus contains a hidden AGN that is driving the outflow.
    \item IRAS 19254-7245: we find line widths $>$2000~\kms\ over 34 contiguous spaxels in C2 and C3, coincident with the southern nucleus \citep[in good agreement with][]{bendo09}. This nucleus is known to harbour an AGN that is presumably driving this outflow.
  \end{itemize}
  The fact that our Sy2 ULIRGs exhibit faster outflows than the pure starburst systems is consistent with previous studies that suggest AGN are required to drive winds to velocities $\gtrsim$500~\kms\ \citep{lipari03, rupke05b}.
  \item Our spatially resolved spectroscopy has allowed us to map the outflows kinematically and morphologically, and in some cases determine for the first time which nucleus is driving the wind and examine the interplay between the outflow and the host galaxy.
  \item Eight systems have clear rotating gaseous disks, and for these we measure rotation velocities, derive virial masses, and calculate the Toomre $Q$ parameter \citep{toomre64}. For these galaxies, masses fall in the range $6\times 10^{9}$--$1.4\times 10^{11}$~\Msol. Three of the eight have Q$<$1 and 5 have Q$>$1. We interpret this variation in terms of the concept of self-regulated star formation where feedback regulates $Q$ to $\sim$1. 
  \item We find [N\two]/H$\alpha$, [S\two]/H$\alpha$ and [O\one]/H$\alpha$ line ratio gradients in a significant number of systems in our study. Some exhibit gradients in just [S\two]/H$\alpha$ and [O\one]/H$\alpha$, and some in all three ratios. Depressed line ratios in the nuclear regions are found in the less disturbed systems (either isolated class 0 or post-coalescence class 2), and a depression in all three ratios is associated with high IR luminosity and a central nuclear concentration. We discuss a number of possible reasons for this, including ionization effects and abundance gradients, and conclude that these gradients are most likely a result of an increase in the hardness of the radiation field responsible for the observed emission lines with radius, most likely due to an increasing contribution of shocks.
  \item Neutral and/or molecular gas outflows have been identified in some galaxies for which we do not see an ionized phase counterpart in our data. The reverse also holds: some galaxies where we find an ionized gas wind have not been found to have a neutral counterpart. Why this might be is a matter of on-going research. It is possible that different wind driving mechanisms (e.g.\ ram pressure vs.\ radiation pressure) may be playing larger or smaller roles in accelerating different material at different radii, times and temperature phases.
  \item In many cases, we observe strong H$\alpha$ emission in tidal features/tails implying vigorous star formation is occurring there. Since it is known that large stellar structures built in tidal debris may end up as tidal dwarf galaxies \citep{weilbacher00}, it is possible that the collisions that form ULIRGs may be a significant source of these types of objects.
\end{itemize}

Because of the stellar and gas dynamics resulting from the major merger (which is the root-cause of the ULIRG state), and the interaction between the different gas phases and feedback-driving mechanisms from the ensuing powerful starbursts and AGN, ULIRGs exhibit extremely complex kinematics and gas properties.  Even in local infrared-luminous mergers such as NGC 6240 or NGC 4038/9, the complex kinematics of the ionized gas make it difficult to distinguish between wind outflows and tidal motions directly produced by the interaction. High spatial resolution, multi-wavelength observations of the launching sites of winds are required to disentangle the complex kinematics produced in ULIRGs \citep[e.g.][]{rupke11}. Furthermore, as this study has shown, any system when looked at in detail shows peculiarities that need detailed study to understand. This makes the case for high resolution follow-up observations even clearer.

The following Appendix contains a discussion of each ULIRG system individually, with reference to the existing literature.

\section*{Acknowledgments}
We thank Santiago Arribas for kindly letting us have the reduced VIMOS data cubes for IRAS 05189-2524 and IRAS 06035-7102. MSW wishes to thank Ana Monreal-Ibero, Dave Rupke, Ric Davies, and Mark Swinbank for discussions on various aspects of this paper. We also thank the referee for their thorough reading and constructive comments and suggestions that led to an improvement in this paper. This research has made use of the NASA/IPAC Extragalactic Database (NED) which is operated by the Jet Propulsion Laboratory, California Institute of Technology, under contract with the National Aeronautics and Space Administration. The research leading to these results has received funding from the European Community's Seventh Framework Programme (/FP7/2007-2013/) under grant agreement No 229517.

\bibliographystyle{mn2e}
\bibliography{/Users/mwestmoq/Dropbox/Work/references}
\bsp


\clearpage

\begin{figure*}
\centering
\includegraphics[width=0.95\textwidth]{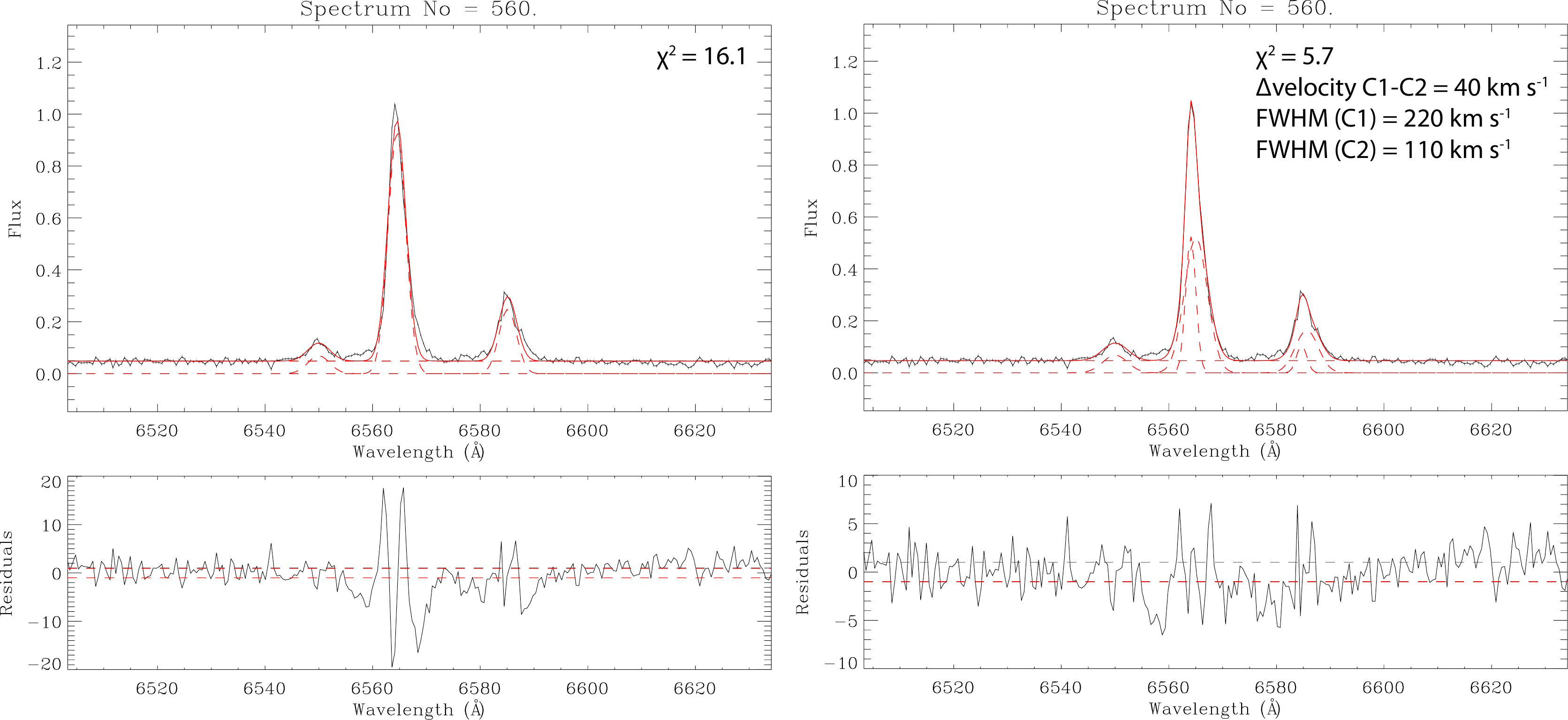}
\caption{Example single and double Gaussian fits to the H$\alpha$ and [N\two]$\lambda$6583 lines from a spaxel in IRAS 20046-0623 (the flux ratio of [N\two]$\lambda$6583 to 6548 is fixed at 3:1 by atomic physics so we do not attempt to accurately model the [N\two]$\lambda$6548 line profile). The $\chi^{2}$ for both fits is shown, together with a plot of the residuals (in units of $\sigma$). For the double-Gaussian fit, the velocity difference between C1 and C2, and the FWHM of each is shown. These plots demonstrate how two components can be needed to fit a line profile, even though their velocity separation is small. In this example hints of a faint blue wing to the lines are visible, although the fitting of a third component would not be accepted by our $\chi^{2}$ ratio tests.}
\label{fig:egfit}
\end{figure*}

\clearpage

\begin{figure*}
\ContinuedFloat*
\centering
\includegraphics[height=0.89\textheight]{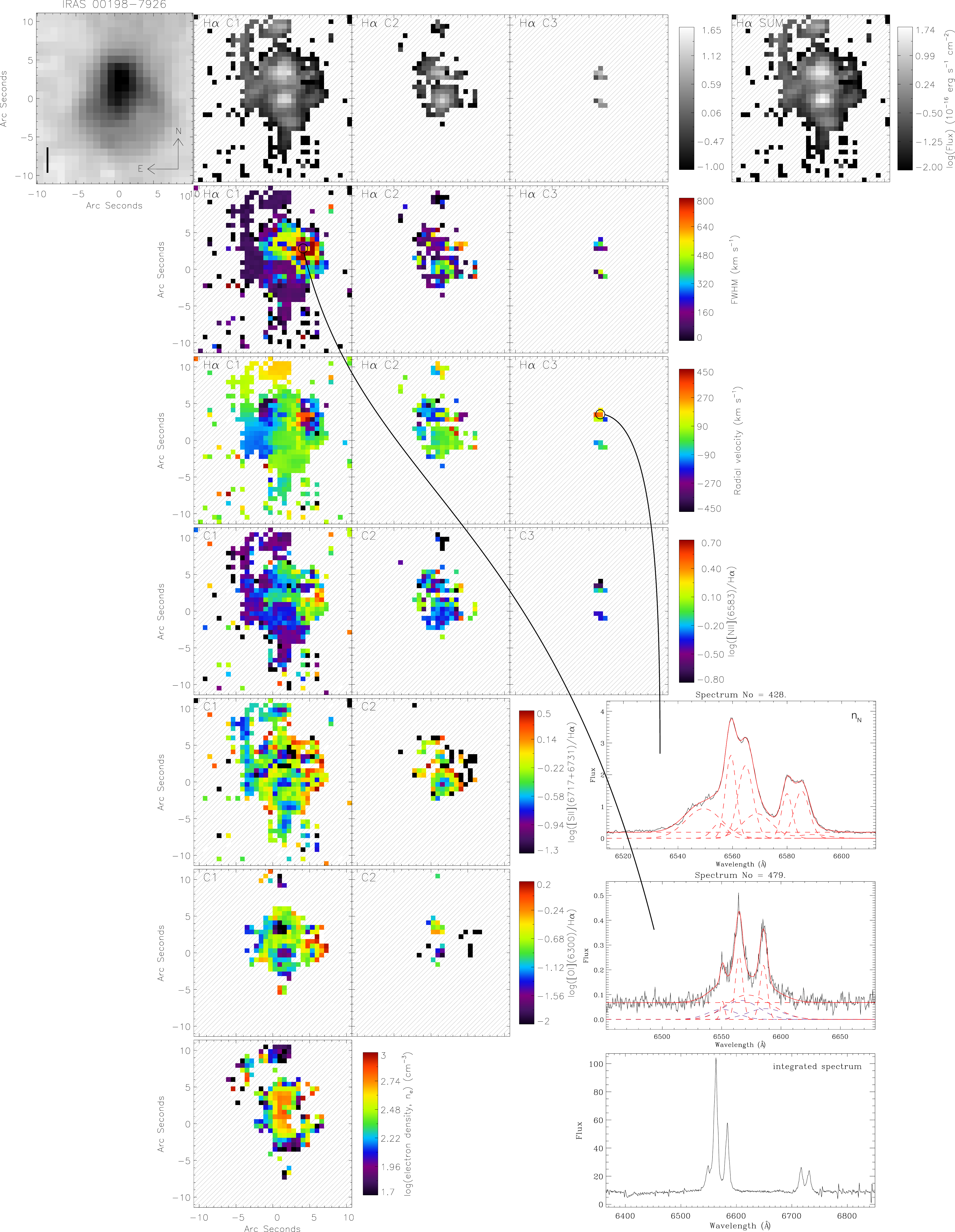}
\caption{Maps showing the H$\alpha$ flux, FWHM and radial velocity, [N\two]/H$\alpha$, [S\two]/H$\alpha$ and [O\one]/H$\alpha$ line flux ratios and electron density distributions for each identified line component in each of the galaxies in our sample (hatched regions represent non-detections). The FWHMs are corrected for instrumental broadening, and the radial velocities are relative to $v_{\rm sys}$. The top-left panel shows the corresponding \textit{HST} F814W image (for IRAS 00198-7926, IRAS 03068-5346, IRAS 20046-0623 and IRAS 21504-0628 no \textit{HST} imaging is available so we instead show the DSS image); the vertical bar shows a linear scale of 5~kpc. The top-right panel shows the summed (C1+C2+C3) H$\alpha$ flux map. The [S\two] and [O\one] datacubes were smoothed in the spatial directions by a 3$\times$3 rolling summation to increase S/N before the emission lines were fit. This also applies to the electron density maps since densities were derived from the summed (all components) [S\two] line ratios. Example H$\alpha$+[N\two] line profiles from selected spaxels are shown as insets, together with their multi-component Gaussian fits (dashed red lines; wavelengths are rest-frame). Nuclear spectra are labelled ``n''. The bottom inset in each case is the integrated spectrum of the whole system. }
\label{fig:iras00198-7926}
\end{figure*}

\begin{figure*}
\ContinuedFloat
\centering
\includegraphics[height=0.95\textheight]{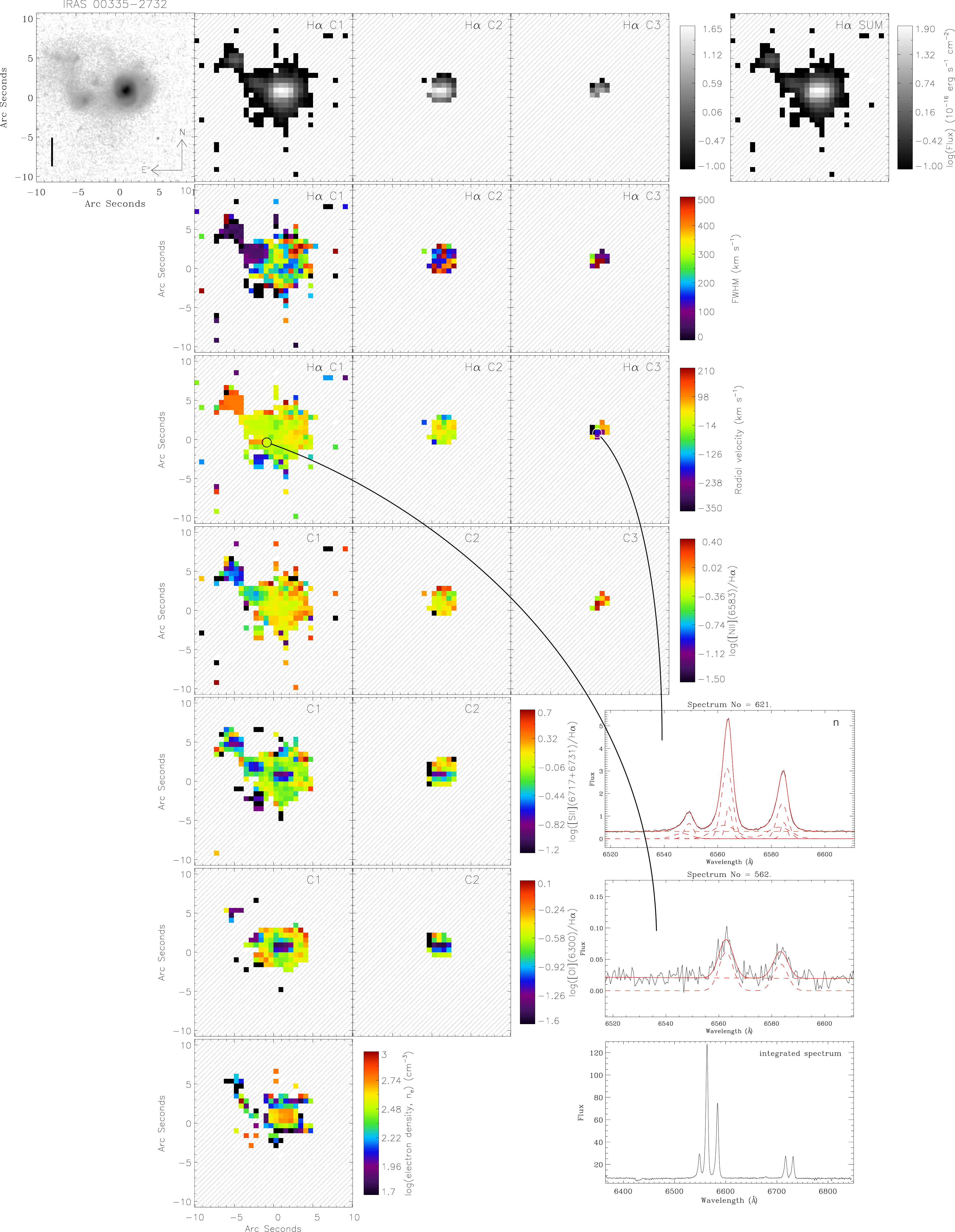}
\caption{}
\label{fig:iras00335-2732}
\end{figure*}

\begin{figure*}
\ContinuedFloat
\centering
\includegraphics[height=0.95\textheight]{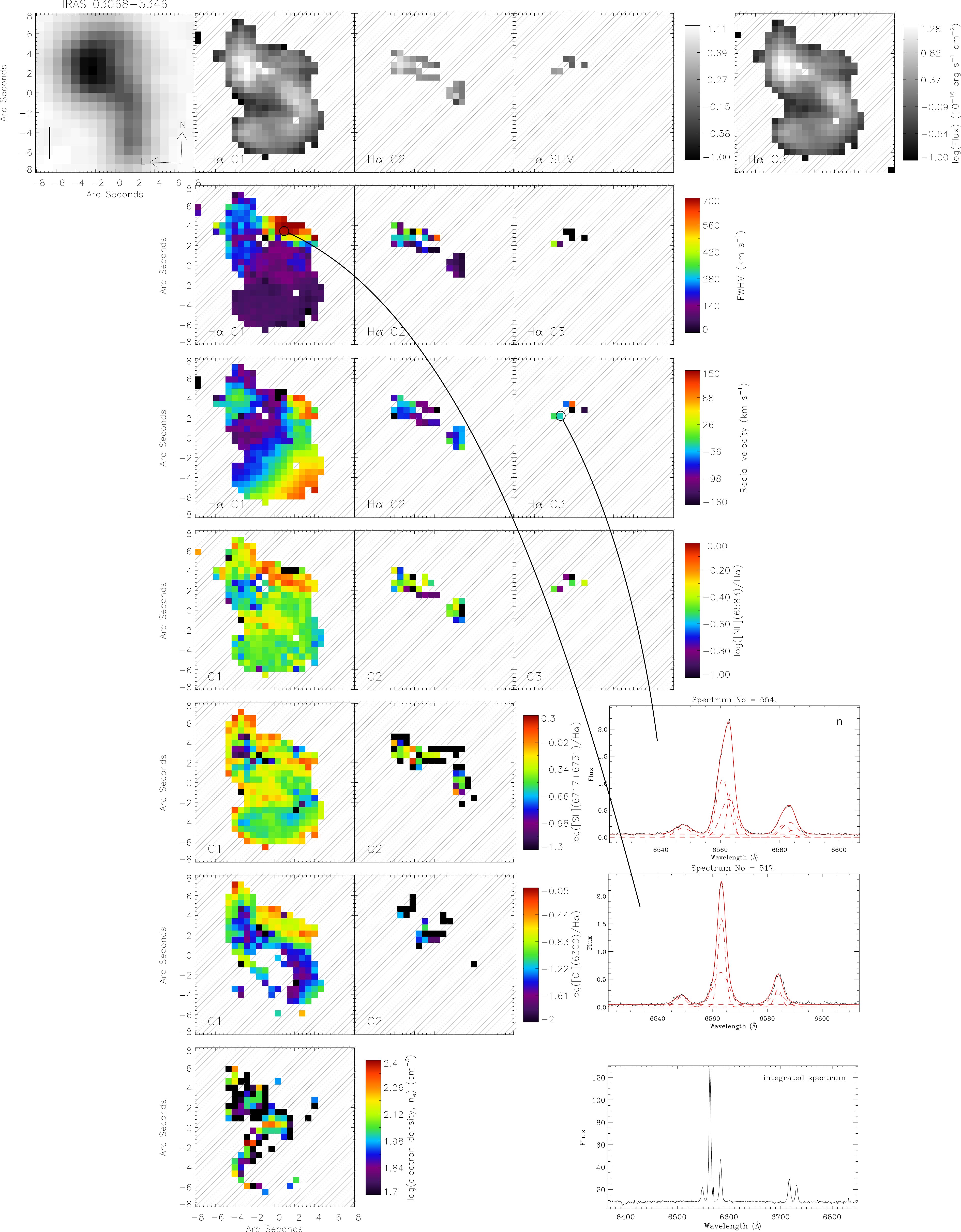}
\caption{}
\label{fig:iras03068-5346}
\end{figure*}

\begin{figure*}
\ContinuedFloat
\centering
\includegraphics[height=0.95\textheight]{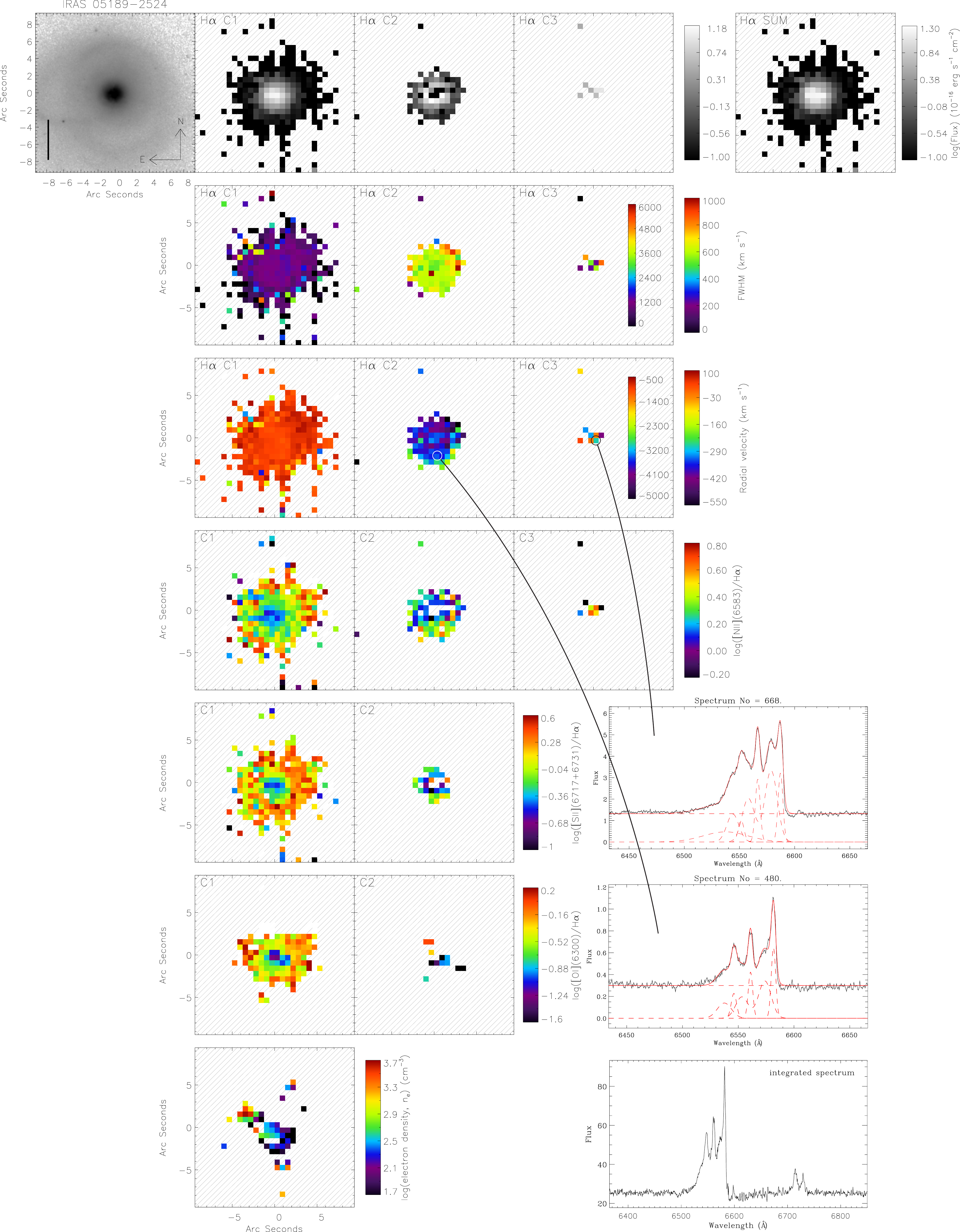}
\caption{The H$\alpha$ FWHM and radial velocity maps for C3 are separately scaled as shown to accommodate the very broad and blueshifted values.}
\label{fig:iras05189-2524}
\end{figure*}

\begin{figure*}
\ContinuedFloat
\centering
\includegraphics[height=0.95\textheight]{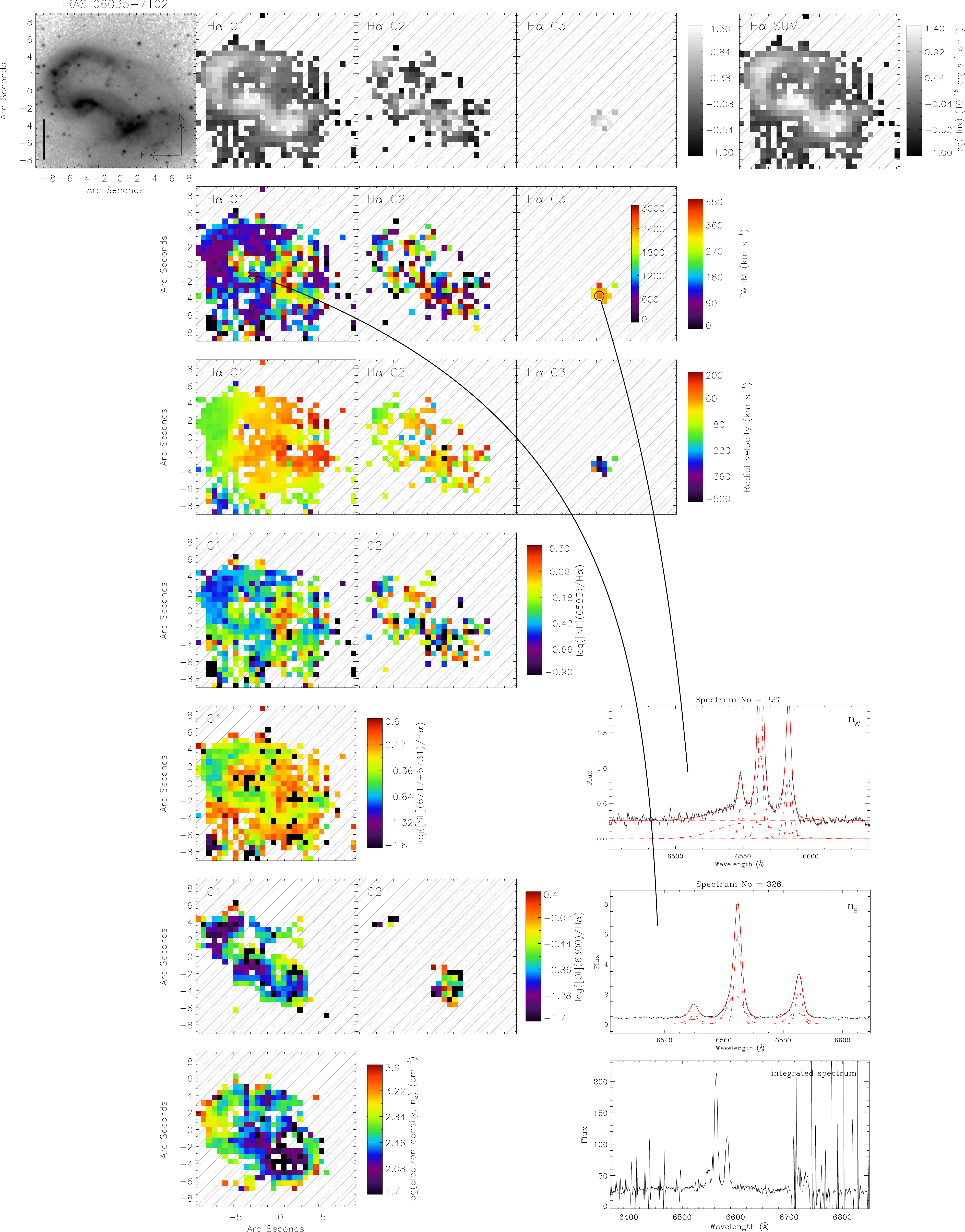}
\caption{The H$\alpha$ FWHM map for C3 is separately scaled as shown to accommodate the very large values. The large number of spikes in the integrated spectrum is due to incompletely subtracted sky emission lines.}
\label{fig:iras06035-7102}
\end{figure*}

\begin{figure*}
\ContinuedFloat
\centering
\includegraphics[height=0.95\textheight]{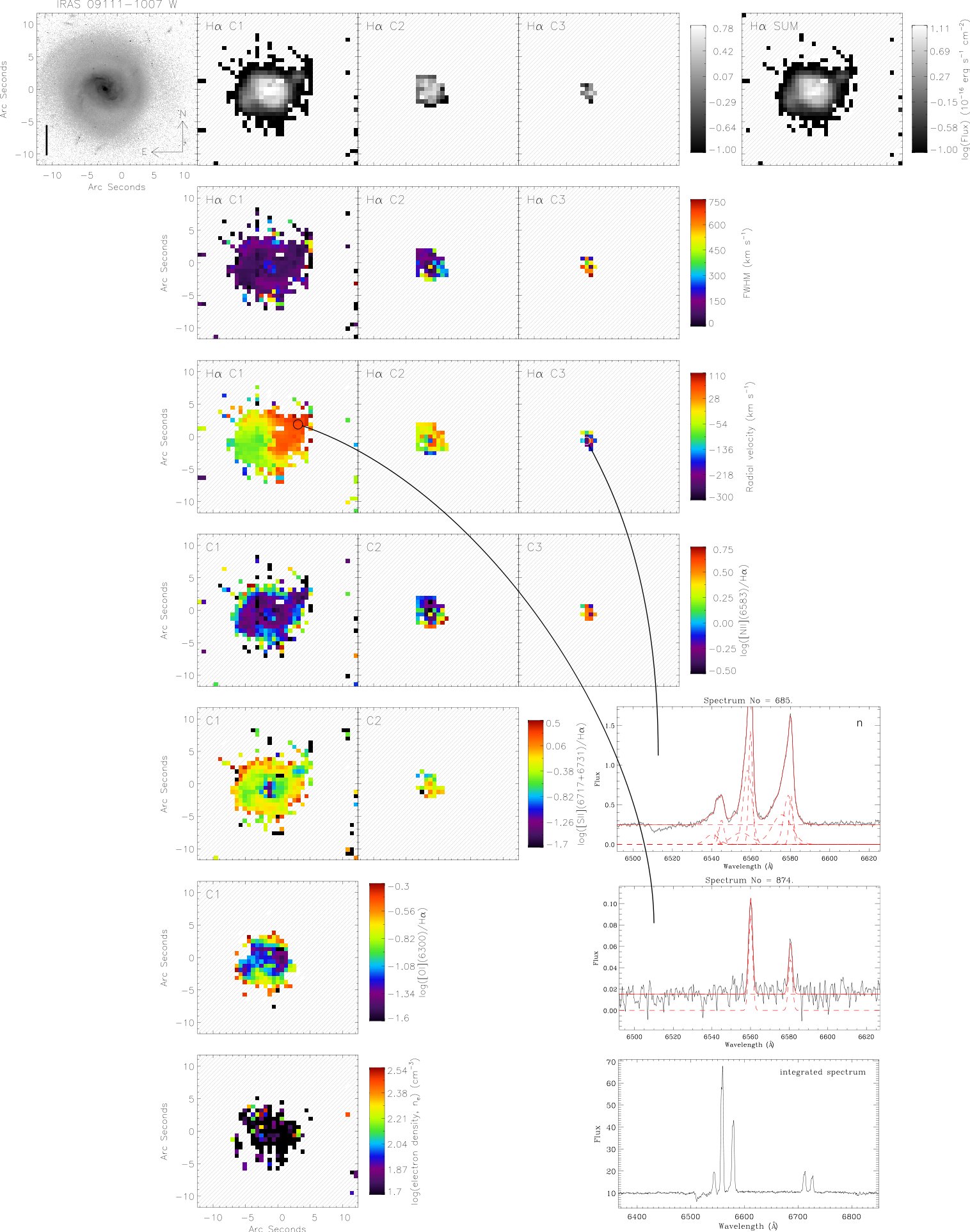}
\caption{}
\label{fig:iras09111-1007}
\end{figure*}

\begin{figure*}
\ContinuedFloat
\centering
\includegraphics[height=0.95\textheight]{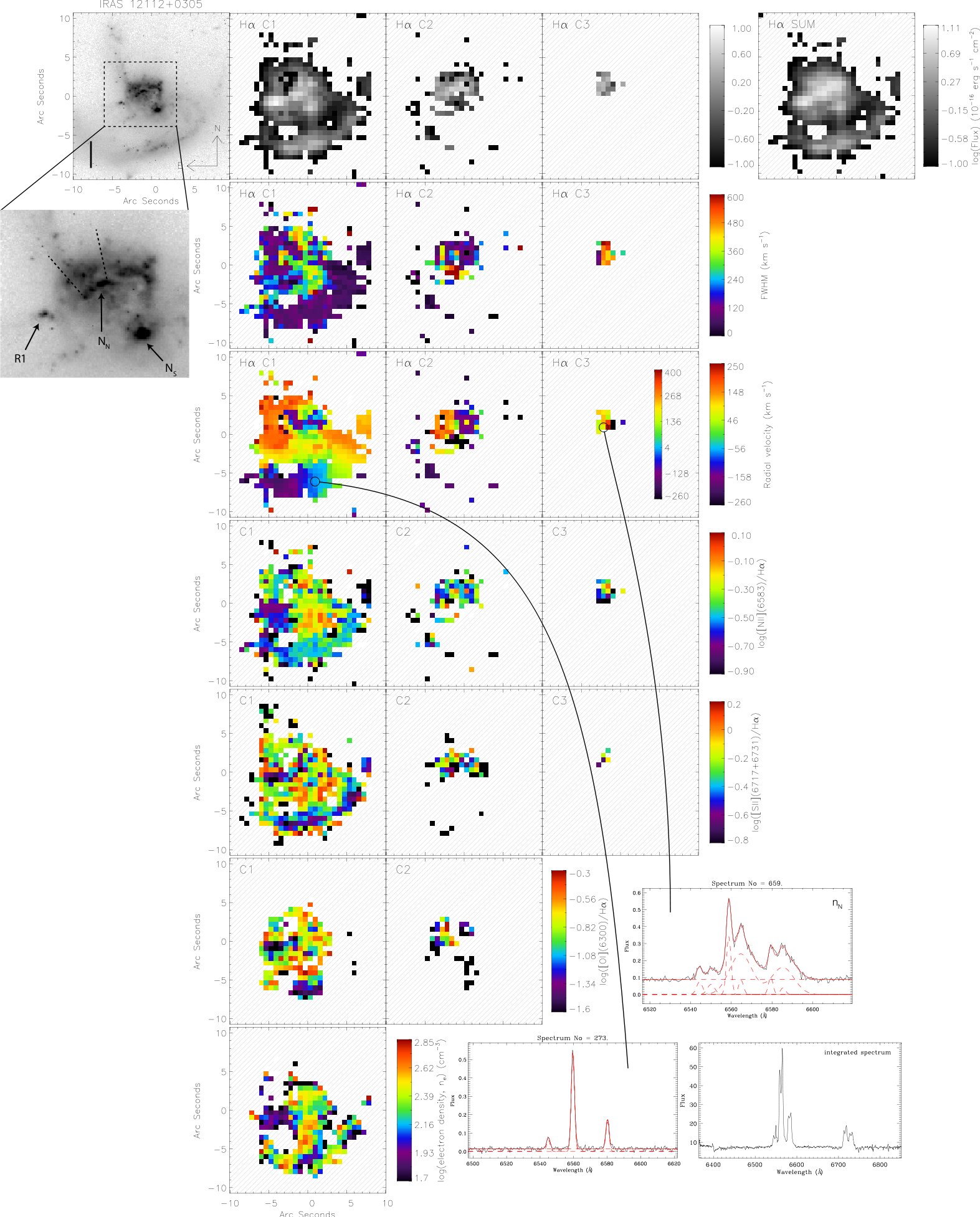}
\caption{The H$\alpha$ radial velocity map for C3 is separately scaled as shown to accommodate the larger values. A blow-up of the region enclosed by the dashed rectangle is shown with various components labelled \citep[N$_{\rm N}$: northern nucleus; N$_{\rm S}$: southern nucleus; R1: H\two\ region 1;][]{colina00}. The dashed lines on the inset indicate the extent and direction of the outflow from N$_{\rm N}$ as found with our data.}
\label{fig:iras12112+0305}
\end{figure*}

\begin{figure*}
\ContinuedFloat
\centering
\includegraphics[height=0.95\textheight]{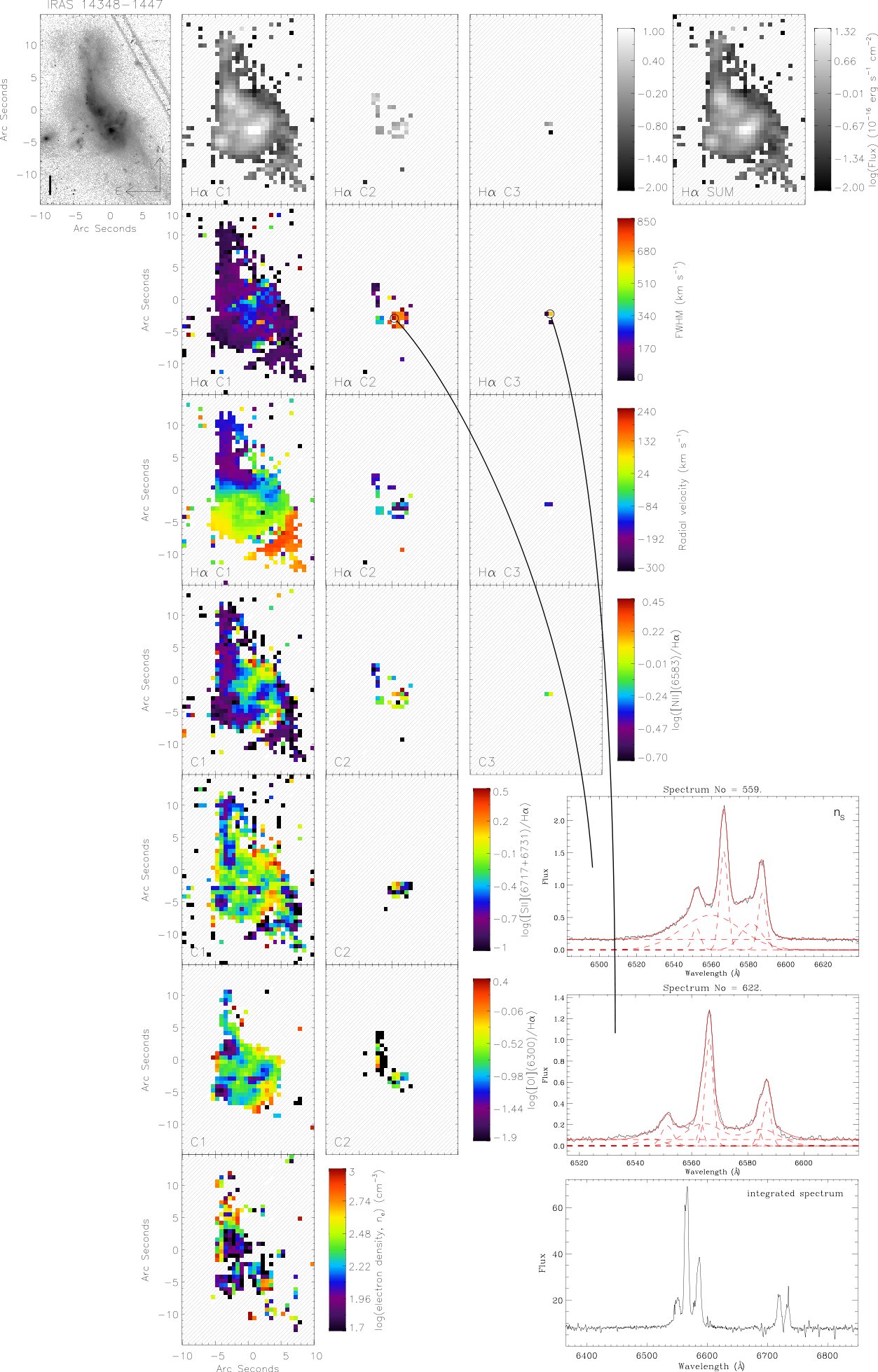}
\caption{}
\label{fig:iras14348-1447}
\end{figure*}

\begin{figure*}
\ContinuedFloat
\centering
\includegraphics[height=0.95\textheight]{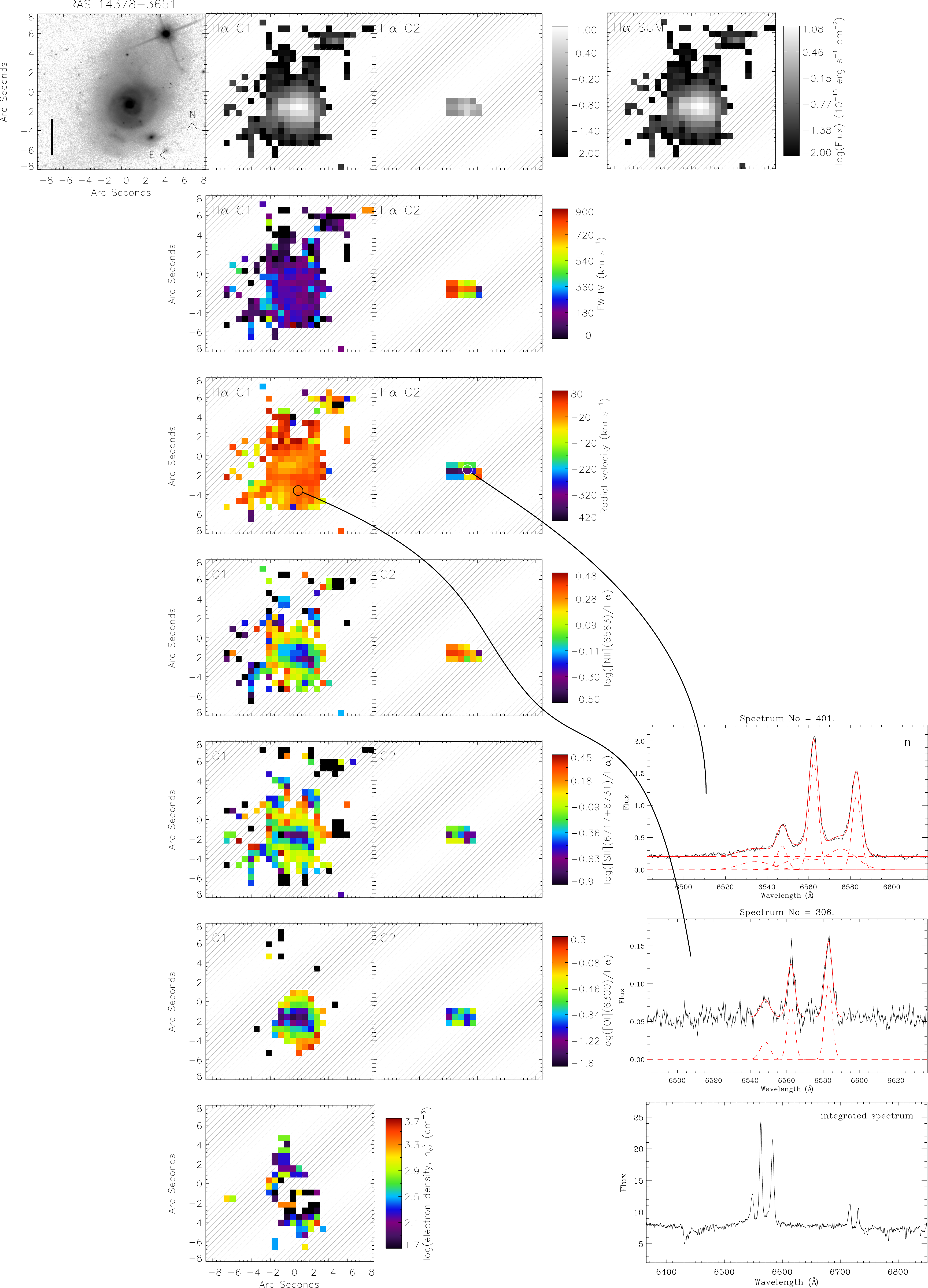}
\caption{}
\label{fig:iras14378-3651}
\end{figure*}

\begin{figure*}
\ContinuedFloat
\centering
\includegraphics[height=0.95\textheight]{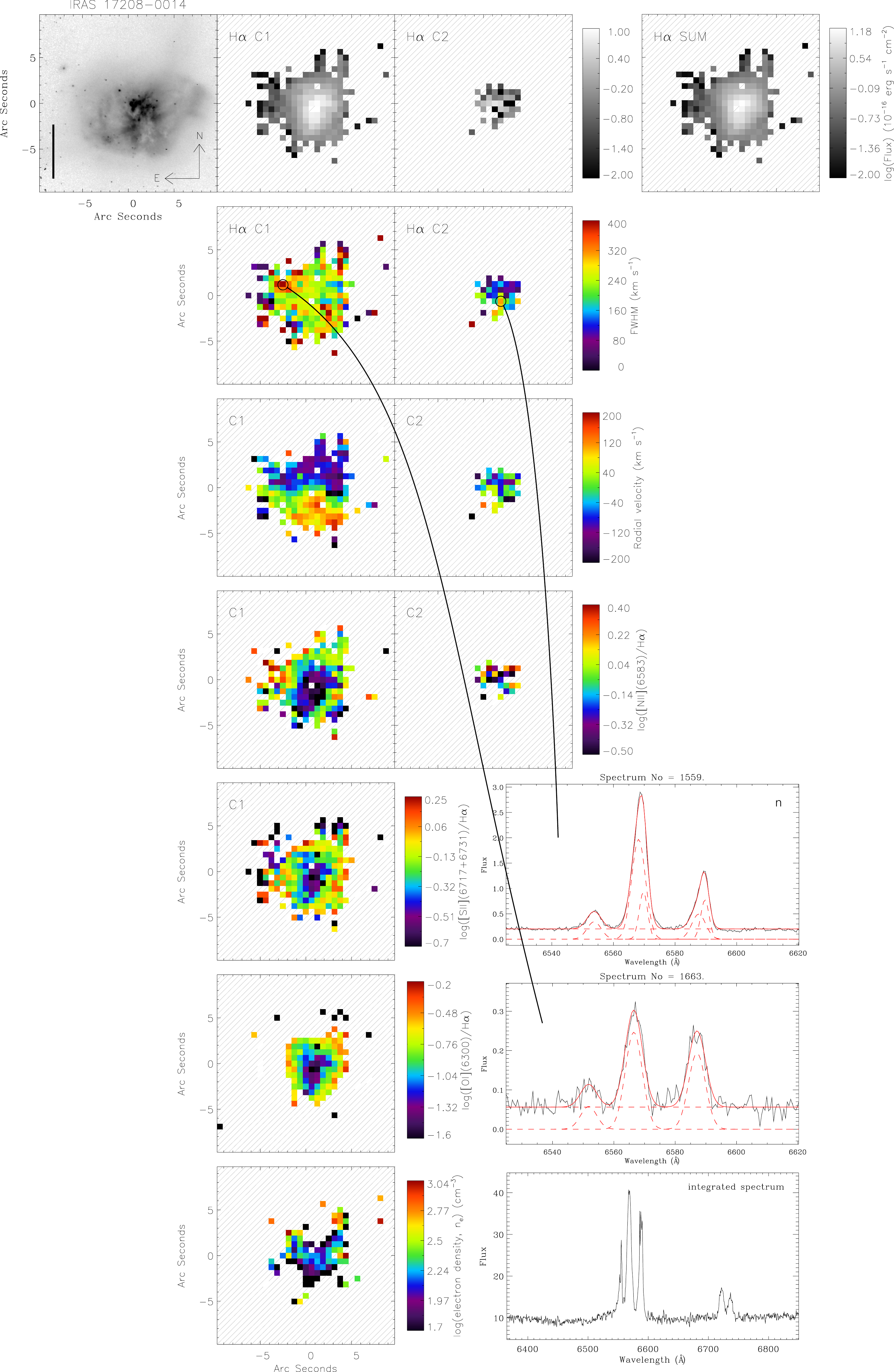}
\caption{}
\label{fig:iras17208-0014}
\end{figure*}

\begin{figure*}
\ContinuedFloat
\centering
\includegraphics[height=0.95\textheight]{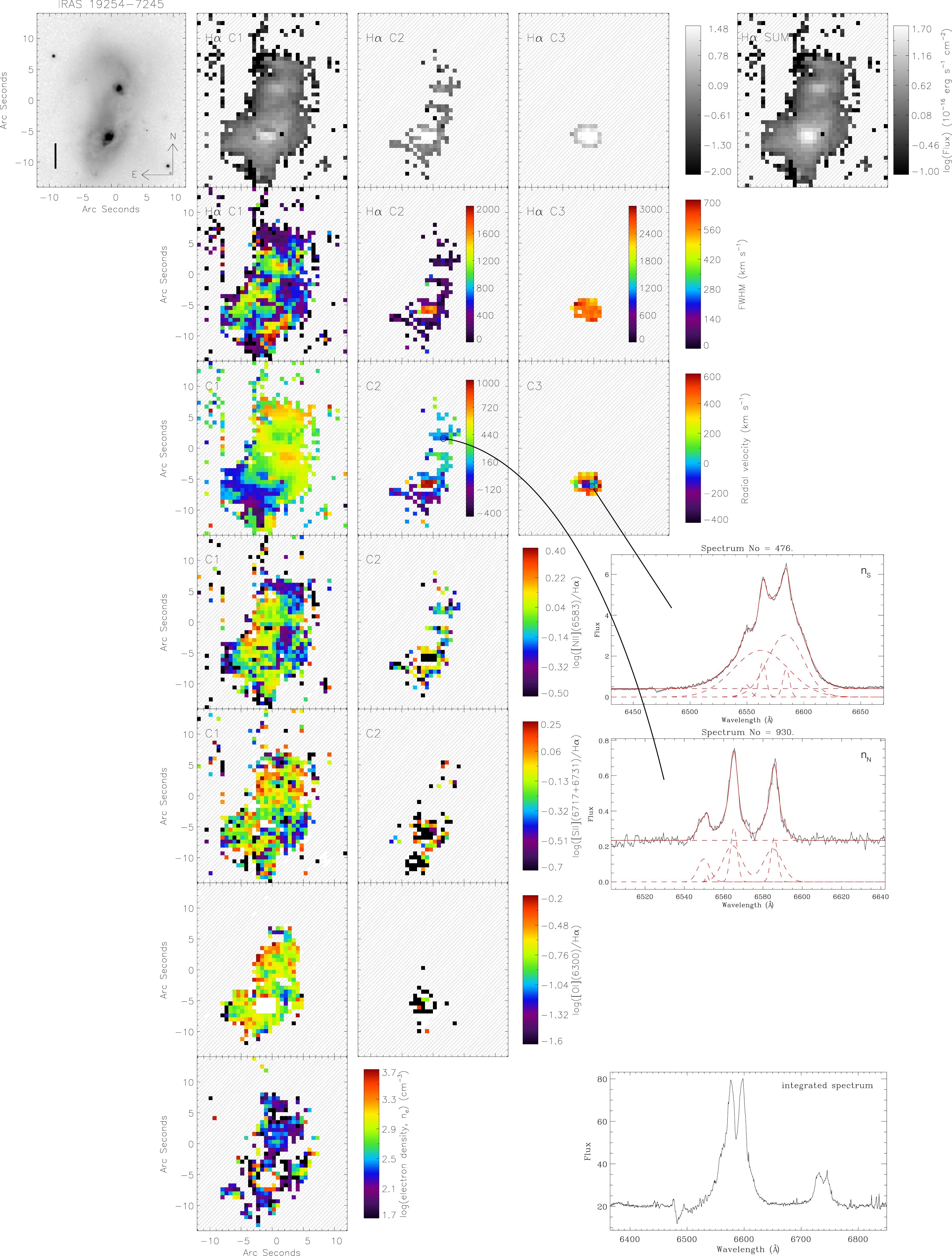}
\caption{The H$\alpha$ FWHM maps for C2 and C3, and the radial velocity map for C2 are separately scaled as shown to accommodate the larger values.}
\label{fig:iras19254-7245}
\end{figure*}

\begin{figure*}
\ContinuedFloat
\centering
\includegraphics[height=0.95\textheight]{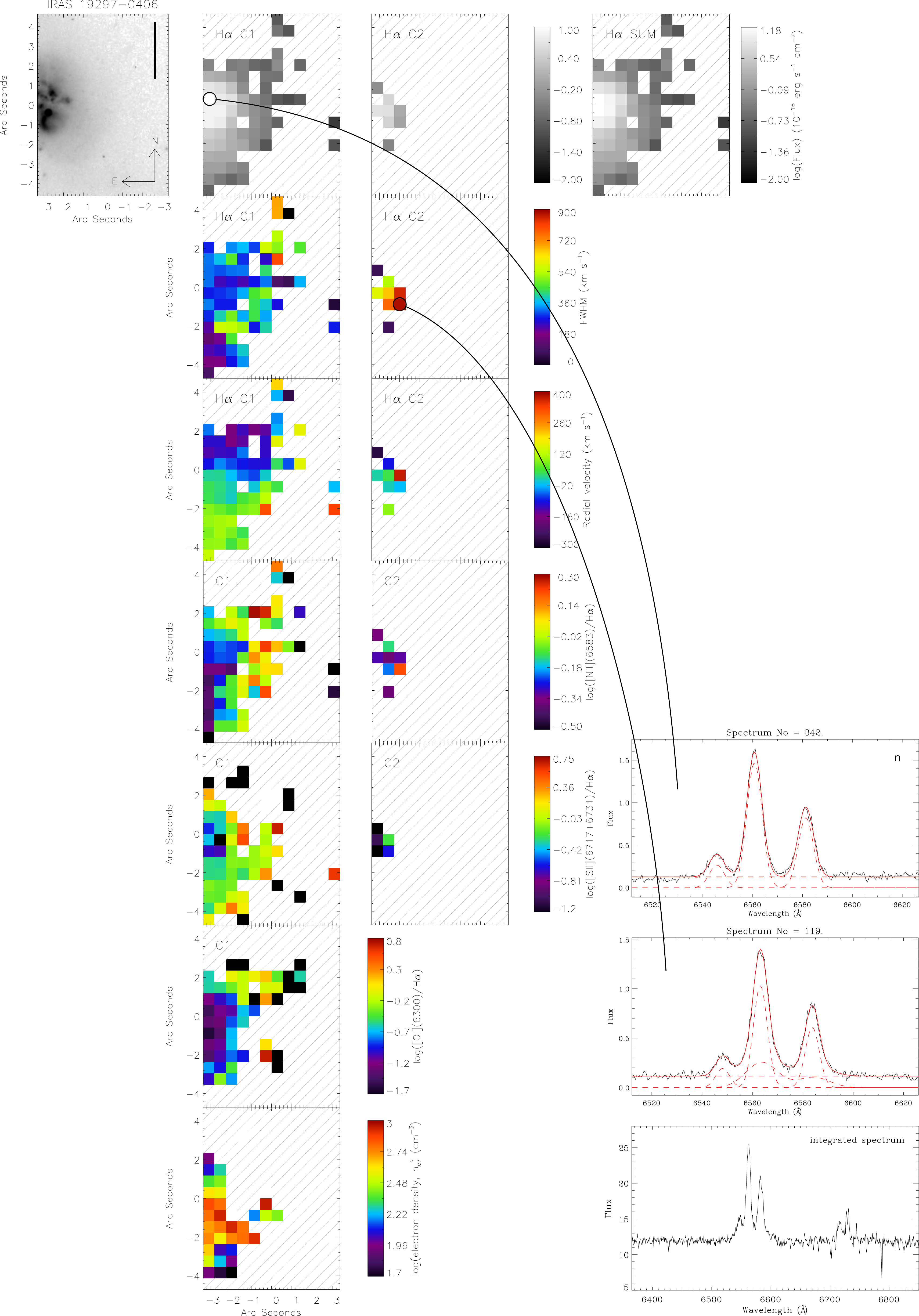}
\caption{}
\label{fig:iras19297-0406}
\end{figure*}

\begin{figure*}
\ContinuedFloat
\centering
\includegraphics[height=0.95\textheight]{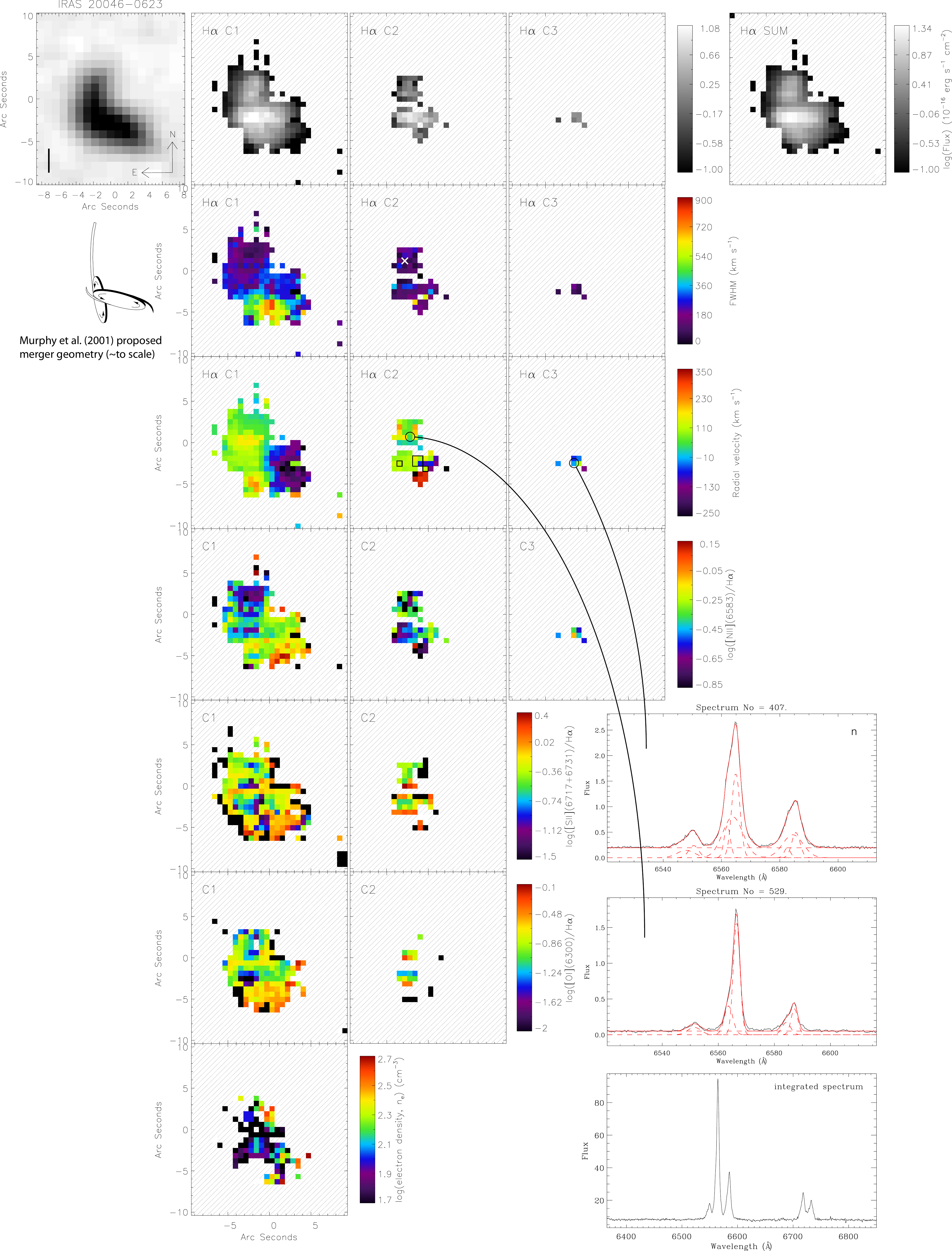}
\caption{Near the top-left we show the \citet{murphy01} proposed merger geometry at approximately the right orientation and scale compared to the DSS image above and the H$\alpha$ maps to the right. For reference, the black outlines on the H$\alpha$ C2 radial velocity map show in which spaxels a third component is also identified. The white cross on the H$\alpha$ C2 FWHM map indicates the spaxel from which the line profile fits in Fig.~\ref{fig:egfit} are from.}
\label{fig:iras20046-0623}
\end{figure*}

\begin{figure*}
\ContinuedFloat
\centering
\includegraphics[height=0.95\textheight]{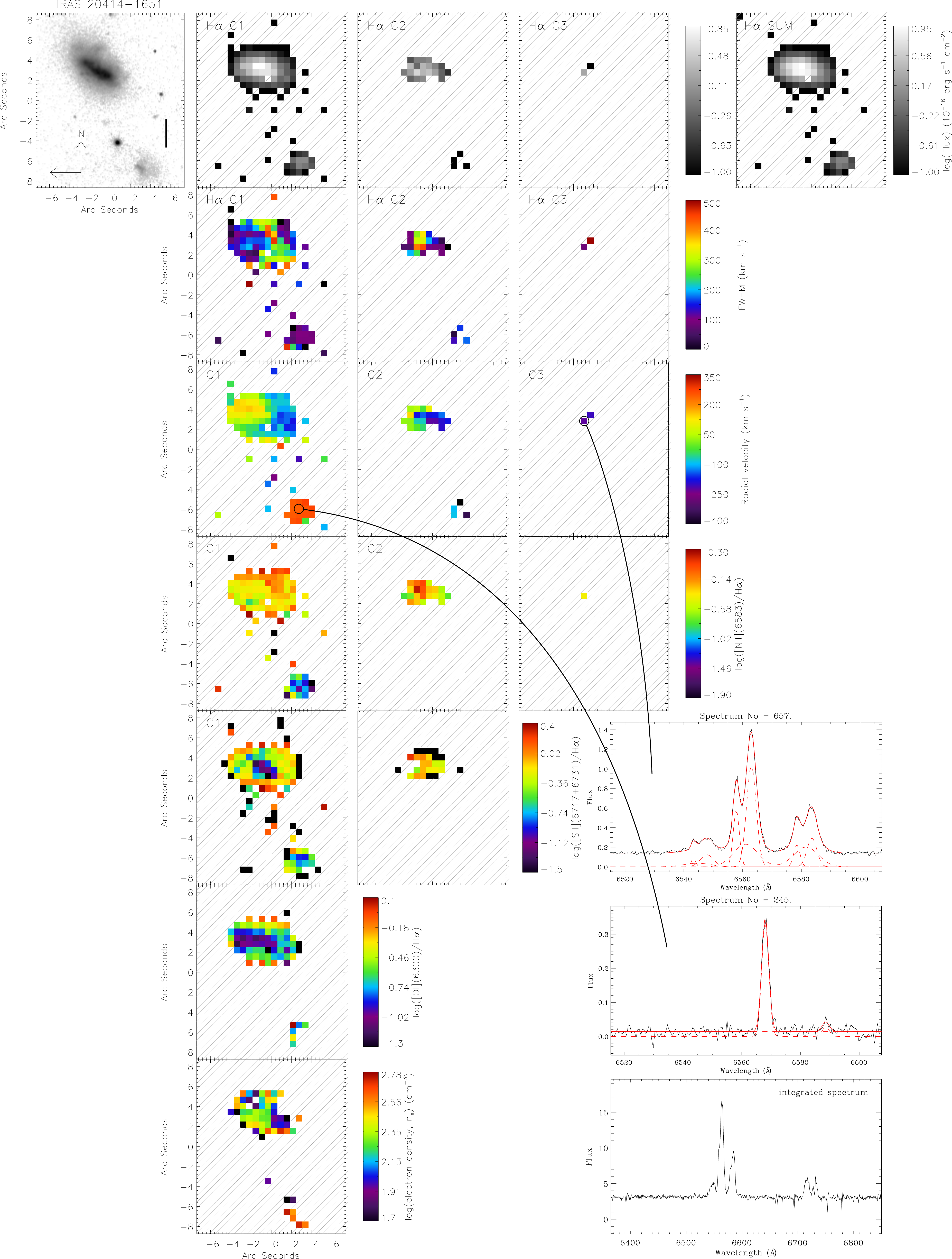}
\caption{}
\label{fig:iras20414-1651}
\end{figure*}

\begin{figure*}
\ContinuedFloat
\centering
\includegraphics[height=0.95\textheight]{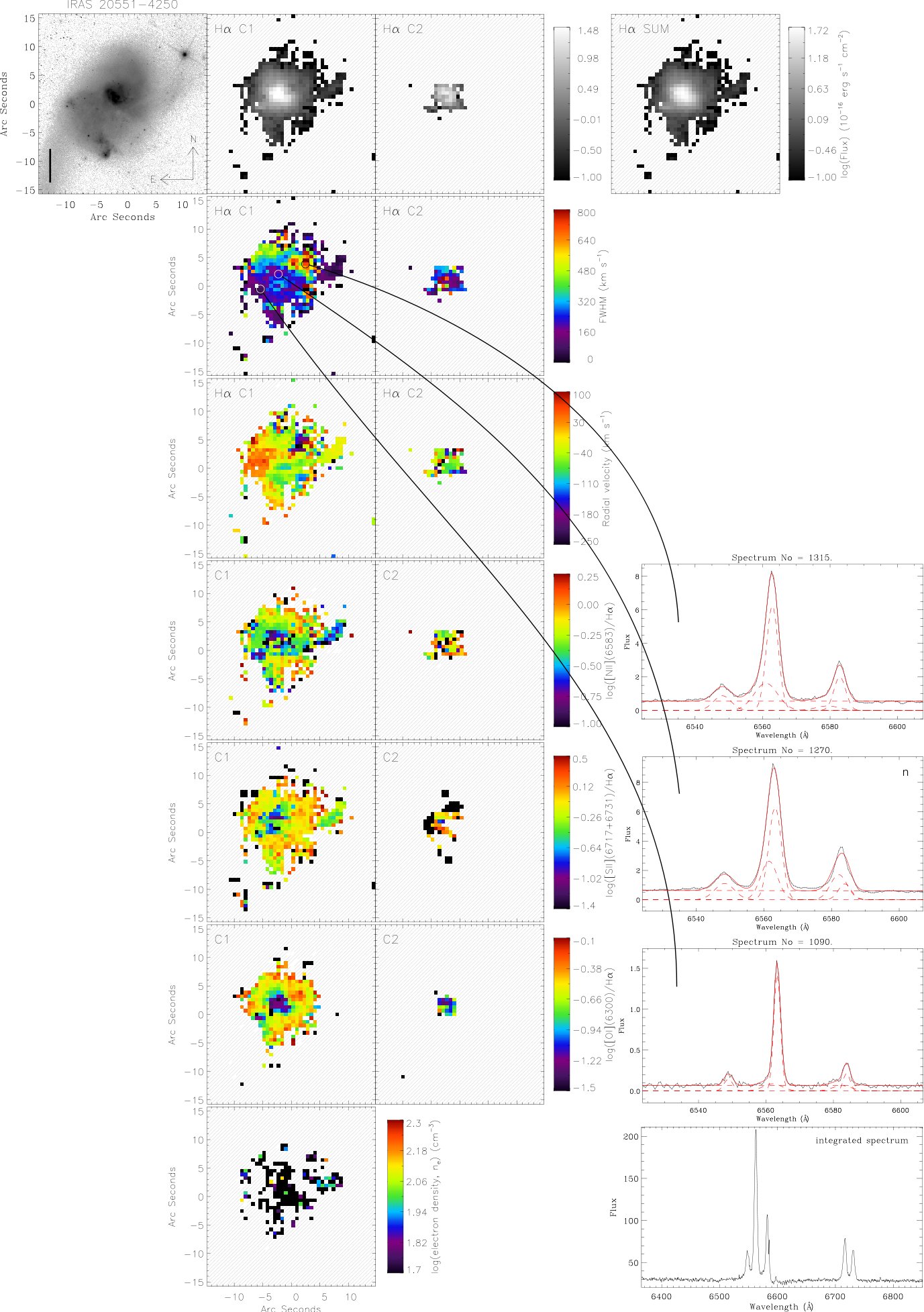}
\caption{}
\label{fig:iras20551-4250}
\end{figure*}


\clearpage

\begin{figure*}
\ContinuedFloat
\centering
\includegraphics[height=0.95\textheight]{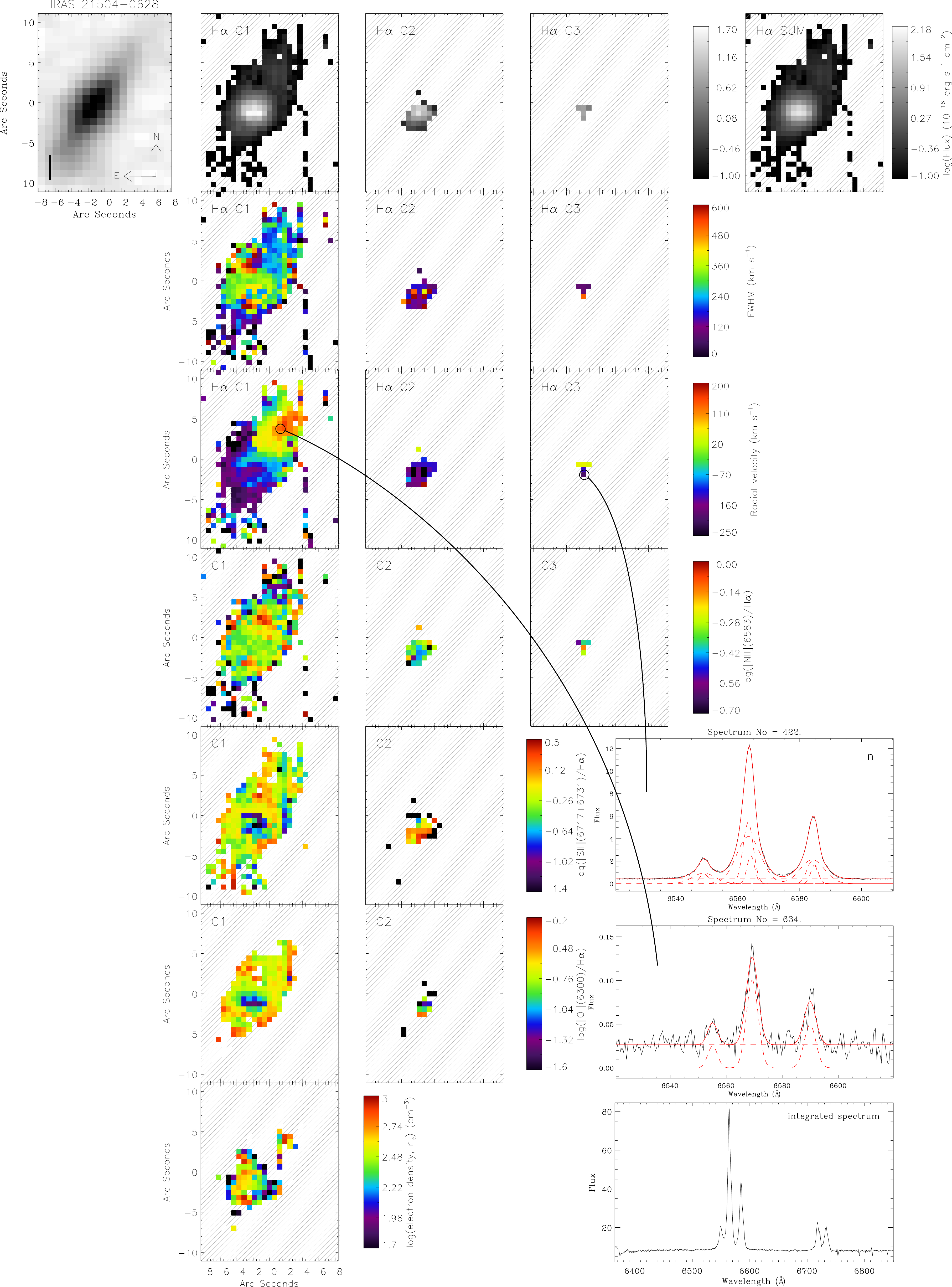}
\caption{}
\label{fig:iras21504-0628}
\end{figure*}

\begin{figure*}
\ContinuedFloat
\centering
\includegraphics[height=0.95\textheight]{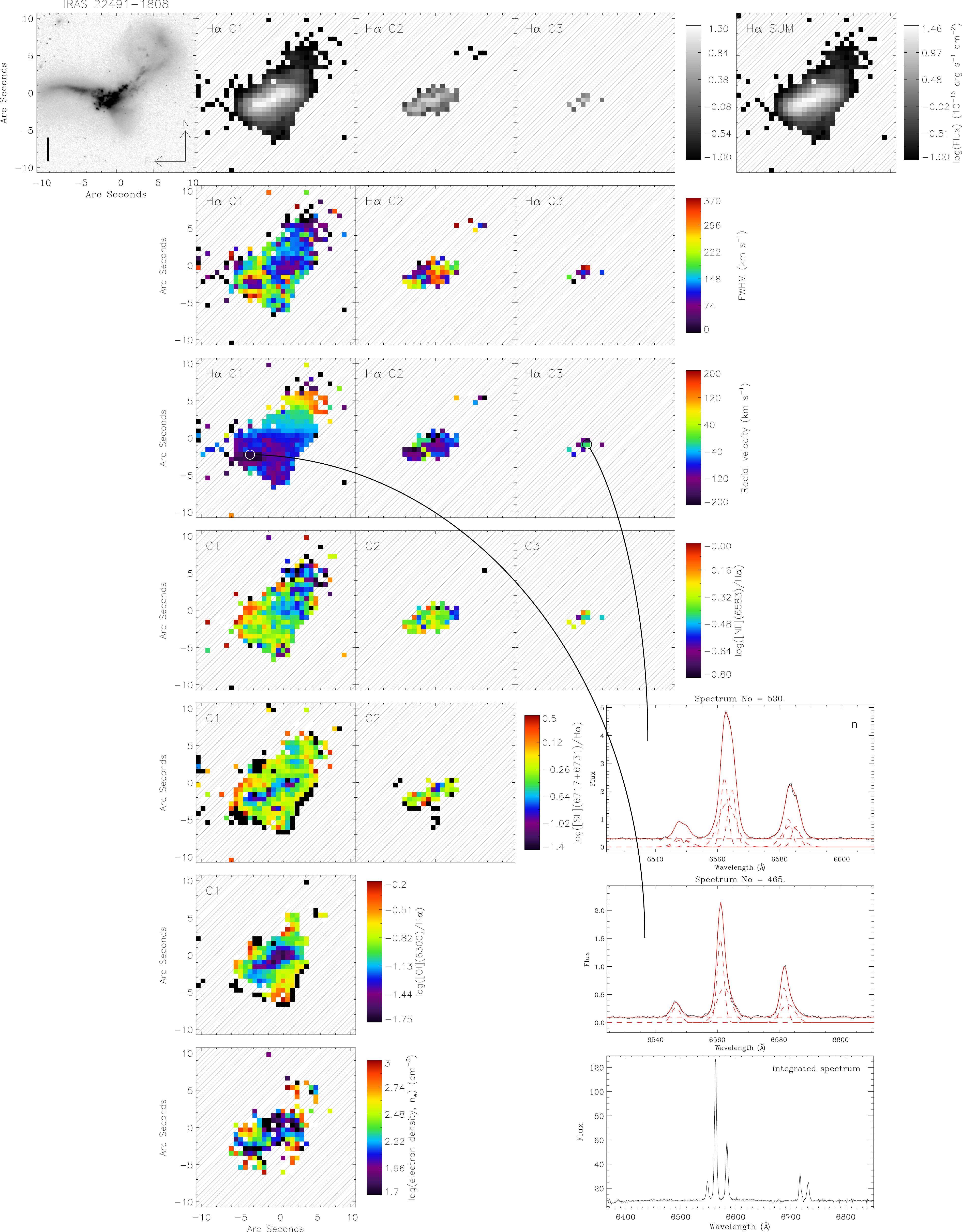}
\caption{}
\label{fig:iras22491-1808}
\end{figure*}

\begin{figure*}
\ContinuedFloat
\centering
\includegraphics[height=0.95\textheight]{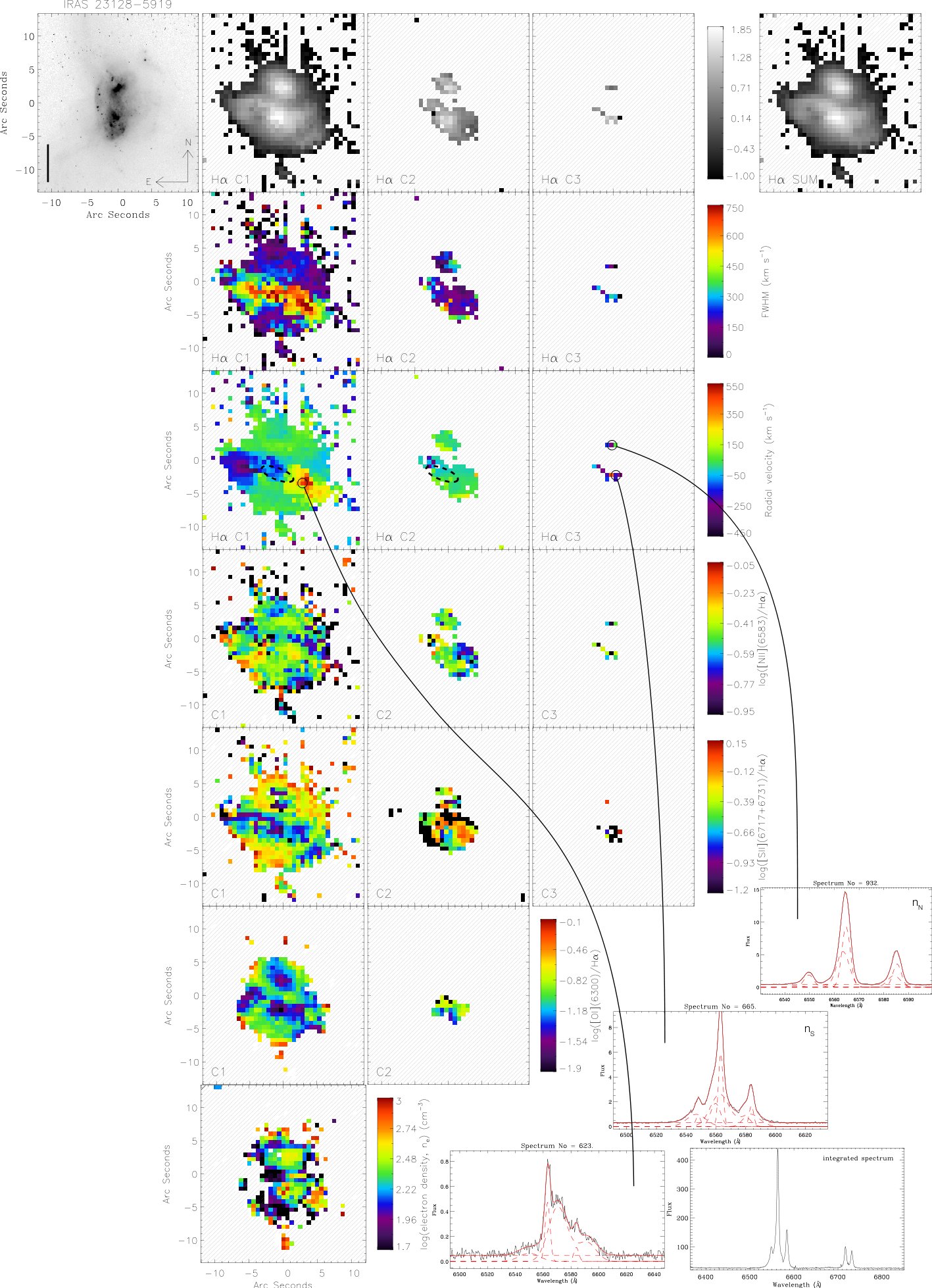}
\caption{The black ellipse on the H$\alpha$ C2 radial velocity map indicates where a third component is also observed.}
\label{fig:iras23128-5919}
\end{figure*}

\begin{figure*}
\ContinuedFloat*
\centering
\begin{minipage}{7cm}
\vspace{0.2cm}
\includegraphics[width=7cm]{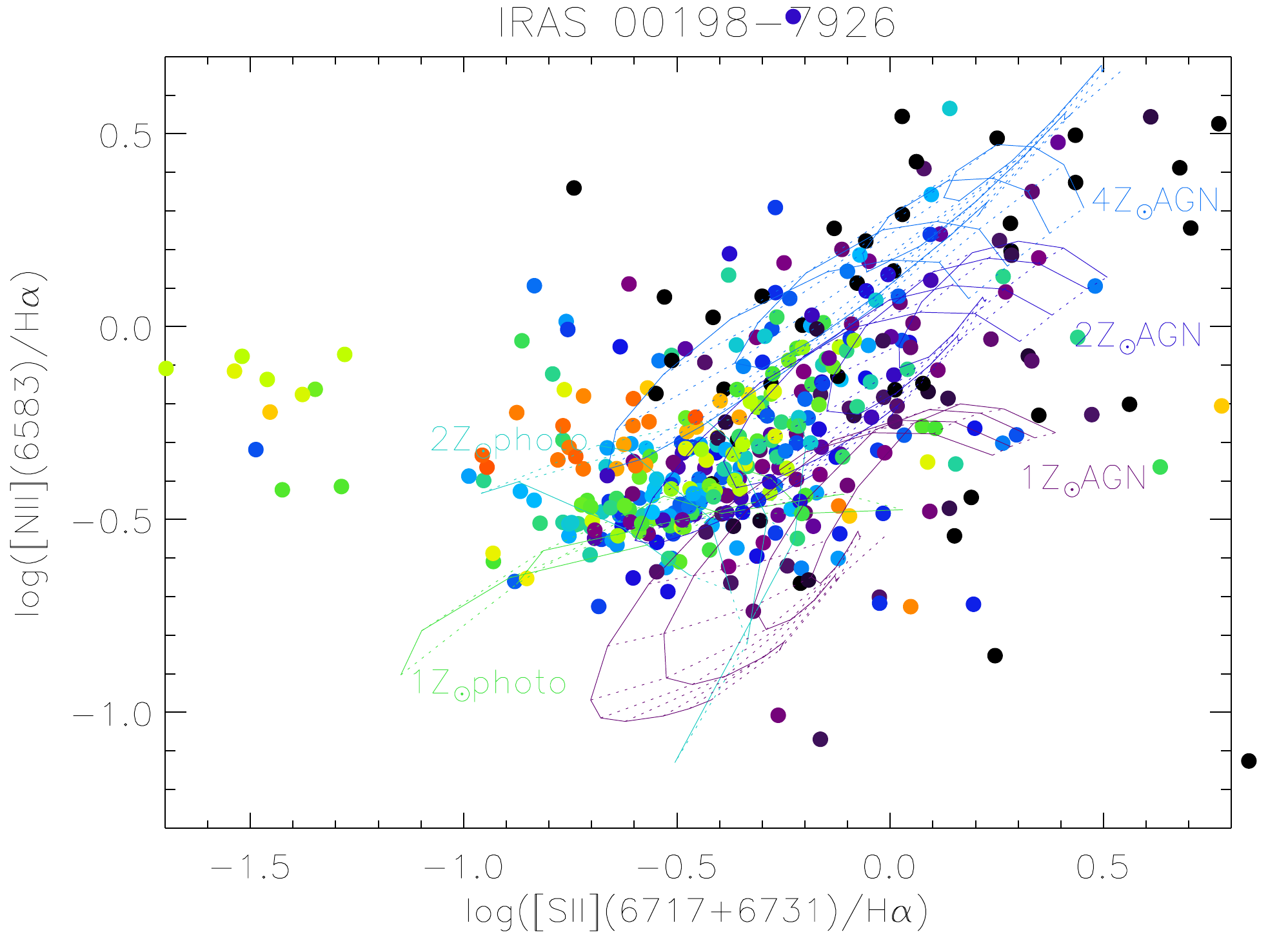}
\end{minipage}
\begin{minipage}{7cm}
\vspace{0.2cm}
\includegraphics[width=7cm]{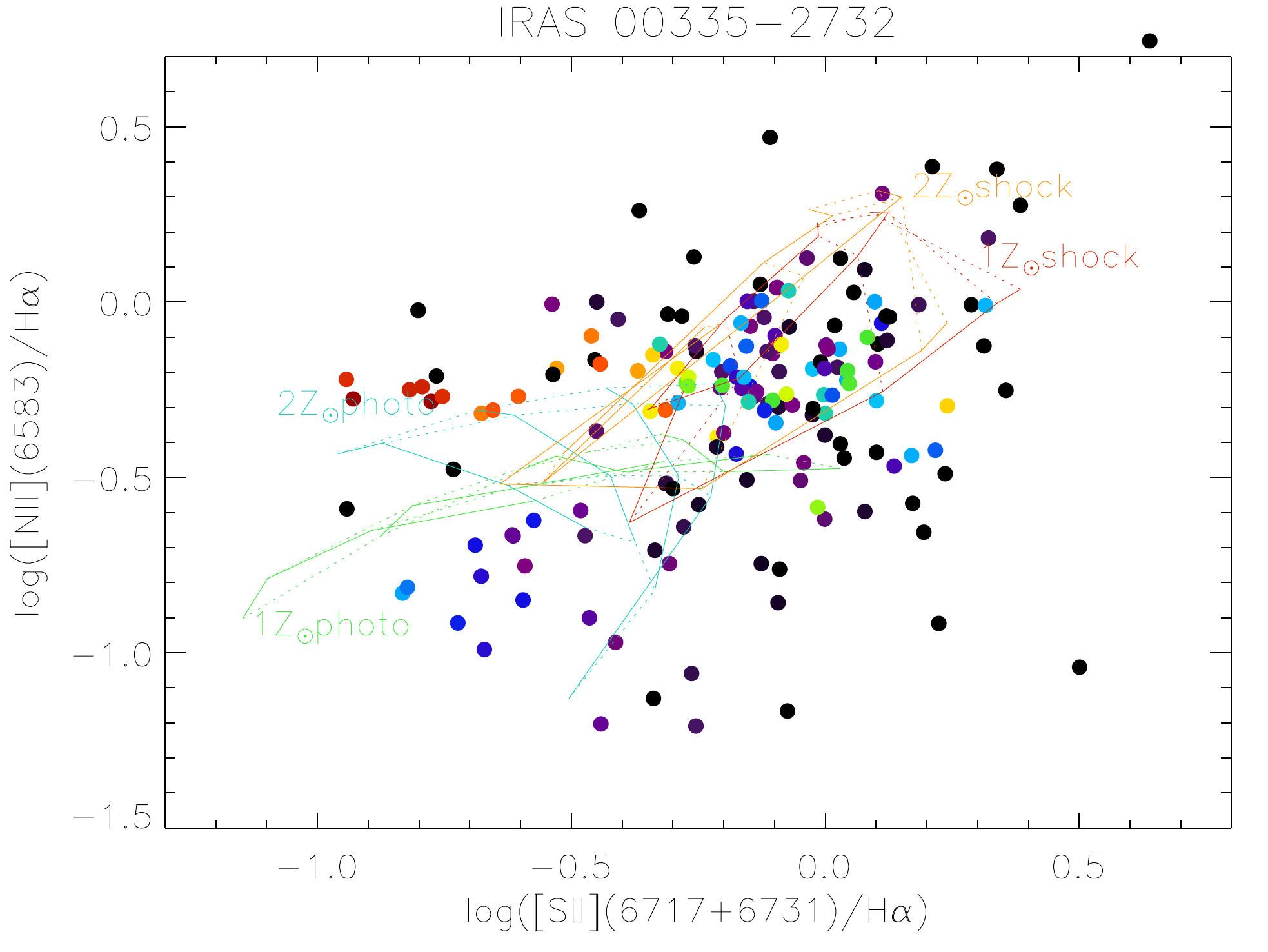}
\end{minipage}
\begin{minipage}{7cm}
\vspace{0.2cm}
\includegraphics[width=7cm]{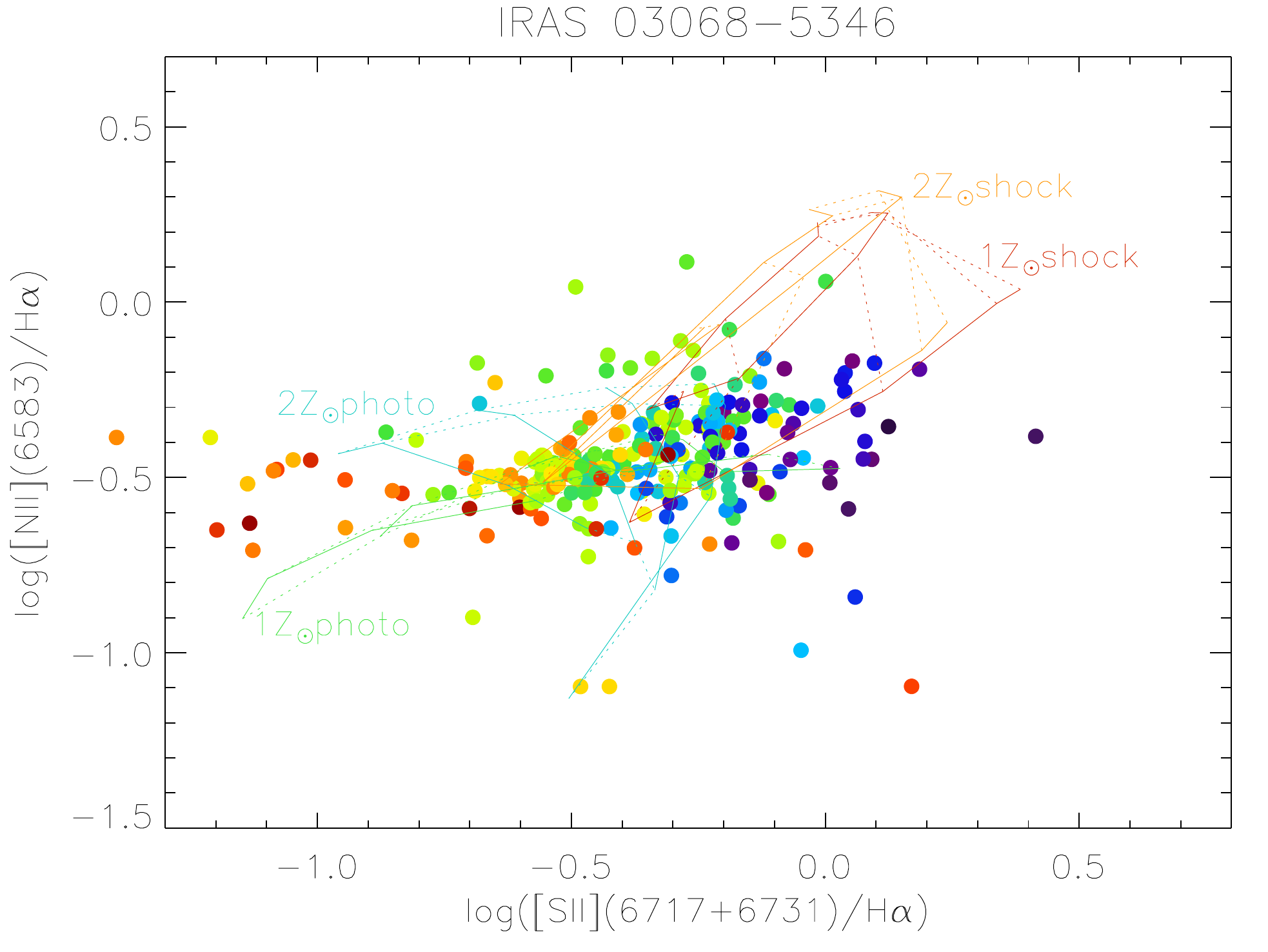}
\end{minipage}
\begin{minipage}{7cm}
\vspace{0.2cm}
\includegraphics[width=7cm]{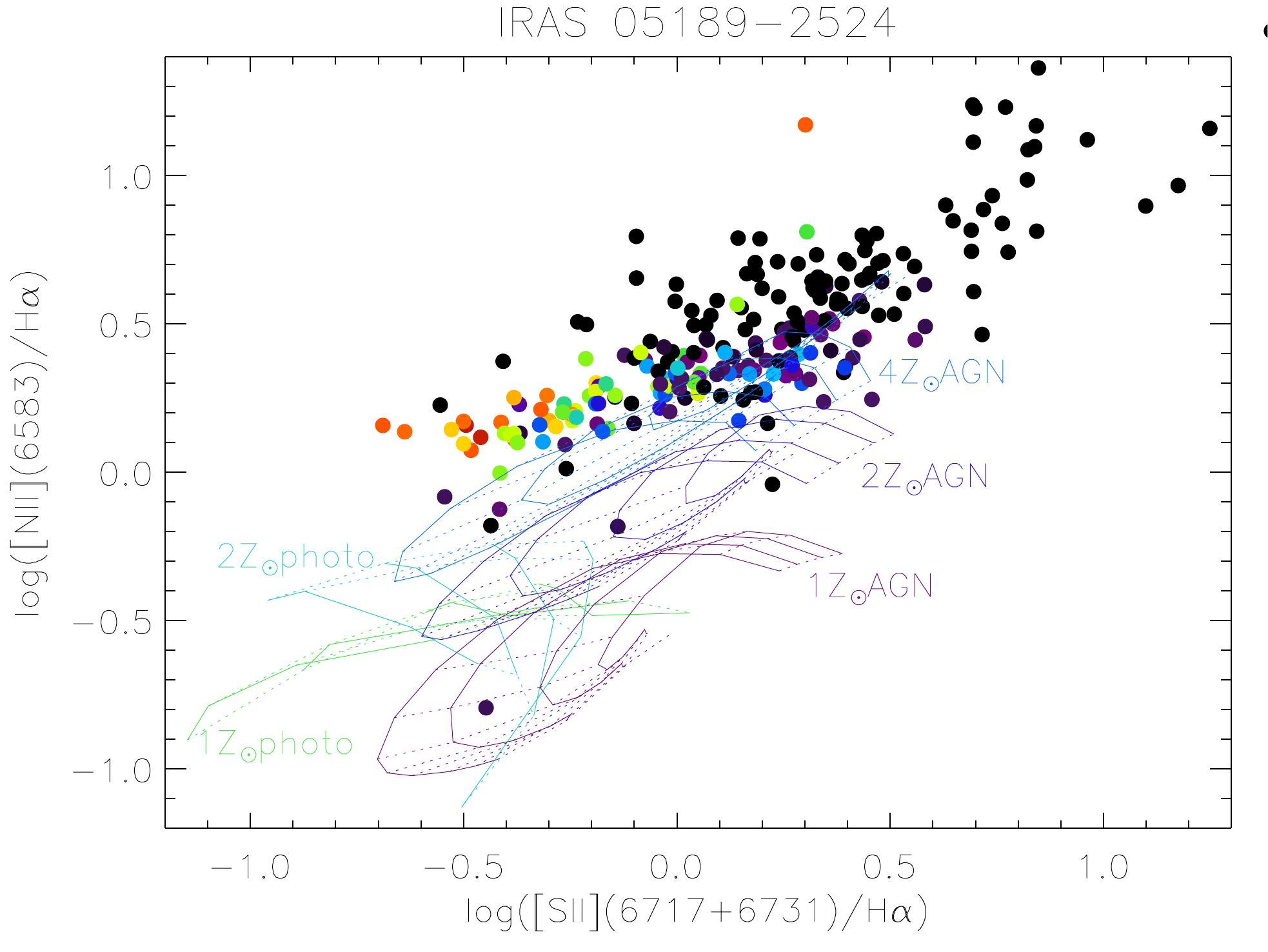}
\end{minipage}
\begin{minipage}{7cm}
\vspace{0.2cm}
\includegraphics[width=7cm]{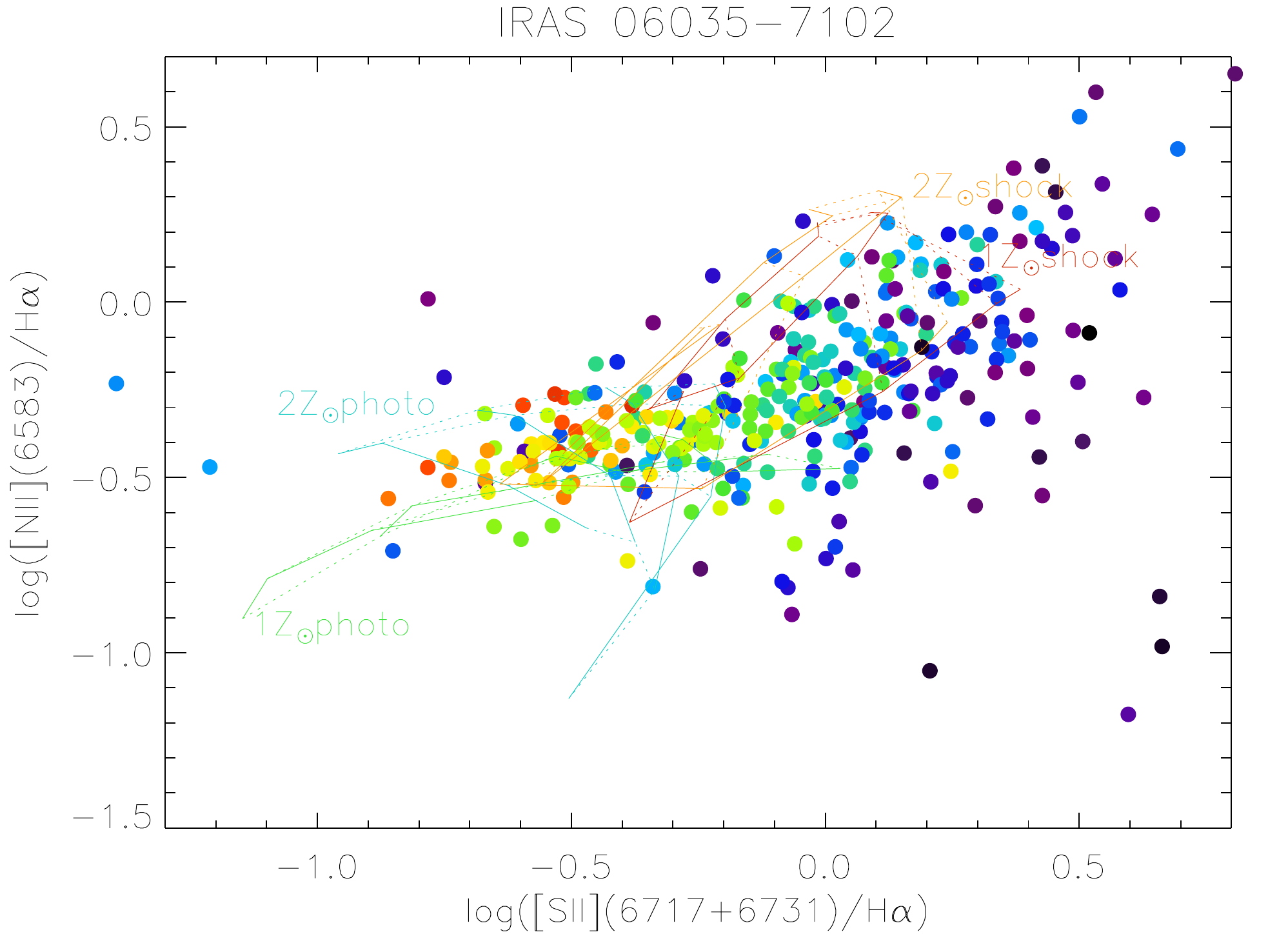}
\end{minipage}
\begin{minipage}{7cm}
\vspace{0.2cm}
\includegraphics[width=7cm]{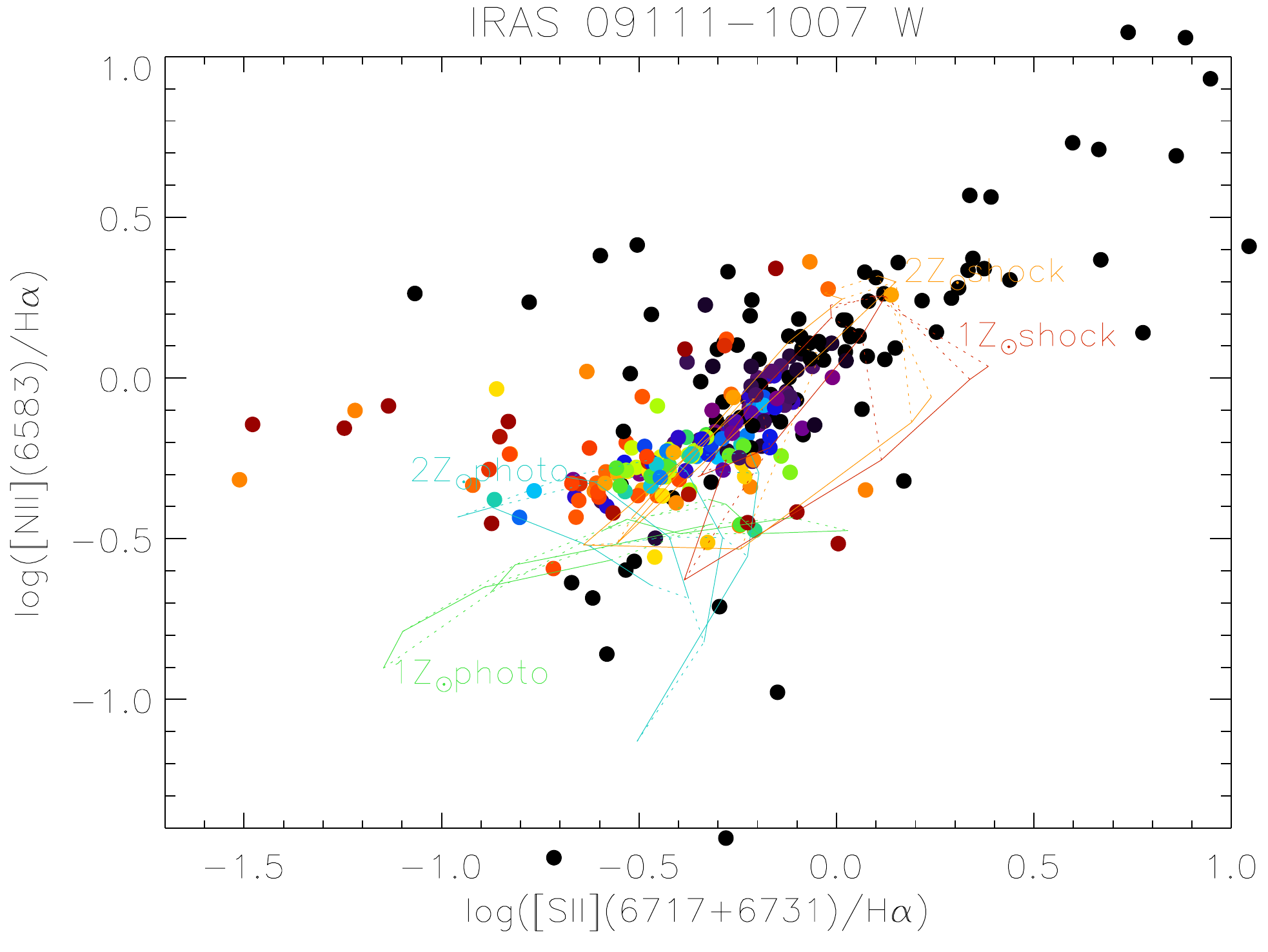}
\end{minipage}
\begin{minipage}{7cm}
\vspace{0.2cm}
\includegraphics[width=7cm]{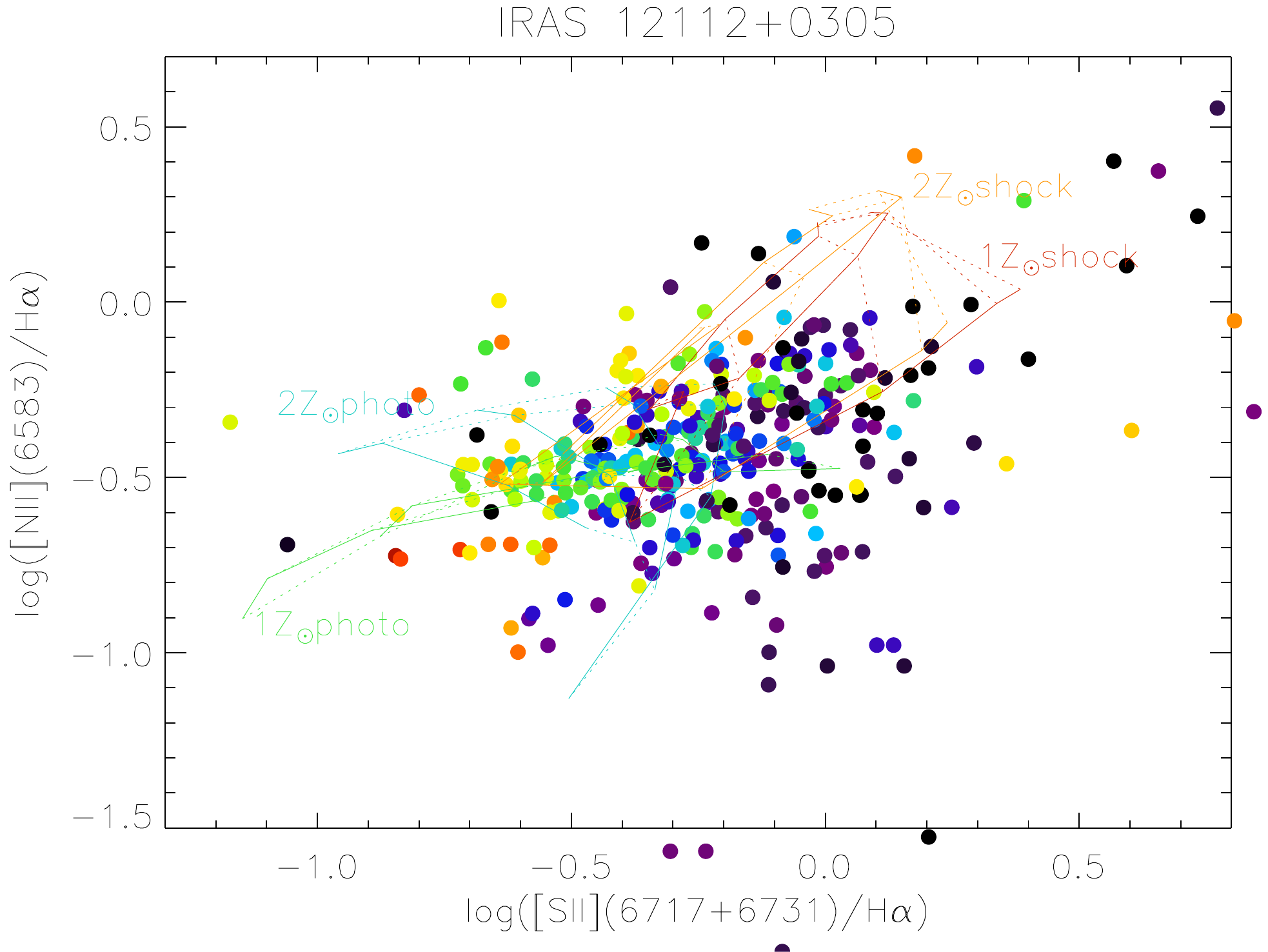}
\end{minipage}
\begin{minipage}{7cm}
\vspace{0.2cm}
\includegraphics[width=7cm]{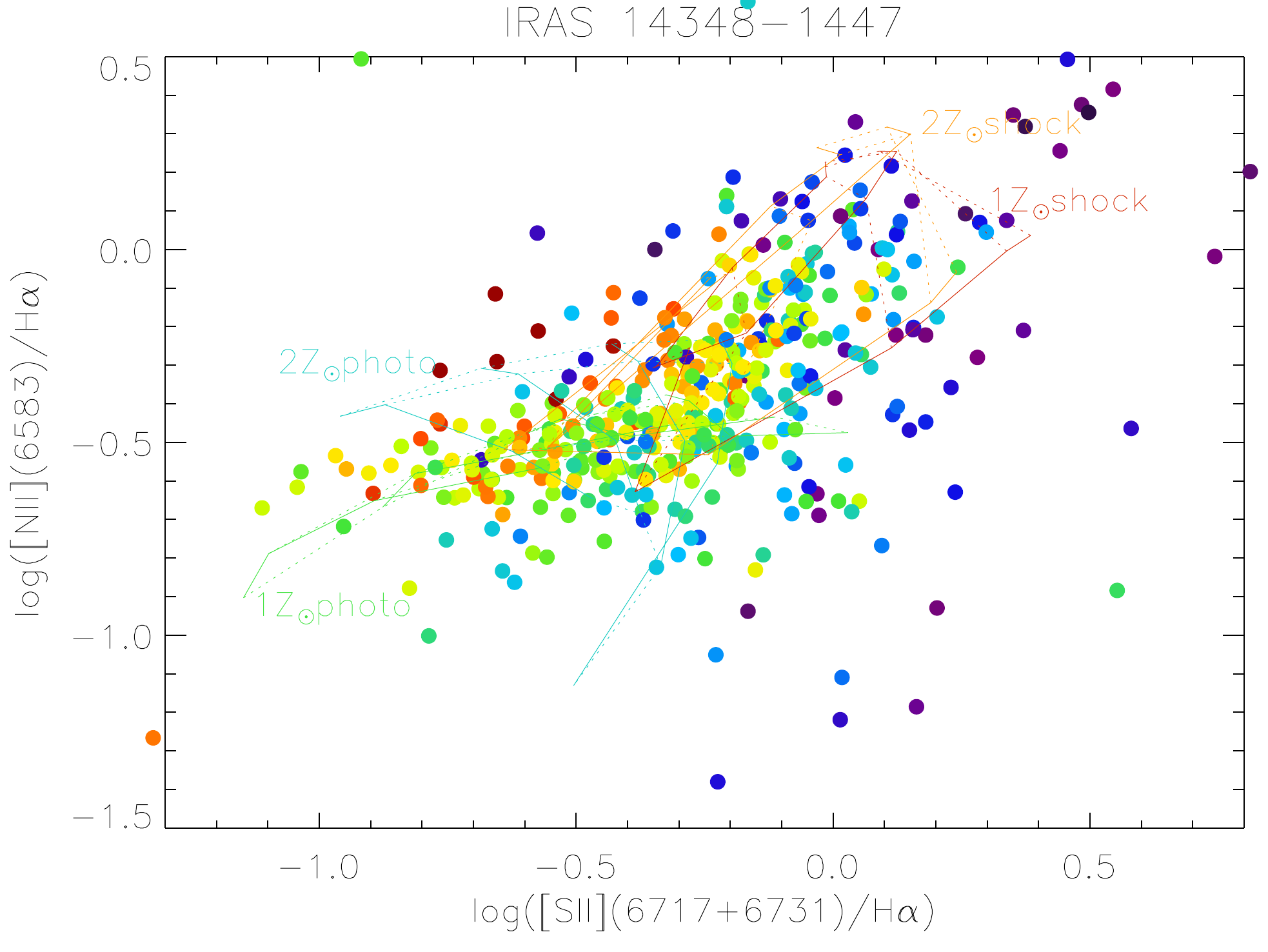}
\end{minipage}
\caption{[N\two]$\lambda$6583/H$\alpha$ vs.\ [S\two]($\lambda$6717+$\lambda$6731)/H$\alpha$ flux ratio plots. Each point is colour-coded according to its H$\alpha$ flux (red=bright, blue=faint). As a guide, the photoionization models of \citet{dopita06b} and either the shock models of \citet{allen08} or the dusty AGN models of \citet{groves04} are overplotted (depending on the spectral type classification of the system). In green (1~\Zsol) and blue (2~\Zsol) are the \citeauthor{dopita06b} photoionization models, with solid lines of constant $R$ (and starburst age ranging between 0.5--3~Myr) and dashed lines of constant age (and $R$ ranging between 0 to $-6$). $R$ is defined as the ratio of the mass of the ionizing cluster to the pressure of the interstellar medium. Overplotted in red (1~\Zsol) and orange (2~\Zsol) are the shock ionization models of \citet{allen08} (without precursor) for $n_{\rm e}=1$~\cmt, with solid lines of constant magnetic parameter (and shock velocities ranging between 100--500~\kms) and dashed lines of constant shock velocity (and magnetic parameter $B/n^{1/2}$ ranging between 0--4~$\mu$G~cm$^{3/2}$). Models for shocks with precursors cover a similar area in this diagram, but are offset by approximately $-0.2$~dex in both line ratios. On selected plots, instead of the shock models we plot the dusty AGN models from \citet{groves04} for a range of metallicities (1, 2, and 4~\Msol, blue to purple) where solid lines represent constant $\alpha$ (AGN spectrum power law index) and dashed lines constant ionization parameter (log\,$U$).}
\label{fig:ratio_plots1}
\end{figure*}

\begin{figure*}
\ContinuedFloat
\centering
\begin{minipage}{7cm}
\vspace{0.2cm}
\includegraphics[width=7cm]{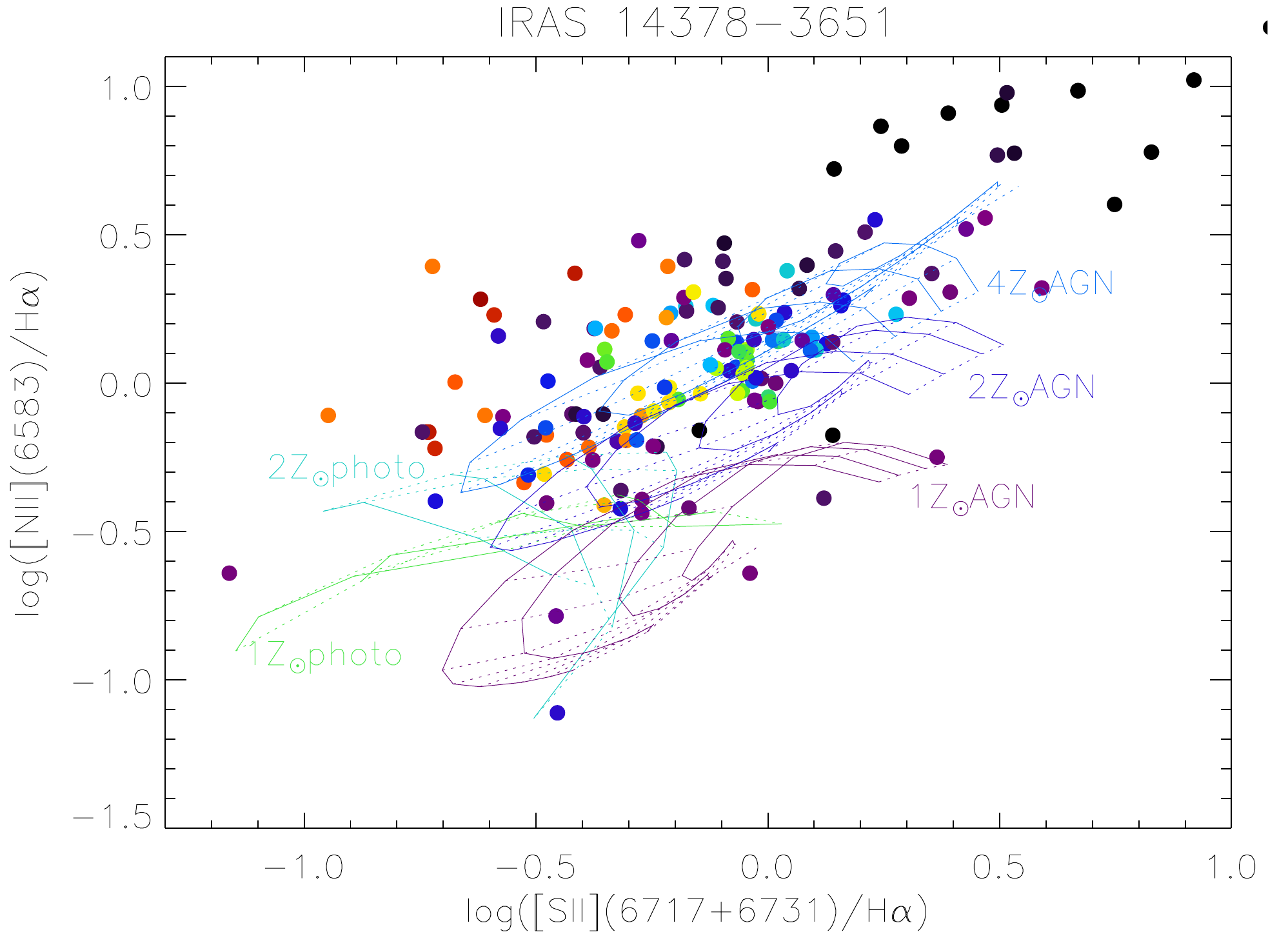}
\end{minipage}
\begin{minipage}{7cm}
\vspace{0.2cm}
\includegraphics[width=7cm]{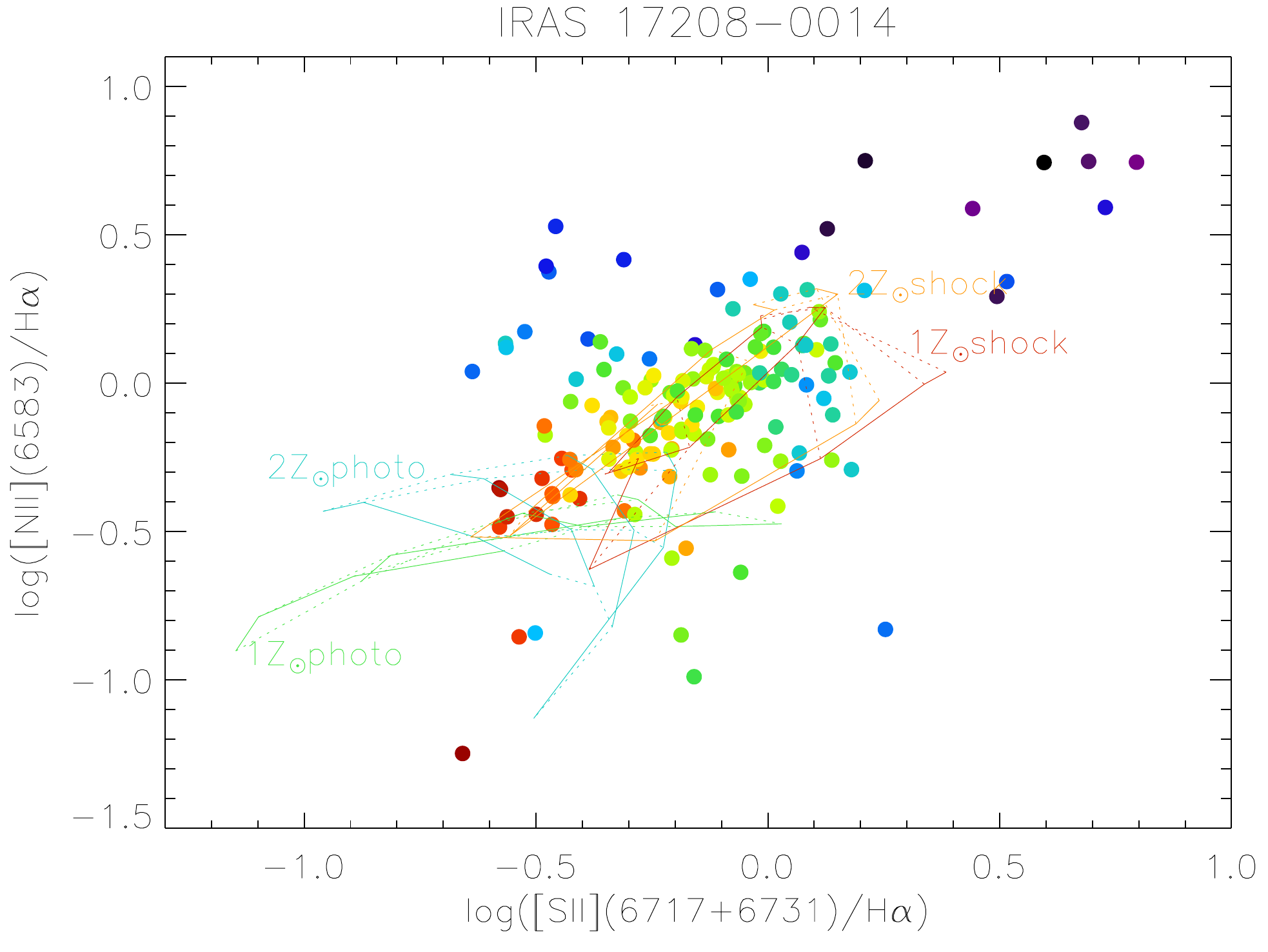}
\end{minipage}
\begin{minipage}{7cm}
\vspace{0.2cm}
\includegraphics[width=7cm]{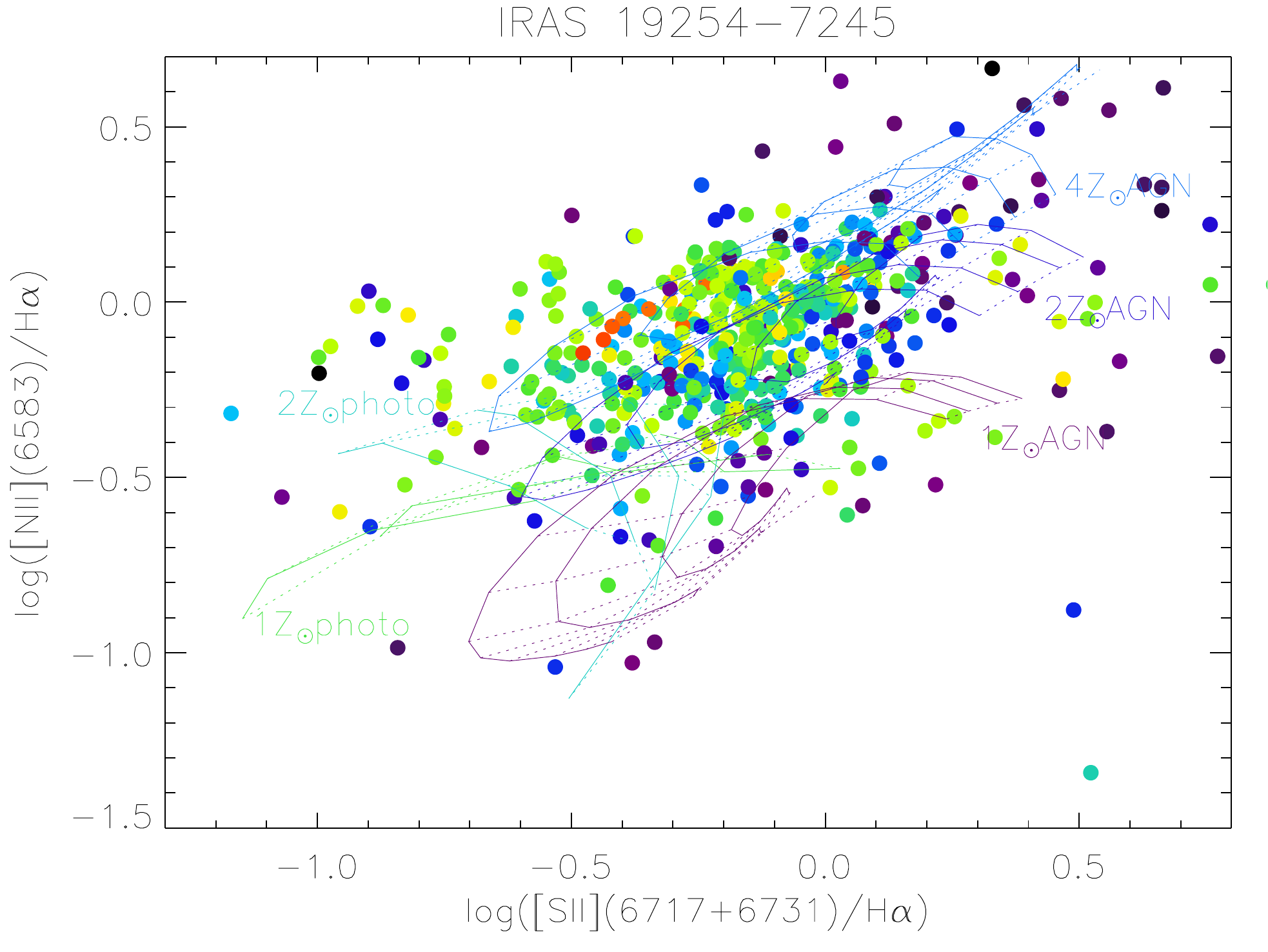}
\end{minipage}
\begin{minipage}{7cm}
\vspace{0.2cm}
\includegraphics[width=7cm]{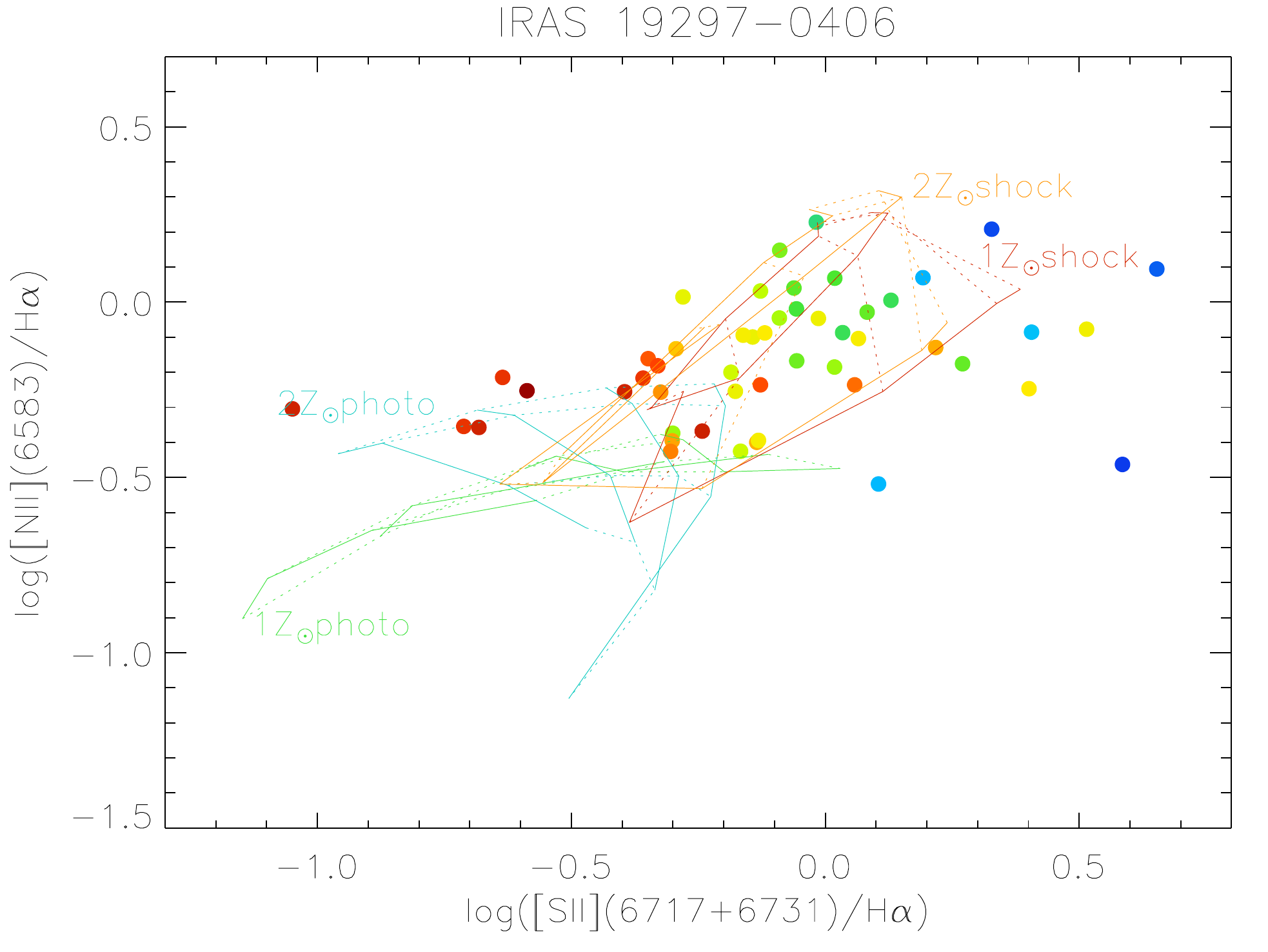}
\end{minipage}
\begin{minipage}{7cm}
\vspace{0.2cm}
\includegraphics[width=7cm]{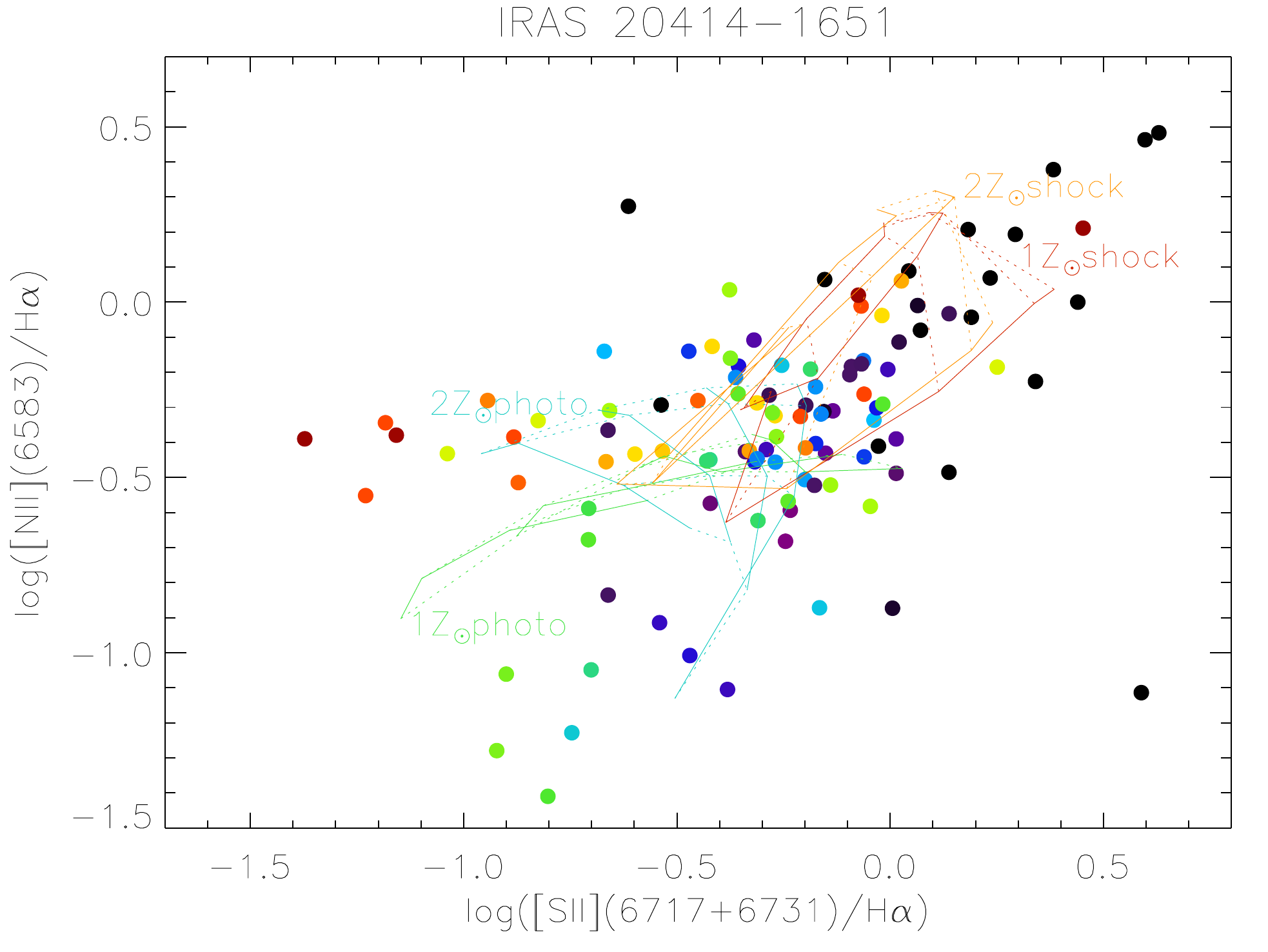}
\end{minipage}
\begin{minipage}{7cm}
\vspace{0.2cm}
\includegraphics[width=7cm]{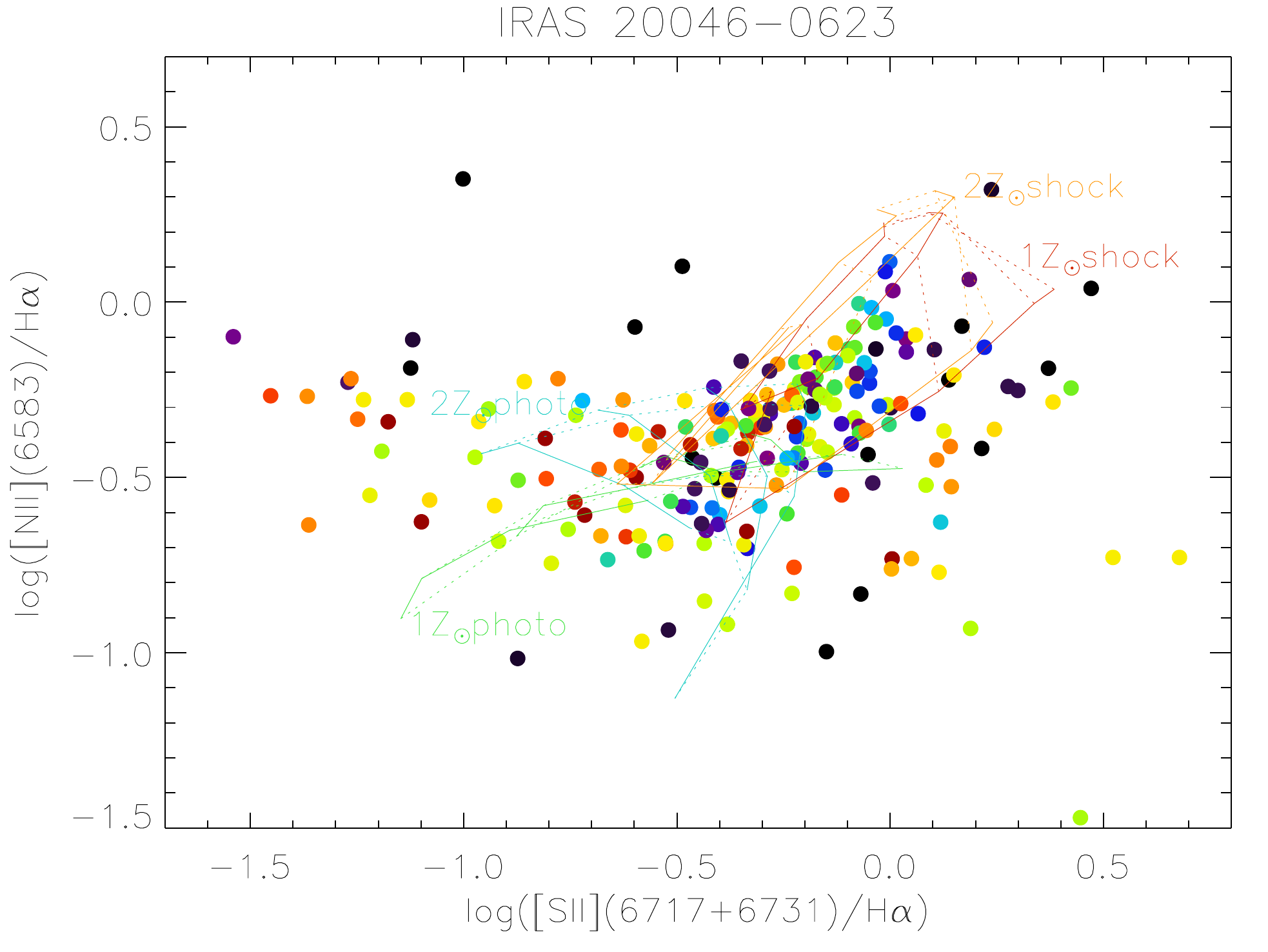}
\end{minipage}
\begin{minipage}{7cm}
\vspace{0.2cm}
\includegraphics[width=7cm]{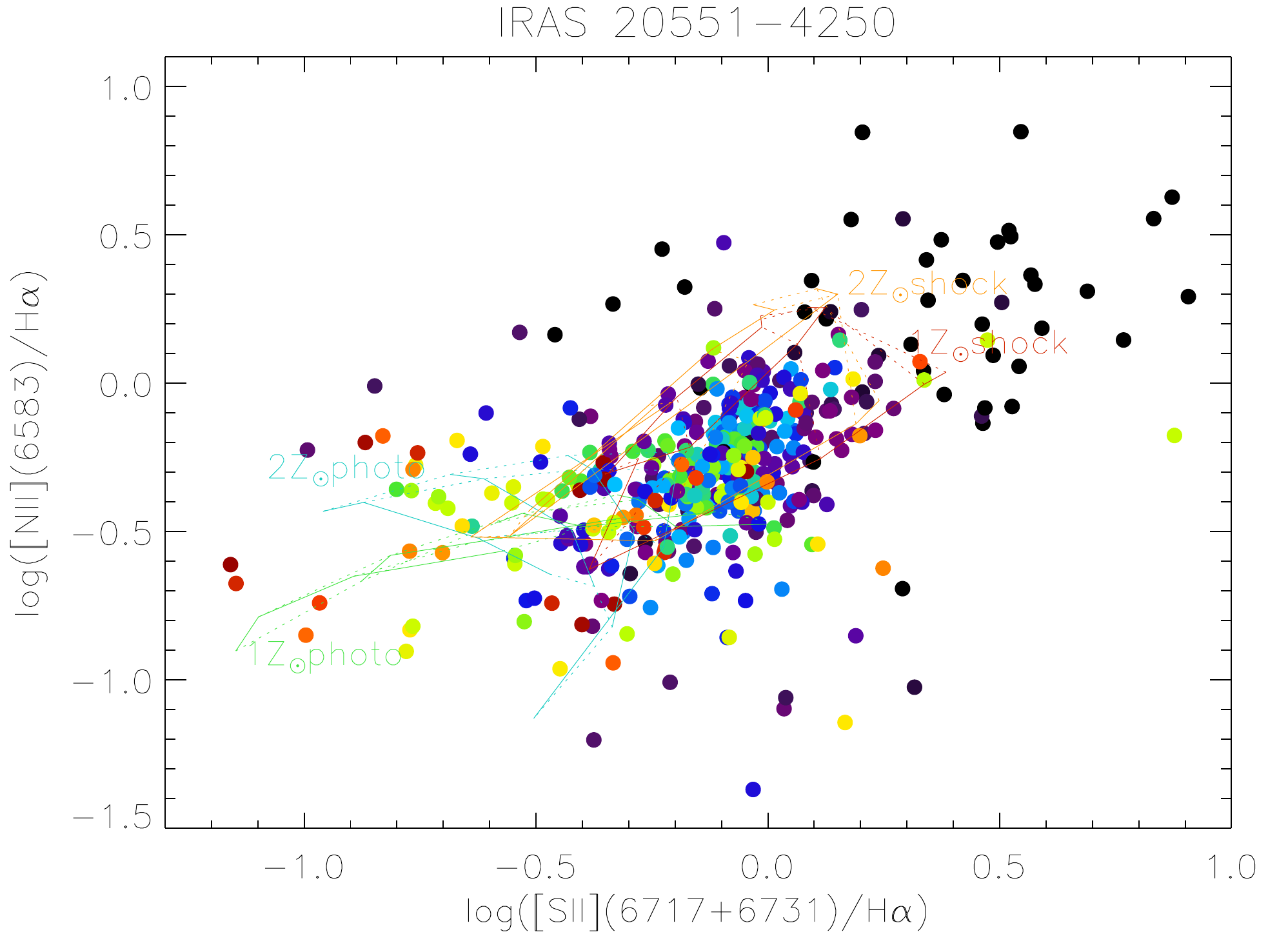}
\end{minipage}
\begin{minipage}{7cm}
\vspace{0.2cm}
\includegraphics[width=7cm]{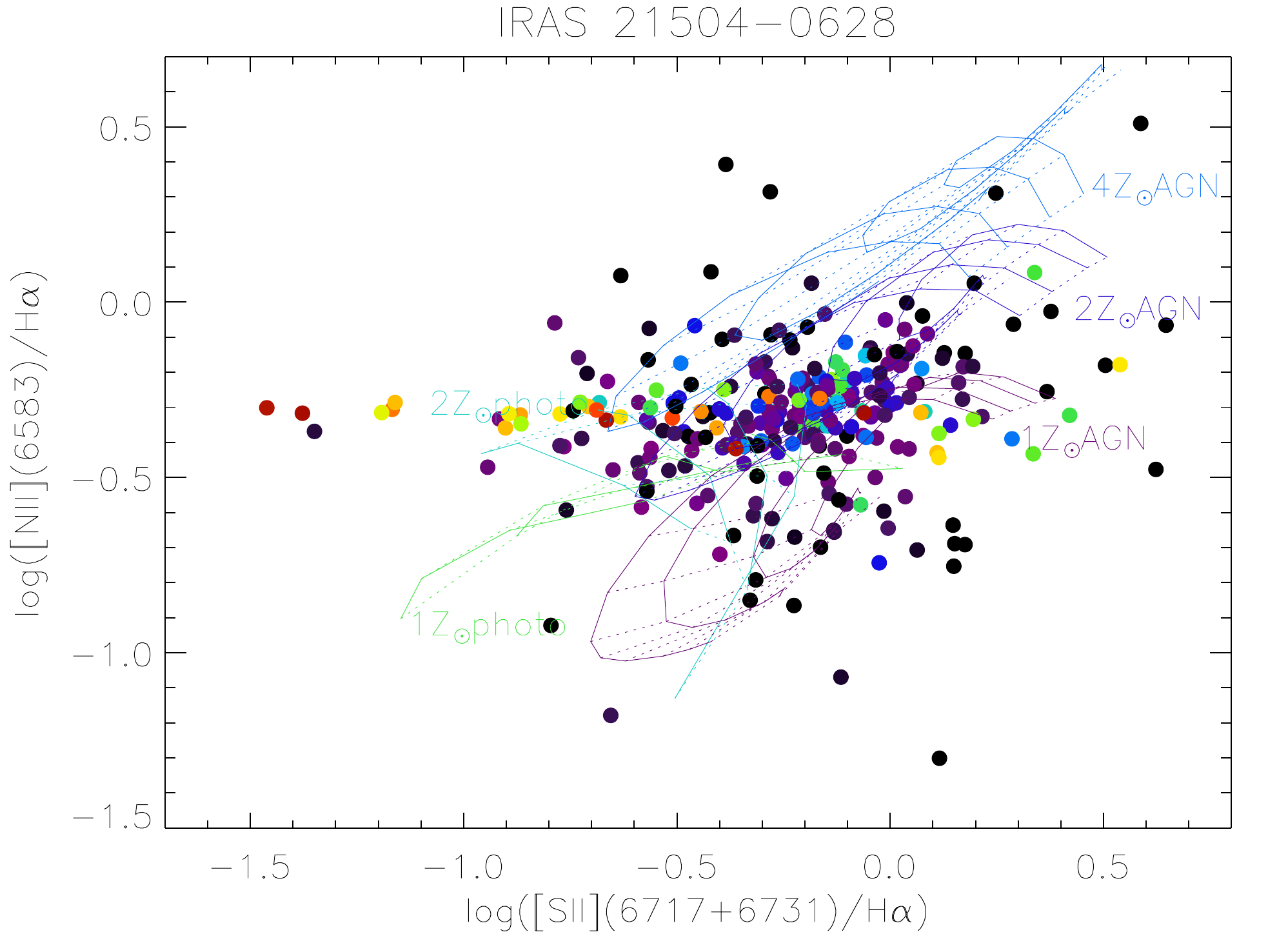}
\end{minipage}
\caption{}
\label{fig:ratio_plots2}
\end{figure*}

\begin{figure*}
\ContinuedFloat
\centering
\begin{minipage}{7cm}
\vspace{0.2cm}
\includegraphics[width=7cm]{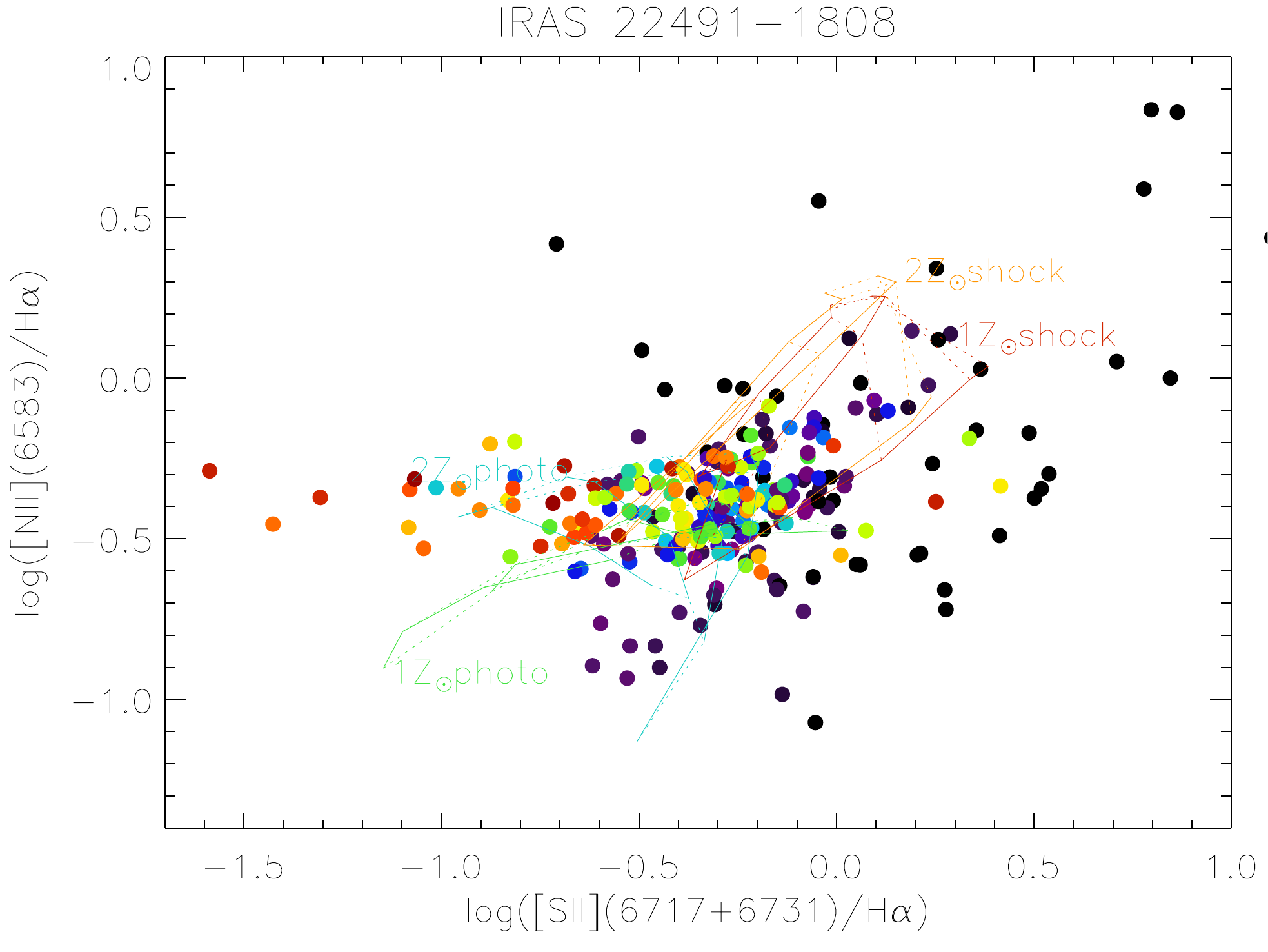}
\end{minipage}
\begin{minipage}{7cm}
\vspace{0.2cm}
\includegraphics[width=7cm]{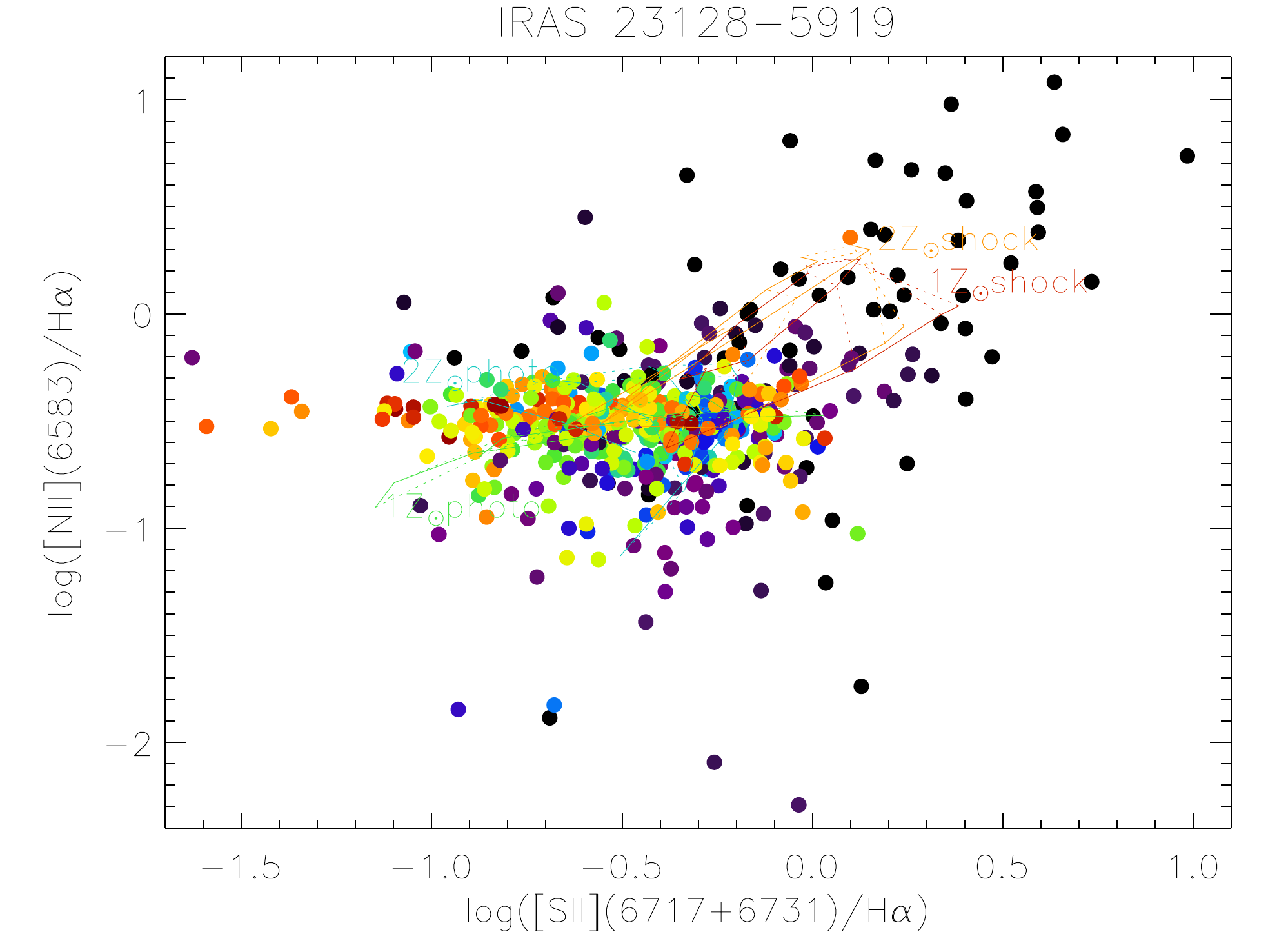}
\end{minipage}
\caption{}
\label{fig:ratio_plots3}
\end{figure*}


\begin{sidewaysfigure*}
\centering
\vspace*{18cm}
\includegraphics[width=0.95\textwidth]{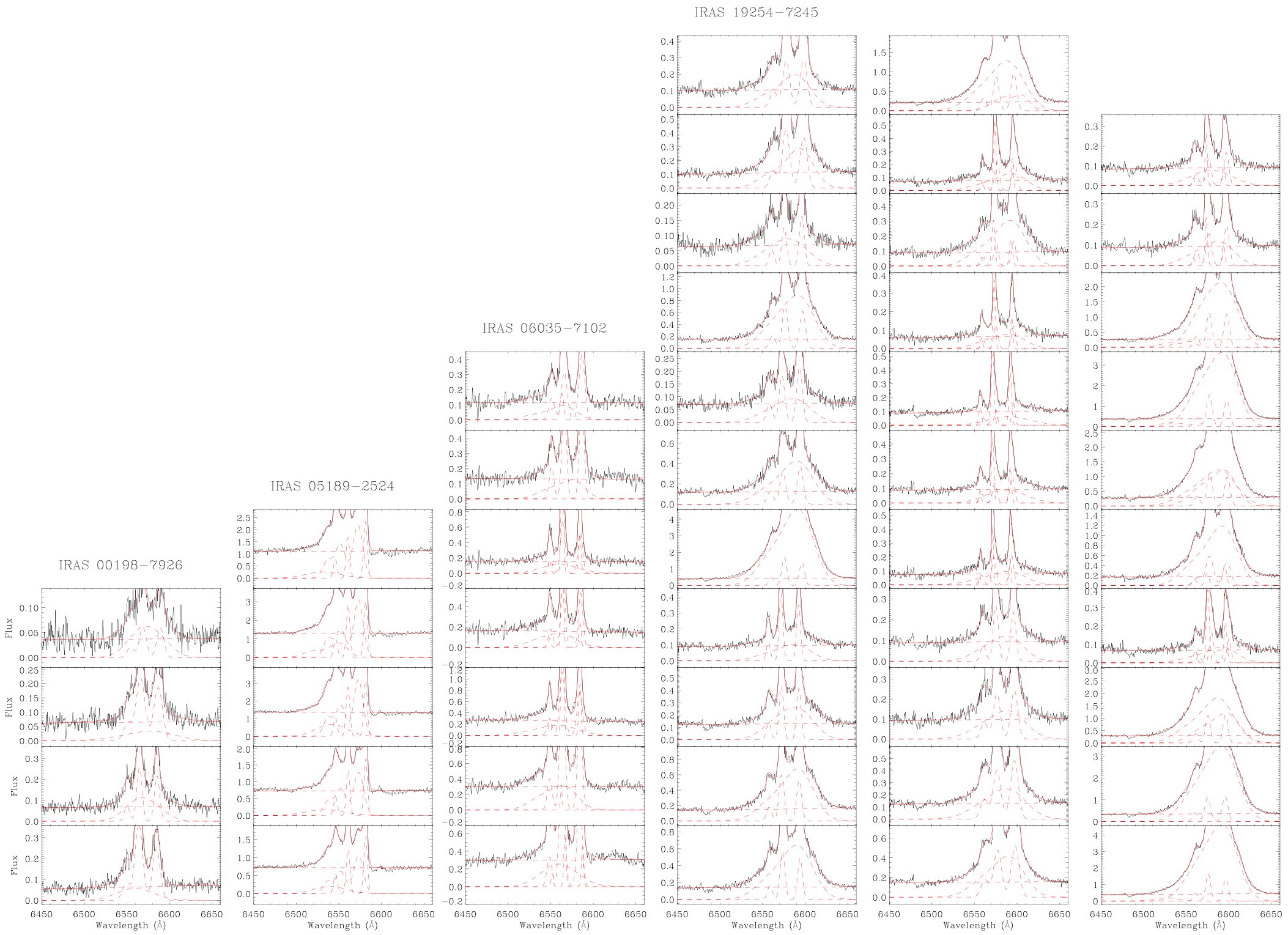}
\caption{H$\alpha$+[N\two] line profiles and best-fitting Gaussian models for all spectra in which a line component has FWHM$>$1500~\kms\ in IRAS 00198-7926 (first column), IRAS 05189-2524 (second column), IRAS 06035-7102 (third column), and 19254-7245 (fourth-sixth columns).}
\label{fig:broad_fits}
\end{sidewaysfigure*}

\begin{figure*}
\begin{minipage}{0.49\textwidth}
\includegraphics[width=\textwidth]{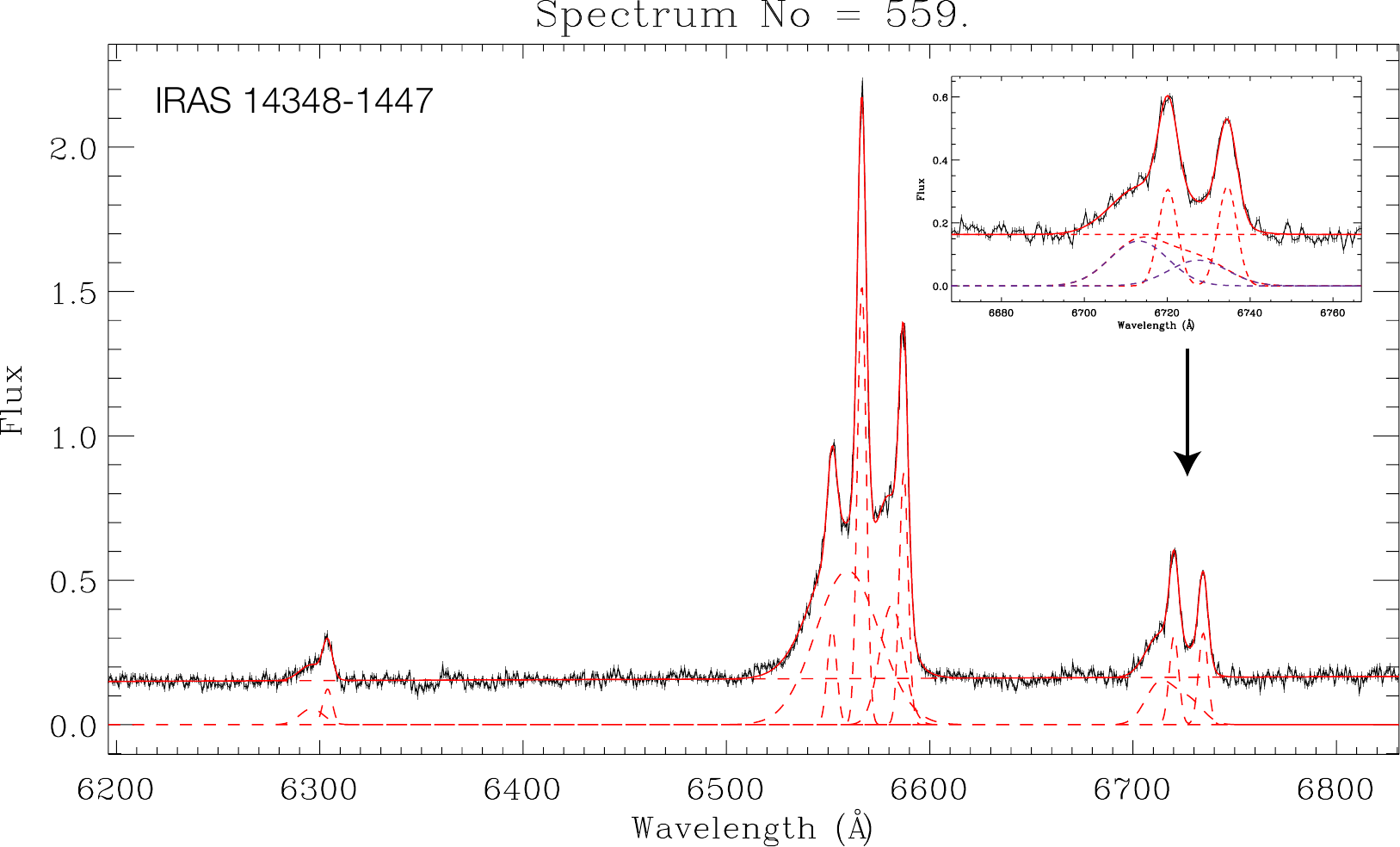}
\end{minipage}
\hspace{0.2cm}
\begin{minipage}{0.47\textwidth}
\includegraphics[width=\textwidth]{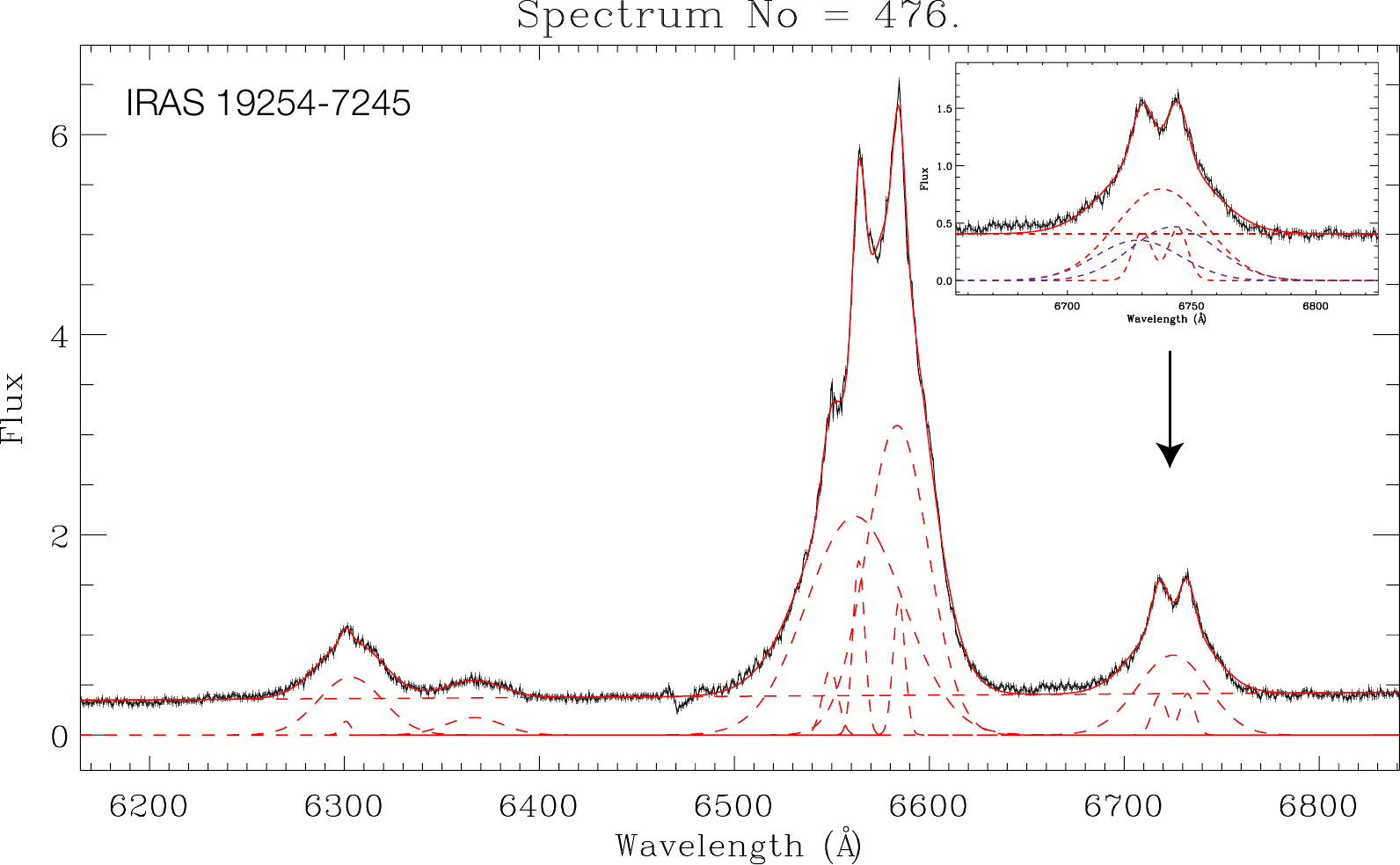}
\end{minipage}
\caption{Simultaneous fits to the [O\one]$\lambda$6300, H$\alpha$, [N\two]$\lambda\lambda$6548,6583, and [S\two]$\lambda\lambda$6717,6731 lines in the nuclear spectra of IRAS 14348-1447 and IRAS 19254-7245, showing how the broad component is present in both the recombination and forbidden lines. (For IRAS 19254-7245 we can also fit the faint [O\one]$\lambda$6360 line.) The insets show an enlargement of the [S\two] region. In purple we show the individual [S\two] 6717 and 6731 elements of the broad component fit since our routine fits the two lines simultaneously and hence displays it as one blended profile. }
\label{fig:fitall}
\end{figure*}

\begin{figure}
\centering
\includegraphics[width=9cm]{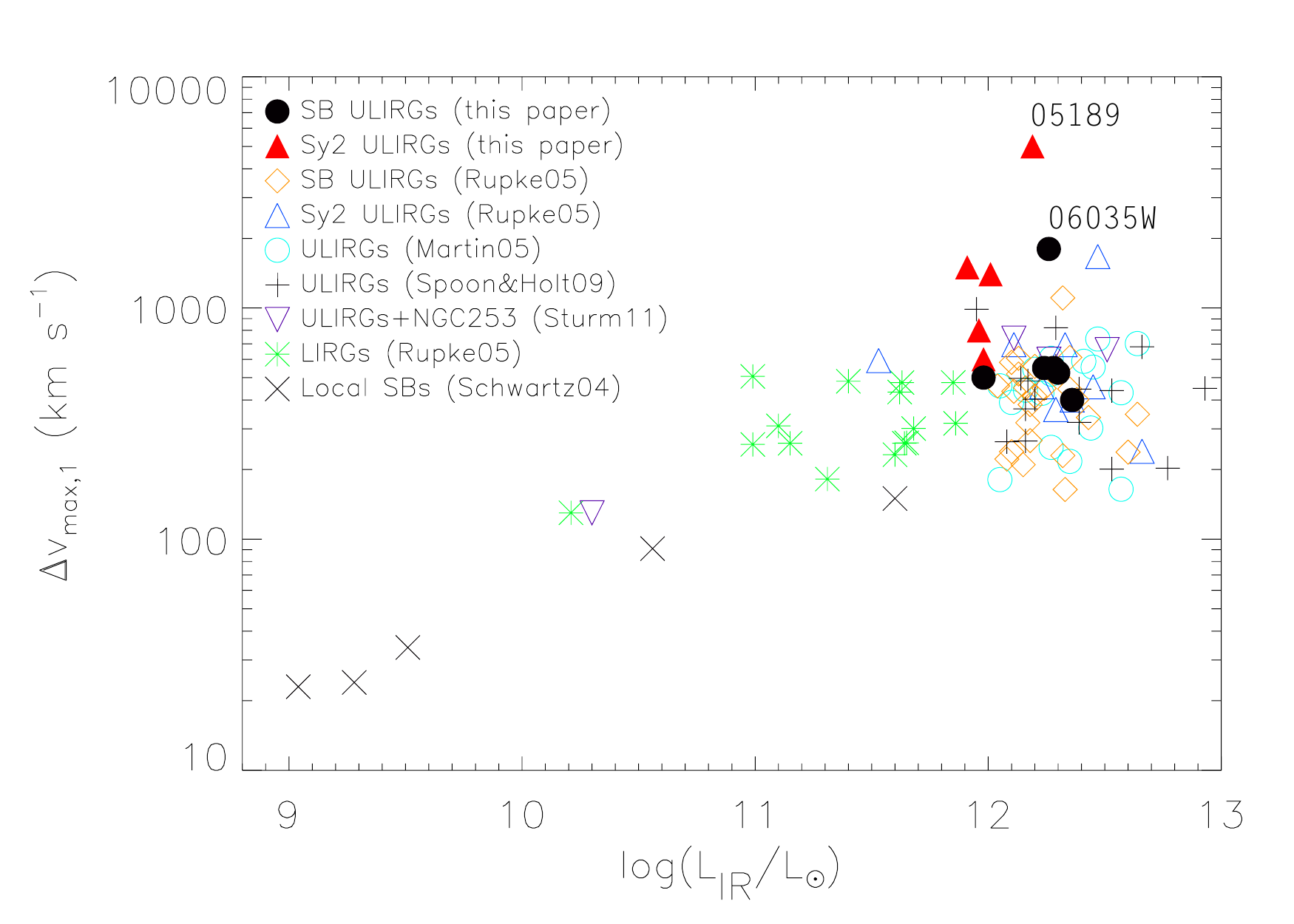}
\caption{Outflow velocity, $\Delta v_{\rm max}$ vs.\ L$_{\rm IR}$ for systems measured in this paper (solid symbols), together with data compiled from Na\one\ absorption line \citep{rupke05a, rupke05b, martin05}, mid-IR [Ne\three] emission line \citep{spoon09}, and far-IR OH~79~\micron\ \citep{sturm11} measurements of ULIRGs, and Na\one\ absorption line measurements of LIRGs \citep{rupke05a} and local starburst galaxies \citep{schwartz04}. Where possible, we have recalculated the measurements given in these papers to correspond to our definition of $\Delta v_{\rm max,1}$. Note that the absence -- or presence -- of an AGN in IRAS 06035-7102W needs further investigation.}
\label{fig:lir_wind}
\end{figure}

\appendix
\section{A case-by-case description of the targets} \label{sect:gal_desc}

ULIRGs are a fairly heterogeneous group of galaxy systems, bound together by their high IR luminosities. This classification derives from the dust re-processing of strong UV radiation produced by a large population of young stars and/or an AGN by dust and gas. Although ULIRGs are all thought to be colliding/merging systems undergoing a transition from gas-rich spirals into low-intermediate mass ellipticals \citep{colina01, genzel01, tacconi02, naab06}, the individual details of each case vary widely in terms of interaction/merger stage, star formation strength, AGN activity, etc., meaning that the reasons for their high luminosities also vary. In the following we therefore describe the VIMOS emission line maps of Figs~\ref{fig:iras00198-7926}--\ref{fig:iras23128-5919} for each galaxy in turn, discussing each in detail with reference to the literature. 

\subsection*{IRAS 00198-7926}
This system, classified as a Sy2 from optical spectroscopy \citep{vader93}, has received very little previous attention in the literature and has never been observed by \textit{HST}. Despite the singular appearance of IRAS 00198-7926 in the DSS image, the H$\alpha$ emission in our VIMOS data is concentrated in two regions implying that this is a colliding galaxy system (the projection separation between the H$\alpha$ peaks is 4$''$ or 6~kpc). We detect multiple H$\alpha$ line components (up to three in the two peaks), the kinematics of which are complex and difficult to disentangle. Line widths of FWHM$>$500~\kms\ are associated with the whole northern sub-system, and in a region just to the west of the northern nucleus we find line components with widths up to 2000~\kms\ (the colours become saturated on the maps due to the upper limit of the scale bar) and redshifts of 300--450~\kms\ with respect to systemic. Fits to the H$\alpha$ and [N\two] lines for these spectra are shown in Fig.~\ref{fig:broad_fits}. Taken together, this suggests we are seeing a fast (possibly AGN-driven) gas flow here. Immediately surrounding this and to the east of the northern nucleus, we also detect reasonably broad line components (FWHM=400--500~\kms) that are blueshifted by a similar amount (300--450~\kms\ compared to systemic). This pattern implies we may be seeing a bipolar wind emerging from behind obscuring material associated with the northern nucleus. We therefore propose that the northern system hosts the Sy2 nucleus, and is deserving of more detailed follow-up observations.

No obvious evidence for disk rotation is apparent in either sub-systems, and there is no radial velocity difference between the two nuclei. An arm of H$\alpha$ emission appears to extend from the internuclear region to the east then to the north around the northern nucleus. A continuous radial velocity gradient along this arm implies that it is a coherent structure, although it begins blueshifted $\sim$150~\kms\ compared to systemic and ends redshifted by $\sim$250~\kms.

The same region surrounding the northern nucleus exhibiting broad lines has generally high [N\two]/H$\alpha$, [S\two]/H$\alpha$ and [O\one]/H$\alpha$ ratios that are consistent with those produced by AGN excitation (Fig.~\ref{fig:ratio_plots1}) The two nuclei show lower [S\two]/H$\alpha$ and [O\one]/H$\alpha$ line ratios than the surrounding gas. Electron densities peak at $\sim$700~\cmt\ in the internuclear region, but remain high ($>$300~\cmt) around both the nuclei.

\subsection*{IRAS 00335-2732}
This starburst dominated galaxy has also received very little previous attention in the literature. In the \textit{HST} imaging it appears to be an asymmetric, post-coalescence phase system with one prominent tail with a number of condensations. This tail is not detected in our emission line data. However, we do detect emission from a much fainter tail seen in the F814W image, extending $\sim$10~kpc to the north-east. This H$\alpha$-bright tail is redshifted compared to the main disk, exhibits low H$\alpha$ line widths and low line ratios. No signature of rotation is seen in the main disk. The [S\two]/H$\alpha$ and [O\one]/H$\alpha$ line ratio maps are depressed in the central region of the disk and higher on the outskirts, whereas the [N\two]/H$\alpha$ line ratios are fairly constant across the disk. Gas densities of $>$500~\cmt\ are observed in the nuclear region.

\subsection*{IRAS 03068-5346}
This is another starburst dominated galaxy that has received very little previous attention, and has never been observed by \textit{HST}. A clear rotation gradient can be seen within the main body of the galaxy implying circular motions of $v \sin i=50$~\kms\ (the inclination is unknown). In the DSS image, a disk and prominent tail extending 10--15~kpc to the south-west can be seen. In our H$\alpha$ emission-line image this south-western tail is also seen, but at a distance of $\sim$10~kpc, a bend in the tail towards the east is observed. In the velocity map, the tail is a coherent structure that appears attached to the galaxy disk and becomes increasingly redshifted towards the south-west. Such stong H$\alpha$ emission in the tail implies significant SFRs ($>$few \Msol~yr$^{-1}$). These high rates could be associated with the crossing orbits in very young tails \citep[cf.][]{toomre72, wallin90}.

 A $\sim$5$\times$3~kpc region to the north-west of our IFU data exhibits very large H$\alpha$ line widths (FWHM $\sim$ 500--1600~\kms), and is roughly coincident with an area of redshifted gas. All three line ratios are relatively high in this whole region. Since this region is on the outskirts of the system ($>$5~kpc from the nucleus), we speculate that here we are seeing the impact site of infalling matter. Of course, higher resolution observations would be needed to investigate this further. Multiple line components can be identified in the main disk and inner parts of the tail, but none are suggestive of outflowing gas. The line diagnostic plot in Fig.~\ref{fig:ratio_plots1} suggests that most of the ionized gas in this system is excited by young stars, with only a small contribution from shocks in the fainter, outer parts, consistent with the proposed infall region.

\subsection*{IRAS 05189-2524}
IRAS 05189-2524 has been classified as a late stage merger with a single, compact and very red nucleus and a Sy2 optical spectrum \citep{veilleux95}. 
The X-ray spectrum \citep{risaliti00, imanishi03}, SED fitting \citep{farrah03}, and \textit{Spitzer} IR spectroscopy \citep{veilleux09a} all suggest this object contains both a starburst and AGN where the AGN provides a significant fraction \citep[$\sim$70\%;][]{veilleux09a} of the total IR luminosity. The detection of N IV]\,$\lambda$1488 in the UV by \citet{farrah05}, however, implies a large population of W-R stars and that a young (3--4 Myr) starburst is certainly present. The \textit{HST} F814W image shows that the compact nucleus of IRAS 05189-2524 is surrounded by a large system of faint concentric annular shells, as is found in $\sim$10\% of elliptical galaxies \citep{schweizer80, malin83}. These shells are believed to be the remnants of a small galaxy after a head-on collision with the larger system \citep{quinn84, merrifield98}.

No evidence of rotation in seen in either the H$\alpha$ C1 or C2 components, implying that we are either seeing a disk fully face-on, or that this system is completely pressure supported. The latter is more likely considering its early-type morphology. A relatively broad (500--800~\kms) and blueshifted ($\sim$350--400~\kms\ with respect to C1) H$\alpha$ C2 line component is identified over a $\sim$5~kpc$^{2}$ circular region centred on the nucleus. These velocities are in good agreement with the blueshifted high-ionization lines of [O\three], [Ar\three]$\lambda\lambda$7137,7752, and [S\three]$\lambda\lambda$9069,9531 measured by \citet{rupke05b} to have velocities of $-500$~\kms\ with respect to systemic, and the similarly blueshifted ($-400$~\kms) Na\one\ absorption. The very large area over which we see this broad, blueshifted emission indicates a significant wind outflow with a wide opening angle that is perhaps directed along our line of sight.

\citet{rupke05b} noted that some of the high ionization lines also have even broader blue wings. We also find five spaxels in the very centre exhibiting very broad wings to the H$\alpha$, [N\two] and [S\two] emission lines. Here we fit a third component with widths $>$1500~\kms\ and centroid radial velocities that are blueshifted by a few 1000~\kms\ (with respect to C1). Fits to the H$\alpha$ and [N\two] lines for these spectra are shown in Fig.~\ref{fig:broad_fits}. \citet{farrah05} also found broad components in H$\beta$ and [O\three]$\lambda$5007 in their \textit{HST}/STIS spectroscopy of the nucleus, with systematic blueshifts compared to the other optical lines of 300--1200~\kms. In their \textit{Spitzer} spectroscopy of this system, \citet{spoon09} identified strongly blueshifted [Ne\five] emission ($\sim$1000--1500~\kms). \citet{dasyra11} also detected blueshifted [O\four] with even higher velocities. \citet{rupke05b} find a very good match between their Na\one\ absorption profile and the [N\two] and [O\three] emission line profiles (all showing blue wings), thus providing very strong support that these features originate in a wind. Strongly blueshifted molecular line emission is also seen in \textit{Herschel}/PACS spectroscopy of this system (E.\ Sturm, priv.\ comm.). That these blueshifted wings are found in the recombination and forbidden emission lines, and absorption lines, rules out an AGN BLR origin. Most of the wing emission in the hydrogen recombination lines must therefore be coming from the wind.\footnote{Results from previous studies \citep{young96, veilleux99, farrah05} imply that the BLR is moderately, although not completely, obscured, meaning that some of the broad recombination line (H$\alpha$, H$\beta$) emission in the central nuclear spaxel may originate here.} Together, the results of \citet{spoon09} and \citet{dasyra11} imply that this wind is accelerating, since higher ionization lines are excited nearer to the ionizing source (the ionization potential of [O\four] is 55\,eV compared to 97\,eV for [Ne\five]).

These findings are reminiscent of those of \citet{rupke11} for Mrk~231, in which they find both a fast blueshifted neutral (Na\one) outflow extending 2--3~kpc from the nucleus in all directions that they associate with the AGN, and a slower, starburst wind arising in the extended galaxy disk. Thus, IRAS 05189-2524 is an ideal candidate for follow-up, spatially resolved observations of the neutral and molecular gas phases.

Finally, the [N\two]/H$\alpha$, [S\two]/H$\alpha$ and [O\one]/H$\alpha$ line ratios are all very high in the outer disk (see Fig.~\ref{fig:ratio_plots1}). These are in fact the highest line ratios found in our whole sample, and are outside the range predicted by the photoionization, shock or AGN models. These results may reflect either a particularly strong influence of the AGN ionization on the whole system, or an unusually high metallicity ($>$4\,\Zsol) All three line ratios are much lower in the very nuclear region, but still remain above those predicted by just photoionization.

\subsection*{IRAS 06035-7102} \label{sect:06035}
IRAS 06035-7102 is a double-nucleus system. The separation of the nuclei (6~kpc) and the strong tidal features suggest that this system is very close to coalescence. This is supported by the smooth H$\alpha$ radial velocity map, indicating the two colliding disks are in the process of forming a single structure. From SED fitting, \citet{farrah03} classified it as a SB/H\two\ system, however \citet{weedman09} included this source in their compilation of obscured AGNs due to the strength of its silicate absorption and weakness of its PAH emission found from \textit{Spitzer}/IRS spectroscopy. Our spatially resolved kinematics show that the western nucleus exhibits very broad, blueshifted wings to the H$\alpha$ line profile that are suggestive of the type of fast outflows usually associated with AGN. The third H$\alpha$ component, identified in nine contiguous spaxels ($\sim$2.5$\times$2.5~kpc) surrounding the western nucleus, has line widths $>$1000~\kms\ and centroid velocities that are blueshifted by 500--600~\kms\ with respect to C1. Those seven with line widths $>$1500~\kms\ are shown in Fig.~\ref{fig:broad_fits}. Using the outflow speed formulation described in Section~\ref{sect:props_sample}, these measurements convert to $\Delta v_{\rm max,1}=1800$~\kms. From Fig.~\ref{fig:lir_wind}, we see that this value puts IRAS 06035-7102 significantly outside the range of outflow velocities measured for other starburst dominated ULIRGs. We therefore conclude that the most likely situation is that the outflow from IRAS 06035W is being driven by a heavily obscured AGN, although we cannot be sure without further observations to investigate the presence of an AGN.

Broad H$\alpha$ C1 line widths (200--450~\kms) are found in the fainter regions just to the north of both nuclei between the tidal tails, and are coincident with elevated line ratios in [N\two]/H$\alpha$, [S\two]/H$\alpha$ (only one component of [S\two] could be identified due to strong sky line residuals at this wavelength). Most of these H$\alpha$-faint regions have line ratios that fall in the shocked region of diagnostic diagram shown in Fig.~\ref{fig:ratio_plots1}. Unlike the majority of the system where low gas densities are measured, large densities (few 1000~\cmt) are found in the north-eastern tidal arm.

\subsection*{IRAS 09111-1007W (SB+Sy2)}
IRAS 09111-1007 consists of two widely spaced but interacting galaxies. The two nuclei have a projected separation of $\sim$40~kpc and a velocity difference of 425~\kms\ \citep{duc97}. Monte Carlo simulations give the probability of the pair being bound as 0.88 \citep{schweizer87}. The western component dominates the far-IR flux at both 60~$\micron$ and 350~$\micron$, carrying $\sim$80\% of the total system luminosity \citep{khan05}. The wide separation and the IR luminosity-to-H$_{2}$ mass ratio suggest that the pair are possibly at an early stage of interaction \citep{khan05}, although the high IR luminosity of the system might suggest otherwise. Compared to the eastern galaxy, the western one seems to have a high dust content since H$\beta$ is invisible and there is strong Na\one\ absorption \citep{duc97}.

Observations of both galaxies in the system were obtained, but here we only discuss the western galaxy since the IRAS 09111-1007E is not classified as a ULIRG. IRAS 09111-1007W is a ULIRG with a SB+Sy2 spectrum. A study of the two together will form the subject of a future paper (Khan et al., in prep).

The \textit{HST} F814W image shows that \mbox{IRAS 09111-1007W} is a barred spiral. Our H$\alpha$ radial velocity map shows a clear disk rotation pattern, and distortions in the central region follow the bar shape of the bar. A second velocity component is observed in the central $\sim$5~kpc, and it is likely that this component is kinematically related to the bar. H$\alpha$ line widths across the whole disk are low ($<$200~\kms), and the distribution of line ratios plotted in Fig.~\ref{fig:ratio_plots1} imply that this galaxy has a high metallicity. In C1, the [N\two]/H$\alpha$ line ratio remains low across the whole disk. This is in contrast to the [S\two]/H$\alpha$ ratio, which shows a strong radial gradient increasing at larger galactocentric distances (which is to some extent mirrored in the [O\one]/H$\alpha$ ratio). A few spaxels covering the nucleus exhibit broad ($\sim$500~\kms) and blueshifted (200--300~\kms\ with respect to C1) emission, which we have assigned to C3. Since these components also have very high [N\two]/H$\alpha$ ratios (but are undetected in [S\two] or [O\one]), we conclude that this component represents emission associated with a nuclear wind outflow. The velocities would imply a starburst-driven wind rather than an AGN outflow.

\subsection*{IRAS 12112+0305}
IRAS 12112+0305 is classified as a LINER \citep{veilleux99}, and through SED modelling \citet{farrah03} were able to say confidently that this object is powered by a starburst with no evidence for an AGN component. Mid-IR spectroscopy from \textit{Spitzer} suggests that the AGN fraction is $\leq$10\% \citep{veilleux09a}. IRAS 12112+0305 was previously studied with integral field spectroscopy by \citet{colina00} using INTEGRAL on the WHT. They found that the observed ionized gas distribution is decoupled from the stellar main body of the galaxy. They also confirmed that the optical properties of the two dust enshrouded ($A_{V}=3$--8~mag) nuclei are consistent with being weak [O\one] LINERs, and measured the integrated star formation rate of the starbursts associated with the two nuclei to be $\sim$80~\Msol~yr$^{-1}$ \citep[this is significantly less than the IR-derived value given in Table~\ref{tbl:sample}, perhaps reflecting that the dust is optically thick at H$\alpha$; e.g.][]{rangwala11}.

Our significantly higher spatial resolution data clearly resolve the two nuclear complexes, the main H$\alpha$-bright H\two\ regions, and the southern arc-like tidal tail. The H$\alpha$ velocity field is quite distorted across the main body of the galaxy. The gas in the southern tidal tail is quite well ordered, and becomes progressively and smoothly blueshifted towards the south-east, showing that this tail is indeed attached to the central regions. It's H$\alpha$ brightness would suggest that stars are forming vigorously here. The patchy observed density distribution shows that some areas outside the main nuclear collision region and H\two\ complex have quite high densities. The young massive H\two\ region R1 \citep[][labelled on the \textit{HST} image]{colina00} appears to be forming within a quiescent region of the collision since it is located in a region of narrow line widths, low [N\two]/H$\alpha$, [S\two]/H$\alpha$ and [O\one]/H$\alpha$ line ratios, and a comparatively low gas density. In contrast, the southern nucleus is embedded in a region of broader line widths (300--400~\kms) and higher line ratios.

Redshifted (200--400~\kms\ relative to $v_{\rm sys}$) and broad (500--600~\kms) H$\alpha$ emission is seen in spaxels extending to the north-east from the northern nucleus (N$_{\rm N}$; shown on the \textit{HST} image with dashed lines). This suggests we are seeing an outflow from the northern nucleus and H\two\ region complex. \textit{Herschel}/PACS spectroscopy also reveals evidence for outflowing molecular gas in the OH lines (E.\ Sturm, priv.\ comm.).

\subsection*{IRAS 14348-1447}
IRAS 14348-1447 consists of two nuclei separated by 4.8~kpc \citep{carico90, scoville99, surace00} with a bright tidal tail to the north and a fainter one to the south-west \citep{sanders88, melnick90}. According to previous work, this system is one of the most luminous and molecular gas rich ULIRGs that show no clear evidence of AGN activity \citep[i.e.\ broad line emission or strong high ionization lines;][]{veilleux97, evans00}. However, SED fitting \citep{farrah03} and hard X-ray observations \citep{franceschini03} imply that this object contains an AGN, but the total luminosity is dominated by the starburst \citep[the AGN contribution is estimated to be $\sim$17\% of the total IR flux;][]{farrah03,veilleux09a}. Indeed, the south-western nucleus has been classified as a Sy2 by some authors \citep{sanders88, nakajima91}. Our H$\alpha$ maps show that broad ($\sim$700~\kms) and blueshifted (100--250~\kms) emission in C2 is associated with the south-western nucleus, implying that this object is driving a wind outflow. An example spectrum and multi-Gaussian fit from this region is shown in the figure inset, and in Fig.~\ref{fig:fitall} we show a simultaneous multi-component fit to all the observed emission lines. The very broad width of this blueshifted component suggests that the obscured AGN may be the powering source. Indeed, similar molecular-phase outflows can be identified in \textit{Herschel}/PACS spectroscopy (E.\ Sturm, priv.\ comm).

The H$\alpha$ intensity map shows extended clumpy emission throughout the body of the system and tidal tails, in good agreement with what was observed by \citep{mihos98}. Strong H$\alpha$ emission is associated both with a number of knots to the south-east of the two nuclei at radii of $\sim$5$''$--15$''$ (7--20~kpc), and with the northern tail, implying that there is vigorous star formation occurring throughout the merging system. An interesting arc-shape string of H$\alpha$ clumps to the south seems kinematically distinct (redshifted) from rest of ionized gas in system, in agreement with what was found by \citet{mihos98}. These authors suggested that these H$\alpha$ bright clumps may represent the early stage of the formation of tidal dwarf galaxies. Elevated line ratios and H$\alpha$ line widths are found in a strip between the two nuclei, that are reminiscent of the shock regions in Stephan's Quintet \citep{pietsch97, guillard09}.

\subsection*{IRAS 14378-3651}
This system appears on the \textit{HST} F814W imaging morphologically similar to IRAS 05189-2524, in that it has a single, compact nucleus surrounded by a system of faint concentric shells. It also displays no signature of rotation in the H$\alpha$ kinematics, implying that it is an early-type system. Through SED fitting, \citet{farrah03} found no evidence for an AGN in this galaxy, although mid-IR line diagnostics imply that there is some AGN component \citep{farrah07, veilleux09a, sturm11}. A molecular gas outflow of 800--1000~\kms\ was observed via the OH 79~$\micron$ and OH 119~$\micron$ emission lines with \textit{Herschel} \citep{sturm11}, and these authors argued that such high velocities implies that an AGN is powering the outflow. In our H$\alpha$ maps we find broad (500--800~\kms) and blueshifted (200--400~\kms\ with respect to C1) emission coincident with nucleus and elongated in an east-west direction. This component exhibits high [N\two]/H$\alpha$ ratios but is not elevated in in [S\two]/H$\alpha$ or [O\one]/H$\alpha$. In the line ratio diagnostic diagram of Fig.~\ref{fig:ratio_plots2} we see that the highest metallicity AGN models from \citet{groves04} still do not cover this part of the plot, suggesting either metallicities $>$4~\Zsol, or some peculiar ionization effects. Of the very few systems that have a measurement of the outflow speeds in both the warm ionized and cold molecular phases, this galaxy presents a very unusual case where the molecular material is outflowing faster than the ionized material.

In C1, lower line ratios are consistently found in the nucleus, whereas much higher values are found on the outskirts. From the diagnostic plot in Fig.~\ref{fig:ratio_plots2}, these values are consistent with high metallicities and AGN/shock excitation.


\subsection*{IRAS 17208-0014}
This object consists of a single nucleus surrounded by a disturbed disk containing several compact H\two\ regions. Its pure SB classification is supported by optical spectroscopy \citep{veilleux99}, near-IR imaging \citep{scoville00}, and SED fitting \citep{farrah03}, which all argue against the presence of an AGN. However, a significant hard X-ray source has been detected by both \textit{XMM-Newton} \citep{franceschini03} and \textit{Chandra} \citep{ptak03}. Mid-IR AGN diagnostics predict that an AGN may contribute $\leq$10\% to the total luminosity \citep{sturm11}. Our H$\alpha$ kinematics show an ordered radial velocity field with a clear signature of rotation, but distortions that are consistent with the disturbed morphology. H$\alpha$ C1 is fairly broad throughout the whole disk (200--350~\kms) implying significant turbulence presumably driven by the intense star formation. A second H$\alpha$ component, seen only in the central regions, is much narrower, but not significantly offset in velocity-space. IRAS 17208-0014 was also observed with IFS by \citet{arribas03}, who also did not detect any evidence for an ionized gas outflow. \citet{martin05}, however, identified a neutral gas outflow via Na\one\ absorption lines with speeds $>$500~\kms, and \citet{sturm11} measured a molecular outflow from \textit{Herschel}/PACS observations with speeds of few 100~\kms. A large area of depressed line ratios consistently seen in all three diagnostics is coincident with the nuclear region. Higher ratios seen in the outer parts of the galaxy are consistent with shocks (Fig.~\ref{fig:ratio_plots2}).

Our higher resolution kinematic results support the conclusions of \citet{arribas03} who suggest that IRAS 17208-0014 is a luminous, cool ULIRG that is in the final coalescence phase. Furthermore, the fact that there is little evidence that an AGN has developed is consistent with the scenario put forward in \citet{colina01} that cool ULIRGs are the merger of two low-mass ($<$m$^{*}$) disk galaxies that will never evolve into a QSO.

\subsection*{IRAS 19254-7245 (the ``Superantennae'')}
IRAS 19254-7245 was the subject of study in \citetalias{bendo09}. Here we present a brief re-analysis of these data for both the purposes of completeness, and to concentrate on the evidence for outflows. In this study we have re-fit the emission lines with the method used for all the other targets to ensure consistency. The results are, or course, in very good agreement with what we found before. The main difference is that with the line-fitting method used here, we were able to identify an additional H$\alpha$ emission line component in some regions that is clearly related to a wind outflow.

The ``superantennae'' nomenclature refers to two extremely long tidal tails, reminiscent of the tails in NGC4038/9 (the ``antennae galaxies''), that extend $>$500~kpc on either side of the colliding nuclei. The nuclei themselves are separated by $8\farcs5$ (10~kpc). To form these features a coplanar collision between two massive disk galaxies is implied \citep{melnick90}. The southern nucleus has been optically classified as a Sy2 and the northern as a pure SB \citep{mirabel91}. \citet{berta03} \citep{franceschini03} found that the near-IR (X-ray) luminosity from the southern nucleus is dominated by the emission from an obscured AGN, and showed that this nucleus is responsible for about 80\% of the total infrared luminosity of the system. Both studies also found that the northern nucleus does not show evidence for AGN emission and appears to be in a post-starburst phase. High angular resolution \textit{HST} imaging shows that a double nucleus may be present in both the north and southern components of the Superantennae \citep{borne99}, suggesting a multiple merger origin of the system.

In our VIMOS data, the strongest H$\alpha$ emission is concentrated around the two nuclei, and in particular the southern Sy2 nucleus. In \citetalias{bendo09} we clearly identified the rotation signature of the two disks. The major axis of rotation of the southern disk is aligned in a NE--SW direction, and the northern disk in a E-W orientation; both disks rotate in the same sense \citep[with their northern sides receding; see also][]{mihos98}. This is consistent with the aforementioned coplanar collision scenario. In \citetalias{bendo09} we identified broad spectral line emission from the core of the southern nucleus, and potential outflows with shock-excited spectral features near both nuclei. We estimated that $<$10\% of the 24~$\micron$ flux density originates from star formation, implying that most of the 24~$\micron$ emission originates from the AGN in the southern nucleus. In this paper we have re-fit the emission lines with the method used for all the other targets to ensure consistency. The results are, or course, in very good agreement with what we found before. With our updated fitting method we identify a third H$\alpha$ line component in a cluster of spaxels covering the nuclear regions of the southern disk. In 8 of these spaxels, two broad components with FWHM of 1500--2500~\kms\ and velocity offsets of 600--800~\kms\ are required to fit the asymmetric wings of the profile. All those spectra in IRAS 19254-7245 exhibiting line widths $>$1500~\kms\ are shown in Fig.~\ref{fig:broad_fits}, and in Fig.~\ref{fig:fitall} we show a simultaneous multi-component fit to all the observed emission lines in the spectrum of the southern nucleus. These line widths are consistent with those found in \citetalias{bendo09}, and as suggested there, the presence of these extremely broad components and their velocity offsets imply the presence of high-velocity outflowing ionized material associated with the southern (Sy2) nucleus. Although we measure a $\Delta v_{\rm max}$ of 1500--2200~\kms\ (Table~\ref{tbl:measured_wind}), emission in the wings of the H$\alpha$-[N\two] complex extends from $-3300$~\kms\ to $+3700$~\kms\ at zero intensity. Both \citet{colina91} and \citet{mihos98} also reported evidence of outflowing material associated with the southern nucleus.

Over the rest of the system, (as we found in \citetalias{bendo09}) the line width of C1 varies from $\sim$100--700~\kms, with the broadest widths seen across the southern disk and close to the nucleus of the northern disk. This implies a very turbulent ISM caused by the intense feedback. Line ratios are generally high, with evidence for ionization from hard radiation sources \citepalias[AGN, shocks; Fig.~\ref{fig:ratio_plots2},][]{bendo09}.

\subsection*{IRAS 19297-0406}
\textit{HST} imaging shows that this object has a complex, compact nucleus, composed of several luminous cores embedded within a common, diffuse envelope \citep{bushouse02}. One obvious large tidal tail extends to the east away from the nucleus. Due to a pointing error, the nucleus of IRAS 19297-0406 fell on the very edge of our VIMOS field-of-view, meaning we only cover the western half of the galaxy (as can be seen in Fig.~\ref{fig:iras19297-0406}). Two H$\alpha$ line components are identified in the nuclear regions. Here line widths of the fainter component (C2) extend up to FWHM=800--900~\kms, but there is little evidence for any radial velocity offset between the line centroids. Nevertheless, these line widths suggest the presence of an outflow, and are surprisingly broad for a system without a previously identified AGN component. \citet{martin05} measured a Na\one\ neutral gas outflow from this system with speeds $>$500~\kms\, and \textit{Herschel}/PACS spectroscopy of the OH 79~$\micron$ line reveals evidence for a slightly slower molecular gas outflow (E.\ Sturm, priv.\ comm).

A radial velocity gradient across the system in the north-south direction implies that it is in the late stages of its merger and is in the process of forming a coherent disk. The measured rotation velocity is $\sim$175~\kms. The [N\two]/H$\alpha$, [S\two]/H$\alpha$ and [O\one]/H$\alpha$ ratios are all significantly depressed in the centre of the disk, with the ratios increasing with radius into the shock region of the diagnostic diagram shown in Fig.~\ref{fig:ratio_plots2}. Electron densities throughout the main disk are high with values $>$300~\cmt, peaking at $\sim$1000~\cmt.

\subsection*{IRAS 20046-0623}
This optically classified SB system was studied in some detail my \citet{murphy01} via optical continuum and H$\alpha$ imaging and Pa$\alpha$ integral field spectroscopy. Through analysis of its morphology and velocity maps, they proposed a merger geometry model, which we have reproduced to scale in Fig.~\ref{fig:iras20046-0623}. They propose that this is a system of two merging disk galaxies: the eastern galaxy is in the background, and it is this galaxy to which the northern tidal tail belongs, while the foreground western galaxy is observed nearly edge-on and could be providing a screen of extinction that obscures emission from the central regions of the eastern galaxy. Indeed, two nuclei are visible in H-band images of this source \citep{dasyra06a}. The tidal tail extending to the north shows appreciable line emission. From their H$\alpha$ imaging, \citet{murphy01} estimate star formation rates of a few solar masses per year. The relatively short length and high surface brightness of the tail support the idea that this system is being viewed soon after its first encounter \citep{murphy01}. 

Our VIMOS data are in agreement with the Pa$\alpha$ data of \citet{murphy01}, although in our observations the H$\alpha$ redshifted radial velocities in the eastern half of the system are not centrally concentrated, but are distributed into two peaks on either side of the disk. One is at the base of the northern tidal tail. The H$\alpha$ line widths are FWHM$<$150~\kms\ in the tail, but increase above 200~\kms\ in the disk and peak at $>$800~\kms\ in a region to the south of the main disk. This is approximately coincident with a peak in all three line ratios. This southern region exhibits a strong radial velocity gradient in the east-west direction in the same sense as the main disk, implying that it is part of the same structure. It is difficult to reconcile the high line widths and their location with the \citeauthor{murphy01} merger model.

In the disk and inner region of the northern tail, two H$\alpha$ line components are observed, and in the central part of the disk, a third component can be identified. An example spectrum and multi-Gaussian fit from this three-component region is shown in the figure inset. An interpretation of what is happening in this region is not straightforward and would require higher resolution observations to disentangle. Components 2 and 3 trace a region of redshifted (C2; 300--350~\kms) and blueshifted (C3; $-50$~\kms) emission with the velocity gradient oriented in the north-south direction. The black outlines on the H$\alpha$ C2 radial velocity map indicate where the third component is identified. At the same location, the C1 component is at its broadest. Taken together, this may suggest the presence of an outflow with velocities of $\sim$200~\kms, which is is consistent with the Na\one\ outflow found by \citet{rupke05a} with a velocity of $\sim$220~\kms. More detailed, higher resolution data are needed to disentangle the complex kinematics of the merger and confirm the presence of an outflow.

\subsection*{IRAS 20414-1651}
This system has been imaged on several occasions from the ground, from optical to near-IR wavelengths \citep{melnick90, duc97, surace00, bushouse02}. \textit{HST} images reveal that this system is composed of a single central body that is highly elongated, with hints of faint, amorphous emission surrounding it in various places. From these previous data it is unclear whether the south-western component, separated from the main galaxy body by $10\farcs5$ ($\sim$18~kpc) is tidally connected or related to the main galaxy disk. Our observations clearly detect this companion in H$\alpha$ (implying strong star formation is taking place), and since the radial velocity of the emission line gas is redshifted by only $\sim$300~\kms\ compared to the main disk, this suggests that the two are connected and interacting. In the I-band the central body has a complex morphology, and no clear sign of any dominant nucleus. In the H-band, however, the morphology is quite regular, showing a distinct central core \citep{bushouse02}. These near-IR observations imply that dust extinction in the main disk is high, and are consistent with the optical starburst classification. Indeed, four out of five mid-IR indicators as seen with \textit{Spitzer} are consistent with there being no AGN contribution \citep{veilleux09a}. 

In H$\alpha$ a clear signature of rotation is seen across main disk. The [S\two]/H$\alpha$ and [O\one]/H$\alpha$ ratios are both significantly depressed in the centre of the disk, but rise into the shock regime (Fig.~\ref{fig:ratio_plots2}) in the outer parts. This is not mirrored in the [N\two]/H$\alpha$ ratio, which remains fairly high across the whole disk. The line ratios in the south-western companion are all low, whereas the gas densities are much higher than in the main galaxy disk.

A region of broad H$\alpha$ emission (250--450~\kms), seen partly in C1 and partly C2, is oriented in a north-south direction across main disk. Just to the west of this, we identify a component blueshifted by 100--200~\kms\ compared to C1 that we have assigned to C2. It is not clear whether this is a signature of outflow or not. Evidence of split line components in the south-western companion are also seen, with velocity separations of 400--500~\kms. This is consistent with the fact that ongoing intense star formation is occurring here, and may imply a superbubble or wind is being blown. \citep{rupke05b} find evidence for a neutral gas outflow from this system via the Na\one\ absorption line, with speeds of $\sim$200~\kms.

\subsection*{IRAS 20551-4250}
IRAS 20551-4250 comprises of a single nucleus, a relatively high surface brightness large-scale tidal tail to the south, and a fainter tidal plume to the north. The quasi-elliptical light profile of the central regions suggests that the violent relaxation of the galaxy collision is almost complete and this system is therefore in an advanced coalescence stage. Ionized gas emission is only detected in the central regions, except for some extended emission to the south-east, at the base of the tidal tail. The disk appears to be in well-ordered rotation with $v\sin i \sim 90$~\kms. \citet{mihos98} suggest that these low rotation velocities indicate that we are viewing the disk largely face-on. All three line ratios are depressed in the central region, as we see in many other systems in our sample.

Much evidence points to this system being at the end of the ULIRG phase: a regular velocity field in the disk, centrally concentrated H$\alpha$, and the fact that population synthesis modelling \citep{johansson91} suggests that the peak in star formation already occurred some $10^7$~yr ago. Point-like hard X-ray sources have been detected with both \textit{XMM-Newton} \citep{franceschini03} and \textit{Chandra} \citep{ptak03}, and those, together with more analyses of \textit{Spitzer}/IRS spectroscopy suggest that this system hosts a highly obscured AGN \citep{farrah07, sani12}, again consistent with the evolved nature of this ULIRG. In our H$\alpha$ maps, we find evidence for fast-flowing gas that is highly suggestive of a turbulent outflow. In a strip across the northern side of the disk, H$\alpha$ line widths range between FWHM = 500--800~\kms, and the region of peak line widths is also blueshifted by 150~\kms\ beyond that expected from pure rotation. The fact that the starburst is no longer active, and the presence of the AGN, strongly implies that the outflow is AGN-powered (despite the velocities being quite low). Indeed, a fast molecular-phase outflow is identified in \textit{Herschel}/PACS spectroscopy (E.\ Sturm, priv.\ comm), supporting the AGN-driven scenario.

The total exposure time allocated for this system was split into two, with a second pointing covering the southern tail obtained with an exposure time of 6000~s. However, no emission lines were detected in the tail region, even after binning 5$\times$5 in the spatial axes.

\subsection*{IRAS 21504-0628}
This object has received very little previous attention in the literature and has never been observed with \textit{HST}. It is classified as a Sy2 although this is uncertain \citep{farrah03}. In optical continuum light, IRAS 21504-0628 appears to have a flattened disk with a very faint tidal arm extending to the north-west and a small extension to the south-west \citep{duc97}. Our H$\alpha$ velocity map indicates ordered disk rotation with $v\sin i \sim 175$~\kms, although there are some obvious distortions in the velocity field. The H$\alpha$ emission in the central disk is much less extended, appearing compact and symmetrical, although both the northern and southern tidal features are also identifiable. The [S\two]/H$\alpha$ and [O\one]/H$\alpha$ line ratios are depressed in the nuclear region, whereas the [N\two]/H$\alpha$ ratio remains fairly high across the whole disk. This is similar behaviour to other galaxies in our sample. Gas densities are also measurably high in the central disk. H$\alpha$ line widths in C1 are generally broad (FWHM = 250--350~\kms), but in one region to the north of the nucleus we find line widths up to $\sim$500~\kms\ where we also see elevated line ratios in all three indicators. In the central region, we identify a cluster of spaxels in which a second (an in the very centre a third) H$\alpha$ component can be fitted. This second component is consistently blueshifted by $\sim$50~\kms\ with respect to C1, but with line widths of $<$100~\kms\ and average line ratios. Although the kinematics and gas properties in this system are clearly complex, none of our findings is obviously suggestive of an outflow.

\subsection*{IRAS 22491-1808}
IRAS 22491-1808 has a complex morphology with tidal features extending towards the north-west, the east and the south. Two nuclei separated by 2'' ($\sim$3~kpc) were distinguished in ground-based K-band imaging \citep{carico90, surace00}. However, \citet{cui01} find three putative nuclei from \textit{HST} I-band imaging, one located 1.2~kpc to the southeast of the central bright nucleus and one 2.4~kpc to the west, and therefore suggest a multiple merger scenario for this system. The many star forming knots found around the central region (preferentially distributed around the western nucleus) contribute significantly to the total bolometric luminosity of the galaxy \citep{surace00}. The eastern nucleus has very warm optical/near-IR colours, suggesting that this nucleus may harbour an AGN \citep{surace00}. Indeed, \citet{farrah03} find evidence in their SED fitting for both a SB and AGN component, where the AGN contributes $>$50\% of total IR luminosity. \citet{franceschini03}, however, find only very weak X-ray emission from this system. Star forming knots are also clearly embedded in the tails.
 
Our H$\alpha$ maps show distorted velocities within the nuclear region, as expected from such a complex merger. The north-western tail is H$\alpha$-bright and shows smooth redward velocity gradient. The southern extension is visible in our H$\alpha$ intensity map, although the eastern tail is not. Broad C2 line widths (FWHM = 300--350~\kms) are found at the base of the north-western tail, suggesting that material here has been compressed during the collision and is now forming stars. The presence of multiple line components in the central regions reflects the complex dynamical state of this system. As we have seen in many of the other objects in our study, the [S\two]/H$\alpha$ and [O\one]/H$\alpha$ line ratios are depressed in the disk but rise on the outskirts. The [N\two]/H$\alpha$ ratio does not follow this trend, and is high on eastern side and low in north-western tail.

\subsection*{IRAS 23128-5919}
This system consists of two merging galaxies, the nuclei of which are separated by a projected distance of 4 kpc \citep[5$''$;][]{zenner93, duc97, charmandaris02}. As a result of the collision, two tidal tails stretch 40~kpc in opposite directions \citep{bergvall85, mihos98}. Based on optical studies, the northern galaxy is classified as a starburst, while it is unclear whether the southern one is a Seyfert, a starburst or a LINER \citep{duc97}. In hard X-rays the southern nucleus is detected as a relatively luminous point-like source therefore indicating the presence of an obscured AGN \citep{franceschini03, ptak03}.

Previous optical spectroscopy of the southern nucleus has shown asymmetric emission line profiles with broad blue wings in a region extending $\sim$5 kpc out from the nucleus \citep{bergvall85}. H$\alpha$ emission associated with the southern galaxy is significantly extended in the east-west direction compared to the I-band morphology. Since spectral features characteristic of Wolf-Rayet stars are also present over the whole central region \citep{johansson88}, these observations have been interpreted as evidence for a starburst driven outflow from the southern nucleus. In their Fabry-P\'erot observations, \citet{mihos98} also found complex H$\alpha$ emission line profiles in the southern galaxy of this pair, with asymmetric shapes and double-line profiles. However, they interpreted the large east-west velocity gradient associated with the southern nucleus as resulting from a disk seen nearly edge-on.

These previous studies, however, did not decompose the emission line profiles into individual components. In our data, the east-west extension associated with the southern nucleus seen in the H$\alpha$ intensity map is characterised by a very broad (FWHM = 600--700~\kms) H$\alpha$ emission component (in our maps assigned to C1). Like \citet{mihos98}, we also find this component exhibits a large east-west velocity gradient spanning $\sim$900~\kms. The corresponding [S\two]/H$\alpha$ and [O\one]/H$\alpha$ line ratios in this strip are low compared to the surrounding material. In the region surrounding the southern nucleus, we also see two narrow components (in our maps assigned to C2 and C3). C2 is at the systemic velocity of the system, while C3 (and C1) are blueshifted. The region in which we see three line components is indicated on the H$\alpha$ C2 radial velocity map with a black ellipse.

Is the region with the large velocity gradient a bipolar wind or a disk? It cannot be a significant disk of stars as it is not visible on the \textit{HST} F814W image, or on older optical continuum imaging \citep{bergvall85, duc97}. It is also very improbable that a large, fast rotating, disk of ionized gas is present, without at least some associated stars visible in the continuum image.
Therefore, we conclude that this is a large bipolar outflow. Evidence in support of this is: (1) large emission line widths coincident with (2) a 900~\kms\ velocity gradient; (3) a lack of continuum emission indicating a stellar disk; and (4) the presence of a narrow line component at the systemic velocity covering the whole inner part of the large velocity gradient region (assigned in our maps to C2) that presumably originates in the diffuse ionized gas associated with the main body of the southern system. Under the outflow interpretation, we are seeing both the red and blueshifted plume of a large-scale bipolar wind with outflow velocities of $>$450~\kms\ (given the unknown inclination). The $\Delta v_{\rm max}$ of 800--1000~\kms\ (Table~\ref{tbl:measured_wind}) and the known presence of an AGN in the southern nucleus together suggest the outflow is AGN driven. Interestingly, no evidence for a molecular gas outflow is seen in \textit{Herschel} OH line spectroscopy (E.\ Sturm, priv.\ comm.).

The aforementioned narrow line component seen in C2 at the location of the outflow plumes matches consistently (in terms of line width and radial velocity) with the C1 component seen away from the bipolar wind. We therefore associate the two as originating in the same material (and were just assigned to the different components as a result of our component assignment convention; see Section~\ref{sect:line_profiles}). Under this assumption, the ionized gas \textit{not} associated with outflow plumes shows no evidence of rotation (even after reexamining the velocity map with a $-100$--+100~\kms\ scaling). The [N\two]/H$\alpha$ and [S\two]/H$\alpha$ line ratios for this systemic velocity component are high, consistent with shock or AGN excitation, and a depression is seen in the [S\two]/H$\alpha$ and [O\one]/H$\alpha$ ratios at the location of both nuclei. Both nuclei also show up in the electron density map with peaks at 400--500~\cmt.

The proximity of the nuclei and complex dynamics led \citet{mihos98} to suggest ``we are most likely witnessing IRAS 23128-5919 in a very short-lived phase where violent relaxation is rapidly altering the structural and dynamical properties of the system''. This system is therefore very worthy of detailed follow-up observations.

\label{lastpage}
\end{document}